\documentclass[floatfix,aps,showpacs,amsmath,amssymb,twocolumn,eqsecnum,superscriptaddress]{revtex4-1}
\let\Twocolumn

\newif\ifTwocolumn
\ifx\Twocolumn\undefined
  \Twocolumnfalse
\else
  \Twocolumntrue
\fi
\ifTwocolumn
\fi
\usepackage{graphicx}
\usepackage{gensymb}
\usepackage{subfigure}
\usepackage{bm}
\usepackage{upgreek}
\usepackage{wasysym}
\usepackage{bibentry}
\usepackage[T1]{fontenc}
\usepackage{textcomp}
\usepackage{mathptmx}
\usepackage[scaled=0.92]{helvet}
\usepackage{courier}
\usepackage{setspace}
\usepackage{color}
\usepackage{epstopdf}
\usepackage{hyperref}
\definecolor{red_n}{rgb}{1.0, 0.0, 0.0}
\definecolor{brown_n}{rgb}{0.6, 0.4, 0.2}
\definecolor{cyan_n}{RGB}{0.0, 255.0, 255.0}
\definecolor{blue_n}{rgb}{0.0, 0.0, 1.0}
\definecolor{green_n}{rgb}{0.0, 0.5, 0.0}
\definecolor{orange_n}{RGB}{255.0, 127.0, 0.0}
\definecolor{magenta_n}{RGB}{255.0, 0.0, 255.0}
\definecolor{purple_n}{RGB}{128.0, 0.0, 128.0}
\definecolor{gray_n}{RGB}{128.0, 128.0, 128.0}
\definecolor{dark_green}{rgb}{0.0, 0.5, 0.0}
\newcommand{\eref}[1]{Eq.~\eqref{#1}}
\ifTwocolumn
  
\else
  
\fi
\begin{document}
\title{Critical and near critical phase behaviour and interplay between the thermodynamic Casimir and van der Waals forces in confined non-polar fluid medium with competing surface and substrate potentials}

\author{Galin~Valchev}
\email[Electronic address: ]{gvalchev@imbm.bas.bg}
\affiliation{Institute of Mechanics-Bulgarian Academy of Sciences, Academic Georgy Bonchev St. building 4, 1113 Sofia, Bulgaria}
\author{Daniel~Dantchev}
\email[Electronic address: ]{daniel@imbm.bas.bg}
\affiliation{Institute of Mechanics-Bulgarian Academy of Sciences, Academic Georgy Bonchev St. building 4, 1113 Sofia, Bulgaria}
\affiliation{Max-Planck-Institut f\"{u}r Intelligente Systeme, Heisenbergstrasse 3, D-70569 Stuttgart, Germany and
IV. Institut f\"{u}r Theoretische Physik, Universit\"{a}t Stuttgart, Pfaffenwaldring 57, D-70569 Stuttgart, Germany}

\date{\today}
%
\begin{abstract}
We study, using general scaling arguments and mean-field type calculations, the behavior of the critical Casimir force and its interplay with the van der Waals force acting between two parallel slabs separated at a distance $L$ from each other, confining some fluctuating fluid medium, say a non-polar one-component fluid or a binary liquid mixture. The surfaces of the slabs are coated by thin layers exerting strong preference to the liquid phase of the fluid, or one of the components of the mixture, modeled by strong adsorbing local surface potentials ensuring the so-called  $(+,+)$ boundary conditions. The slabs, on the other hand, influence the fluid by long-range competing dispersion potentials, which represent irrelevant interactions in renormalization group sense. Under such conditions one usually expects {\it attractive} Casimir force governed by universal scaling function, pertinent to the extraordinary surface universality class of Ising type systems, to which the dispersion potentials provide only corrections to scaling. We demonstrate, however, that below a given threshold thickness of the system $L_{\rm crit}$ for a suitable set of slabs-fluid and fluid-fluid coupling parameters the competition between the effects due to the coatings and the slabs can result in {\it sign change} of the Casimir force acting between the surfaces confining the fluid when one changes the temperature $T$, the chemical potential of the fluid $\mu$, or $L$. The last implies that by choosing specific materials for the slabs, coatings and the fluid for $L \lesssim L_{\rm crit}$ one can  realize {\it repulsive} Casimir force with {\it non-universal} behavior which, upon increasing $L$, gradually turns into an {\it attractive} one described by an {\it universal} scaling function, depending only on the relevant scaling fields related to the temperature and the excess chemical potential, for $L\gg L_{\rm crit}$. We presented arguments and relevant data for specific substances  in support of the experimental feasibility of the predicted behavior of the force. It can be of interest, e.g., for designing nano-devices and for governing behavior of objects, say colloidal particles, at small distances. We have formulated the corresponding criterion for determination of $L_{\rm crit}$. The universality is regained for $L\gg L_{\rm crit}$. We have also shown that for systems with $L \lesssim L_{\rm crit}$ the capillary condensation phase diagram suffers modifications which one does not observe in systems with purely short-ranged interactions.
\end{abstract}
\maketitle
%
\section{Introduction}\label{sec:Inrtoduction}
%
When a fluctuating field is confined by material bodies, effective forces arise on them. This is due to the fact that the bodies impose boundary conditions on the medium, depending  on their geometry, mutual position and material properties, which leads to a modification of the allowed fluctuations in the medium. The last leads to a dependence of the ground state, or the thermodynamic potential of the system (say the free energy) on the geometry of the system and on the distances between its (macroscopic) components. In order to change these distances one has to apply a force that depends on the induced change of the allowed fluctuations. If the fluctuations are long-ranged the corresponding forces are also long-ranged. The existence of such long-ranged fluctuation mediated forces is called the Casimir effect and the corresponding forces -- Casimir like forces \cite{C48,C53,CP48,FG78}, after the Dutch physicist Hendrik Casimir who in 1948 predicted an attractive force between two parallel perfectly conducting metal plates \cite{C48} separated by a finite gap $L$ in vacuum at zero temperature. In order the force to be long-ranged (i.e., to decay in a power-law and not in an exponential with the distance way) the interactions in the system have to be mediated by
massless excitations -- photons, Goldstone bosons, acoustic phonons, etc. Considered in this general form the Casimir effect is a subject of investigations in condensed matter physics, quantum electrodynamics, quantum chromodynamics and cosmology. The results are summarized in an impressive number of reviews \cite{PMG86,MT88,LM93,MT97,M94,KG99,BMM2001,M2001,M2004,L2005,KM2006,GLR2008,BKMM2009,KMM2009,
FPPRJLACCGKKKLLLLWWWMHLLAOCZ2010,OGS2011,KMM2011,RCJ2011,MAPPBE2012,B2012,Bo2012,C2012,CP2011,K94,BDT2000,K99,G2009,TD2010rev,GD2011,D2012}.

When the fluctuating field is the electromagnetic one the effect is known as the quantum \textit{electro-dynamical} (QED) Casimir effect.  There the Casimir force is caused by zero-point and thermal fluctuations of the electromagnetic field. In first approximation, it depends only on the velocity of light $c$, Planck's constant $\hbar$, the temperature $T$, and the separation distance between the bodies $L$, i.e., this force to a great extend is {\it universal}. A more advanced theory, the so-called Lifshitz theory, reveals the dependence of the Casimir force on the material properties of the bodies \cite{DLP61,LP80,L56} and geometry of their boundary surfaces \cite{EGJK2007,KK2008}.

Thirty years after Casimir's prediction, M. E. Fisher and P. G. de Gennes suggested that the fluctuating medium, confined between the bodies can be a fluid, the fluctuating field being the field of its order parameter, in which the interactions in the system are mediated not by photons but by different type of massless excitations like critical fluctuations or Goldstone bosons (spin waves). The corresponding Casimir effect is known as the \textit{thermodynamic} Casimir effect \cite{FG78}. When the confined fluid approaches it's critical point, the corresponding fluctuations are the critical fluctuations of the order parameter and then the effect is usually called \textit{critical} Casimir effect. In first approximation, the thermodynamic Casimir effect depends only on the gross features of the system -- its dimensionality $d$ and the symmetry of the ordered state $n$ (both defining the so-called bulk universality class of the system) and on the boundary conditions (determined by the surface universality classes). Therefore, to a great extend the thermodynamic Casimir force is also \textit{universal}. So far the critical Casimir effect has enjoyed two general reviews \cite{K94,BDT2000} and some concerning specific aspects of it \cite{K99,G2009,TD2010rev,GD2011,D2012}.

Currently the Casimir effect is an object of intensive studies both in its original formulation due to Casimir as well as in its thermodynamic manifestation. In the present article we will report theoretical results dealing with the critical Casimir effect. Let us note, that the critical Casimir effect has been already directly observed, utilizing light scattering measurements, in the interaction of a colloid spherical particle with a plate \cite{HHGDB2008} both of which are immersed in a binary liquid mixture. The effect has been also studied in $^4$He \cite{GC99},\cite{GSGC2006}, as well as in $^3$He--$^4$He mixtures \cite{GC2002} in the context of forces that determine the properties of a film of a substance in the vicinity of its bulk critical point. In Ref. \cite{FYP2005} and Ref. \cite{RBM2007}  one has performed measurements of the Casimir force in thin wetting films of binary liquid mixture. One the theoretical side, the effect has been studied via exact calculations in the two-dimensional Ising model \cite{ES94,NN2008,NN2009,AM2010,RZSA2010,DM2013,I2011,WIG2012,DrzMaBa2011,MaZuDrz2013}, the three dimensional spherical model \cite{D96,D98,DG2009,DDG2006,CD2004,DR2014,DBR2014,DGHHRS2012,DGHHRS2014}, via conformal-theoretical methods \cite{A86,BCN86,BE95,ER95,HSED98,BEK2015,DSE2015}, within mean-field type calculations on Ising type \cite{K97,GaD2006,DSD2007,PE92,VMD2011} and $XY$ models \cite{BDR2011}, through renormalization-group studies via $\varepsilon$-expansion \cite{KD92a,KD92b,DGS2006,GD2008,SD2008,DS2011,D2009,D2013} and via fixed dimension $d$ techniques \cite{D2009,Do2011,D2013} of $O(n)$ models, as well as via Monte-Carlo calculations \cite{KL96,DK2004,H2007,H2009,H2010,H2011,H2013,Has2015,VED2013,VGMD2007,VGMD2009,MoMaDi2010,HGS2011,H2012,VasDit2013}. The fluctuation of importance in all of the above mentioned models are of thermal origin since all these models possess  non-zero critical temperature. In some systems, however, certain quantum parameters govern the
fluctuations near their critical point which is usually close to or at the zero temperature \cite{SK2011,S2008,S2000,Sa2011}.   In this particular case one speaks of a \textit{quantum critical} Casimir effect \cite{CDT2000,BDT2000,PCC2009}.

The rapid progress in nanotechnology has resulted in the growth of interest in fluctuation-induced phenomena, which play a dominant role between neutral non-magnetic objects at short separation distances (below a micrometer). The van der Waals and Casimir forces, both known under the generic name \textit{dispersion forces}, play a key role in Micro- and Nano-Electromechanical Systems (MEMS/NEMS) \cite{CAKBC2001a,DBKRCN2005,BEBDPS2012} operating at such distances. Indeed, upon scaling down devices, the dispersion forces can induce some usually undesirable non-linear behaviors in such systems \cite{CAKBC2001}. Irreversible phenomena appear such as stiction (i.e., irreversible adhesion) or pull-in due to mechanical instabilities \cite{BR2001,BR2001a}. Therefore the ability to modify the Casimir interaction can strongly influence the development of MEMS/NEMS. Several theorems seriously limit, however, the possible search of repulsive Casimir forces \cite{KK2006,S2010,RKE2010}. Currently, apart from some suggestions for achieving Casimir repulsion in systems out of equilibrium \cite{APSS2008,B2009,BEKK2011,KOS2013,MA2011,KOS2014,KEBK2011,BCCKMM2012,CBKMM2011,KMM2011}, the only experimentally well verified way to obtain repulsive Casimir force is to have interaction between two different materials characterized by dielectric permittivities $\varepsilon_1$ and $\varepsilon_2$ such that \cite{L56,DLP61,LP80}
\begin{equation}
\label{permit}
\varepsilon_1<\varepsilon_M<\varepsilon_2
\end{equation}
along the imaginary frequency axis, with $\varepsilon_{M}$ being the dielectric permittivity of the medium in between them. In Refs. \cite{MML96,MLB1997,LS2001,LS2002,MCP2009,IIIM2011} Casimir repulsion was indeed observed experimentally for the sphere-plate geometry.

In the current article we study  the interplay between the critical Casimir force and the van der Waals one in a system composed out of two flat parallel slabs both immersed in a critical fluid. Let us note that both the critical Casimir and van der Waals forces are fluctuation induced ones but due to the fluctuations of different entities. For terminological clarity, let us also remind that in colloid sciences fluid mediated interactions between two surfaces or large particles are usually referred to as solvation forces \cite{E90,E90book,ES94}. Thus, we study here a particular case of such a force when the fluid is near its critical point. In our system we suppose that the slabs are coated by thin layers of some substances, confining either a non-polar one-component fluid or a non-polar binary liquid mixture. We suppose that the liquid phase of the one-component fluid or one of the components of the binary liquid mixture are strongly adsorbed by {\it both} coating layers, i.e., they ensure the so-called $(+,+)$ boundary conditions. The slabs, on the other hand, influence the fluid by long-range competing dispersion potentials. In the case of a simple fluid these potentials increase the adsorption of the fluid near one of the surfaces, leading to preference there of its liquid phase, and decrease it near the other one. In the case of a binary fluid mixture the substrates prefer one of the components near the top and the other one near the bottom of the system. We will demonstrate that this  experimentally realizable competition between the effects due to the coatings and the slabs can result in interesting effects like sign change of the Casimir force, acting between the surfaces confining the fluid when one changes $t$, $\mu$ or $L$. The last facts can potentially be used in designing nanodevices and for governing the behavior of objects at small, below micrometer, distances.

If a fluid system possesses a surface it breaks the spatial symmetry of the bulk system. The quantitative effects of the presence of a surface on the thermodynamic behavior of the system depends on the penetration depth of this symmetry breaking effect into the volume. There are two phenomena which increase the surface effects: long-range interactions and long-range correlations. They can act separately, or simultaneously, which leads to an interesting interplay of the effects due to any of them \cite{DRB2007,DRB2009,DSD2007,MDB2004,BHLD2007}. The penetration depth due to the correlations is set by the correlation length $\xi$ of the order parameter of the system; $\xi$ becomes large, and theoretically diverges, in the vicinity of the bulk critical point $(T_c,\mu_c)$: $\xi(T\to T_c^{+},\mu=\mu_c)\simeq \xi_0^{+}t^{-\nu}$, $t=(T-T_c)/T_c$, and $\xi(T=T_c,\mu\to\mu_c)\simeq \xi_{0,\mu} |\Delta\mu/(k_B T_c)|^{-\nu/\Delta}$, $\Delta \mu=\mu-\mu_c$, where $\nu$ and $\Delta$ are the usual critical exponents. If the system is made finite, e.g., by the introduction of a second wall the behavior of the fluid is further enriched. When $\xi$ becomes comparable to the characteristic system size, say $L$, the size dependence of thermodynamic functions enters through the ratio $L/\xi$, i.e., takes a scaling form given by the finite-size scaling theory \cite{Ba83,Bb83,C88,Ped90,PE90,BDT2000} that incorporates, {\it inter alia}, shift of the critical point of the system \cite{NF83,FN81,BLM2003,NAFI83,REUM86}. Below $T_c$ if the confining walls of the film geometry consist of the same material, one encounters the phenomenon of capillary condensation \cite{E90,REUM86,OO2012,DSD2007,YOO2013} where the confinement of the fluid causes, e.g., the liquid vapor coexistence line to shift away from the coexistence line of the bulk fluid into the one-phase regime.
We will demonstrate in the current article that in the envisaged realization of our system depending on the material properties of the slabs the phase diagram of the finite system might essentially differ from that one of the well studied case of a system with short-ranged type interactions and strongly adsorbing surfaces.

The article is arranged as follows. In Sec. \ref{sec:ThermalCas} we recall and comment on the finite-size behaviour of systems with dispersion forces extending the known facts to the expected behavior of the Casimir and the net forces when they act between walls being semi-infinite slabs coated by some thin substances. By doing so, we especially pay attention to the conditions under which the effects stemming from these interactions are relevant. Section \ref{sec:Model} presents  the corresponding lattice gas models suitable for the investigation of fluid media with account of the long-ranged van der Waals interactions. Here we identify the main coupling parameters characterizing the interactions in the systems and in Sec. \ref{sec:FinSizeBehav} the equation for the equilibrium profile of the finite-size order parameter is obtained, which we later use to calculate the forces of interest. Section \ref{sec:Results} presents the numerical results for the behaviour of the investigated forces followed by Sec. \ref{sec:PhaseBehaviour} where the phase behaviour of the considered type fluid system is briefly discussed. The experimental feasibility of the predicted effects is discussed in Sec. \ref{sec:ExperimentalReal}. Here we also comment on the possible application of our findings in design of nanodevices and for governing the behavior of objects, say colloidal particles, at small distances. The article ends with a summary and discussion section -- Sec. \ref{sec:DisandConcRem}. Important technical details concerning the Hamaker term for a van der Waals system of two different substances separated by a fluid medium are presented in Appendix \ref{sec:HamakerTerm}.
%
\section{The thermodynamic Casimir force in a non-polar fluid film systems with dispersion forces}\label{sec:ThermalCas}
%
\begin{figure}[t]
\centering
\includegraphics[width=\columnwidth]{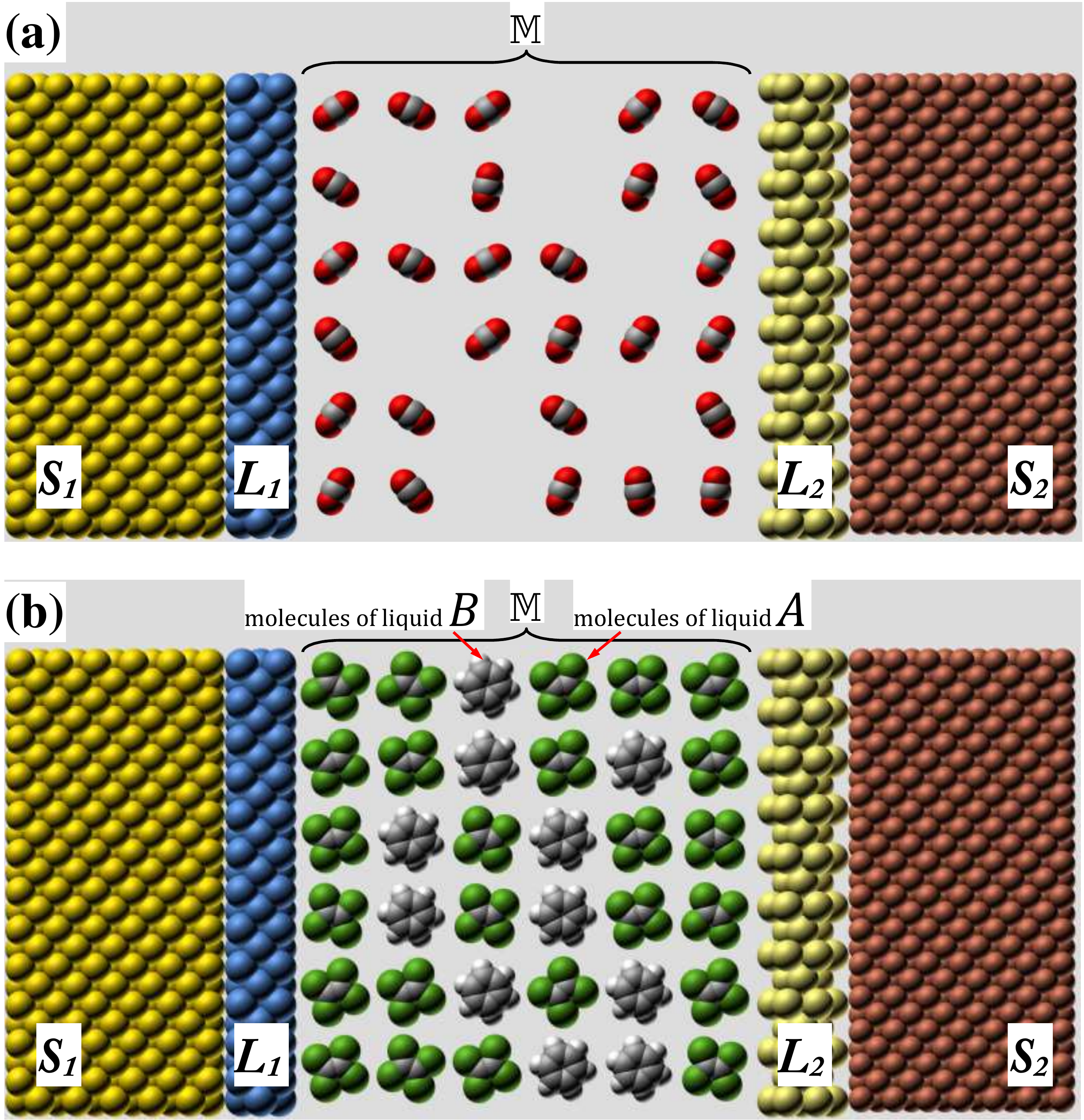}
  \caption{Schematic depiction of a finite-size fluid system, consisting of two parallel slabs of some substances $\mathbb{S}_{1}$ and $\mathbb{S}_{2}$, coated by thin layers of some other substances $\mathbb{L}_{1}$ and $\mathbb{L}_{2}$, respectively, confining some fluid medium $\mathbb{M}$ -- a non-polar one-component fluid [as an example we depicted the carbon dioxide molecules $(\mathrm{CO}_{2})$] $(\mathbf{a})$ or a binary mixture $(\mathbf{b})$ composed out of the molecules of the non-polar liquids $A$ and $B$ [the depicted examples include tetrachloroethene $(\mathrm{C}_{2}\mathrm{Cl}_{4})$ -- as the substance $A$, and benzene $(\mathrm{C}_{6}\mathrm{H}_{6})$ -- as the substance $B$]. The confined fluid medium is considered embedded on a lattice in which $(\mathbf{a})$, some nodes are occupied by a particle and others are not -- thus depicting the "liquid" and "gas" states respectively at some values of fluid temperature $T$ and chemical potential $\mu$, or $(\mathbf{b})$ some of the nodes are occupied by a molecule from the substance $A$ (the "liquid" state) and the rest are occupied by the molecules belonging to the species $B$ ("gas" state). The confining walls impose on the fluid medium boundary conditions of strong adsorption on the coating layers, i.e., the nearest to the coating substances layers are entirely occupied by the particles of the one-component fluid or if the medium is a binary liquid mixture -- by the particles of one of its components (in the presented figure we choose the molecules of the specie $A$).}
  \label{fig:ocf_and_blm}
\end{figure}
Let us consider some fluid medium confined between two parallel slabs of some materials $\mathbb{S}_{1}$ and $\mathbb{S}_{2}$. Any of the slabs is coated by thin solid films of some other substances $\mathbb{L}_{1}$ and $\mathbb{L}_{2}$, respectively (see Fig. \ref{fig:ocf_and_blm}). Let the slabs are situated at some distance $L$ from each other. We suppose that the thicknesses of the coating films are negligible. In the remainder of the text we are going to designate each slab and the thin solid film that coats it as a "wall", and refer to any of its two components separately only when this is necessary.

If the fluid medium is in contact with a particle reservoir with a chemical potential $\mu$, the grand canonical potential $\Omega_{\mathrm{ex}}^{(\tau)}(L|T,\mu)$ of this medium in excess to its bulk value $\mathcal{A}L\omega_{\mathrm{bulk}}(T,\mu)$ depends on $L$ and, thus, one can define the effective force $F_{\mathrm{tot}}^{(\tau)}(L|T,\mu)$ per cross sectional area $\mathcal{A}$ and $k_{B}T$, due to the fluctuations of the medium and dispersion interactions in it as
\begin{equation}\label{def}
\beta F_{A,{\rm tot}}^{(\tau)}(L|T,\mu)\equiv f_{\mathrm{tot}}^{(\tau)}(L|T,\mu)=-
\beta\dfrac{\partial\omega_{\mathrm{ex}}^{(\tau)}(L|T,\mu)}{\partial L},
\end{equation}
where the superscript $\tau$ designates the boundary conditions which the confining walls impose on the fluid medium (see above), $\omega_{\mathrm{ex}}^{(\tau)}(L|T,\mu)=\omega^{(\tau)}(L|T,\mu)
-L\omega_{\mathrm{bulk}}(T,\mu)=\Omega_{\mathrm{ex}}^{(\tau)}(L|T,\mu)/\mathcal{A}$ is the excess grand canonical potential per unit area $\mathcal{A}$, $\Omega^{(\tau)}(L|T,\mu)=\mathcal{A}\omega^{(\tau)}(L|T,\mu)$ is the total grand canonical potential, $\omega_{\mathrm{bulk}}(T,\mu)$ is the density of the bulk grand canonical potential, and $\beta=1/(k_{B}T)$ \cite{note1}.  Let us stress that, as pointed out in Ref. \cite{DSD2007}, one should keep in mind that the force $f_{\mathrm{tot}}(L|T,\mu)$, see \eref{def},  depends on how one defines the thickness of the film. This implies that a quantitative comparison between experimental data and theory is only possible if the data are accompanied by a precise definition of what $L$ is.

Away from the critical temperature of the system it is customary to write the force acting between the plates of the fluid system in the form
\begin{equation}
\label{way}
f_{\mathrm{tot}}(L|T,\mu) \simeq (\sigma-1)\beta H_A(T,\mu) L^{-\sigma}\xi_{\mathrm{ret}}^{\sigma-d},
\end{equation}
where one normally considers the case $d=\sigma$ and omits the apparent dependence on the so-called {\it retardation length} \cite{GC99,DV2012} $\xi_{\mathrm{ret}}$. Here $H_{A}$ is the Hamaker term, whose dependence from the temperature and chemical potential is given by the so-called Hamaker constant \cite{P2006,I2011}
\begin{equation}\label{HamConstant}
A_{\rm Ham}(T,\mu)=-12\pi H_{A}(T,\mu).
\end{equation}
The Hamaker constant, as it is clear from above, is a constant only in the sense that it is $L$-independent. It depends on the temperature, chemical potential and on the material properties of the fluid medium and the walls. The factor $12\pi$ in \eref{HamConstant} is introduced there due to historical reason, according to which the interaction energy between two substrates at a finite separation $L$ in the case of standard van der Waals interaction (i.e., $d=\sigma=3$), away from any phase transition region, is \cite{P2006,I2011}
\begin{equation}\label{deltaomegaincomprfluid}
\omega_{\rm{ex}}(L|T,\mu)=-\dfrac{1}{12\pi}A_{\rm Ham}(T,\mu) L^{-2}.
\end{equation}
The Hamaker term takes into account the leading $L$-dependent parts of the {\it i)} direct interaction between the slabs $A_{s_1,s_2}$, {\it ii)} between each slab and the fluid medium -- $A_{s_{1},l}$, and $A_{s_{2},l}$, as well as the {\it iii)} interactions between the portion of constituents of the fluid medium situated within the cavity bounded by the substrates -- $A_l$, i.e.,
\begin{eqnarray}
H_A(T,\mu)=A_l(T,\mu)&&+A_{s_1,s_2}(T)\nonumber\\&&+A_{s_{1},l}(T,\mu)+A_{s_{2},l}(T,\mu).\label{Ham_l}
\end{eqnarray}
Note that in Eqs. (\ref{way}) -- (\ref{Ham_l}) both the slabs and the fluid medium are characterized by their {\it bulk} properties at the given temperature and chemical potentials.

Near the critical temperature $T_c$ of the bulk system \eref{way} is no longer valid since the critical fluctuations of the order parameter lead to new contribution to the total force called thermodynamic (critical) Casimir force (see below). For such a system, following Ref. \cite{DSD2007}, near the bulk critical point \eref{def} can be written in the form
\begin{eqnarray}\label{CasimirF_l}
f_{\mathrm{tot}}(L|T,\mu) &&\simeq L^{-d} X_{\rm crit}\left[x_{t},x_{\mu},
x_{l},\left\{x_{s_{i}},i=1,2\right\},x_{g}\right]\nonumber\\
&&+(\sigma-1)\beta H_A(T,\mu) L^{-\sigma}\xi_{\mathrm{ret}}^{\sigma-d}.
\end{eqnarray}
In \eref{CasimirF_l} $X_{\rm crit}$ is dimensionless, universal scaling function, $x_{t}=t\left(L/\xi_{0}^+\right)^{1/\nu}$ and $x_{\mu}=\beta\Delta\mu\left(L/\xi_{0,\mu}\right)^{\Delta/\nu}$ are the  temperature and field relevant scaling variables, respectively, while $x_{l}=\Lambda\left(L/\xi_{0}^+\right)^{-\varpi_l}$, $x_{s_{i}}=s_{i}\left(L/\xi_{0}^+\right)^{-\varpi_{s}},\ i=1,2$ and $x_{g}=g_{\varpi}\left(L/\xi_{0}^+\right)^{-\varpi}$ are the irrelevant in renormalization group sense scaling variables associated with the interactions in the system. The freedom of choosing the precise definition of what $L$ is in systems with boundaries leads to the formal necessity to write $L$ as $L+L_0$, with $L_0$ being a microscopic length. That will lead to further scaling corrections proportional to $L^{-1}$. Since in the current article we will keep in all calculated quantities only their leading $L$ dependence, we refrain from further refinement of the scaling Ansatz  \eqref{CasimirF_l}.   By comparing \eref{way} and \eref{CasimirF_l} one immediately concludes that $X_{\rm crit}$ tends to zero away from the critical point, i.e., when at least one of the relevant scaling parameters $|x_t|$ and $|x_\mu|$ becomes large, i.e., when $|x_t|\gg 1$ and/or $|x_\mu|\gg 1$.

As it is well known, the critical behavior of simple fluids and of binary liquid mixtures is described within the Ising universality class which determines the values  of the critical exponents $\Delta\equiv\beta+\gamma$ and $\nu$ ones the dimensionality of the system $d$ is fixed. When $d=3$ this universality class is characterized by critical exponents \cite{ZJ2002}
\begin{equation}\label{3dIsing_exponents}
\nu=0.630, \beta=0.327, \gamma=1.237,\theta=\varpi\nu= 0.524.
\end{equation}
In order to better reflect the actual properties of the non-polar fluids, instead of considering nearest-neighbour interactions we assume long-ranged pair ones between the fluid particles, decaying asymptotically $\sim J^{l}r^{-d-\sigma}$ for distances $r$ between each other, and substrate potentials $\sim J^{s_{i},l}z^{-\sigma},\ i=1,2$ acting on the fluid particles at a distance $z$ from each of the two slabs. We recall that when $\sigma>2$ systems governed by such long-range interactions, usually termed subleading long-ranged interactions \cite{DR2001,D2001}, also belong to the Ising universality class characterized by short-ranged forces \cite{PT77}. The last implies, among the others, that the critical exponents, e.g., do not depend on $\sigma$ for such type of interactions. An important representative of such type of interactions are the non-retarded dispersion interactions with $d=\sigma=3$, which are one of the three types of van der Walls interactions. By varying the ratio between the strengths of the long-ranged -- $J^{l}$ and the short-ranged -- $J_{{\rm sr}}^{l}$ contributions one can quantitatively probe the importance of the long-ranged parts of the interactions within the fluid medium and study potential experiments in colloidal systems which allow for a dedicated tailoring of the form of the effective interactions between colloidal particles.

In \eref{CasimirF_l}, $\varpi$ is the standard correction-to-scaling exponent for short-range systems, while $\varpi_l=\sigma-(2-\eta)$ and $\varpi_{s}=\sigma-(d+2-\eta)/2$ are the correction-to-scaling exponents due to the long-range parts of the interaction potentials between the constituents of the fluid medium and those of the confining walls. Further $L-$dependent contributions to the total forces $f_{\rm tot}$ such as next-to-leading order contributions to the Hamaker terms or higher order corrections to scaling are neglected because they are smaller than those captured in \eref{CasimirF_l}. The exponent $\eta$, which appears in the expressions for $\varpi_l$ and $\varpi_{s}$, is the standard one characterizing the decay of the bulk two-point correlation function at the critical temperature, $g_{\varpi}$ is the (dimensionless) scaling field associated with the Wegner-type corrections, while $\Lambda$ and $s_{i},\ i=1,2$ are dimensionless non-universal coupling constants: $\Lambda$ is proportional to the strength $J^l$ of the long-range part of the interaction potential between the particles of the fluid, whereas $s_{i},\ i=1,2$ are proportional to the contrast between the potentials of the bounding slabs and those in the fluid medium (see below). For systems belonging to three-dimensional Ising universality class  with "genuine" non-retarded van der Waals interaction one has $d=\sigma=3$, and $\eta=0.03627(10)$ \cite{ZJ2002}.  This leads to $\varpi_{l}\simeq 1.03$, $\varpi_{s}\simeq0.52$, and $\varpi=0.832(6)$ \cite{ZJ2002}. Within the mean-field theory with $d=\sigma=4$ and $\eta=0$ one has, instead, $\varpi_{l,\mathrm{MF}}=2$ and $\varpi_{s,\mathrm{MF}}=1$. One then has
\begin{equation}\label{totalforcemeanfieldgeneral}
f_{\rm tot}(L|T,\mu)L^{4}\simeq X_{\rm crit}(\cdot)+3\beta H_{A}(T,\mu),
\end{equation}
where all critical exponents take their mean-field values $\beta=\nu=1/2$, $\gamma=1$, $\Delta=3/2$.

\begin{figure*}[t!]
\centering
\mbox{\subfigure{\includegraphics[width=\columnwidth]{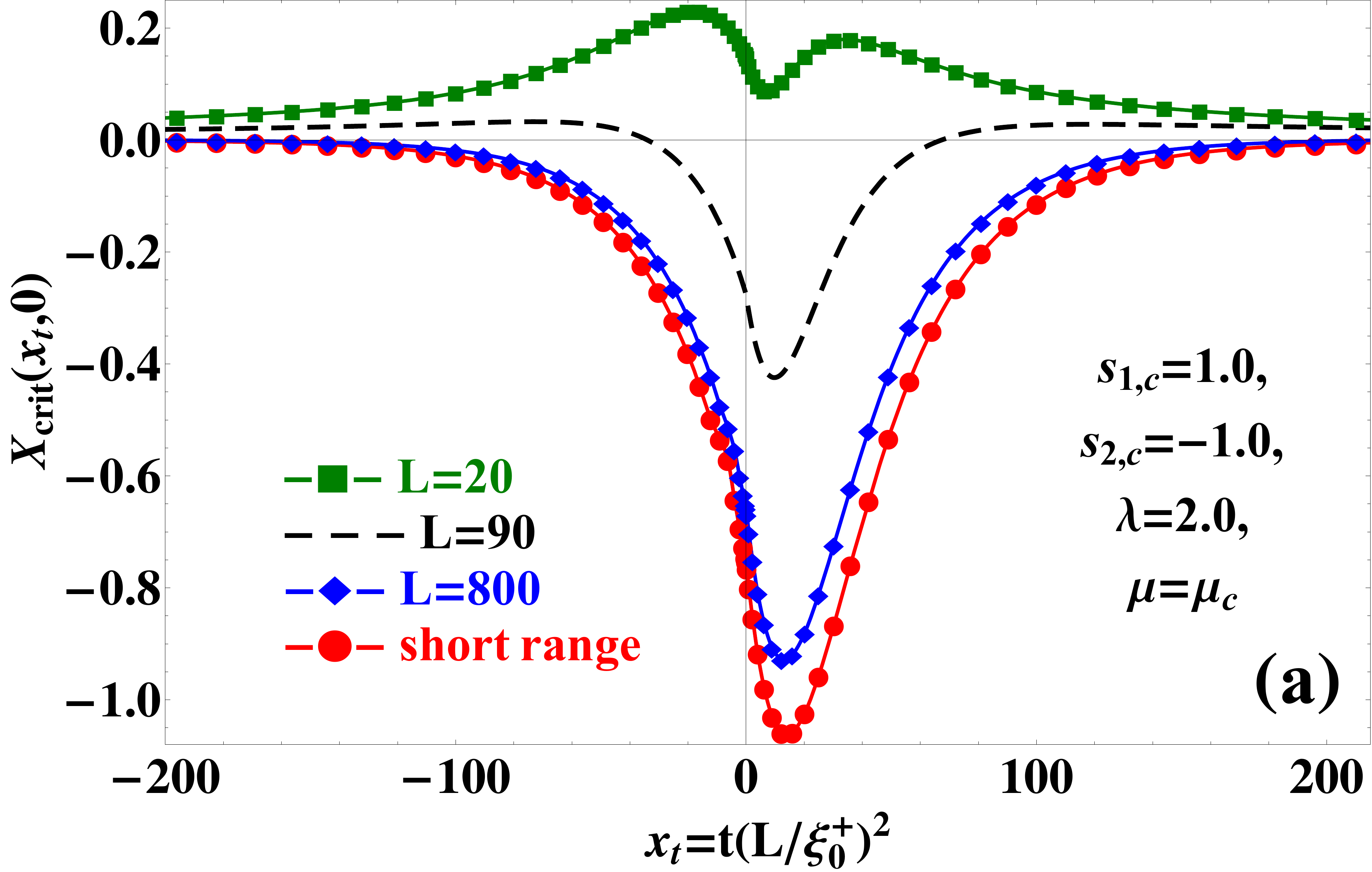}}\quad
      \subfigure{\includegraphics[width=\columnwidth]{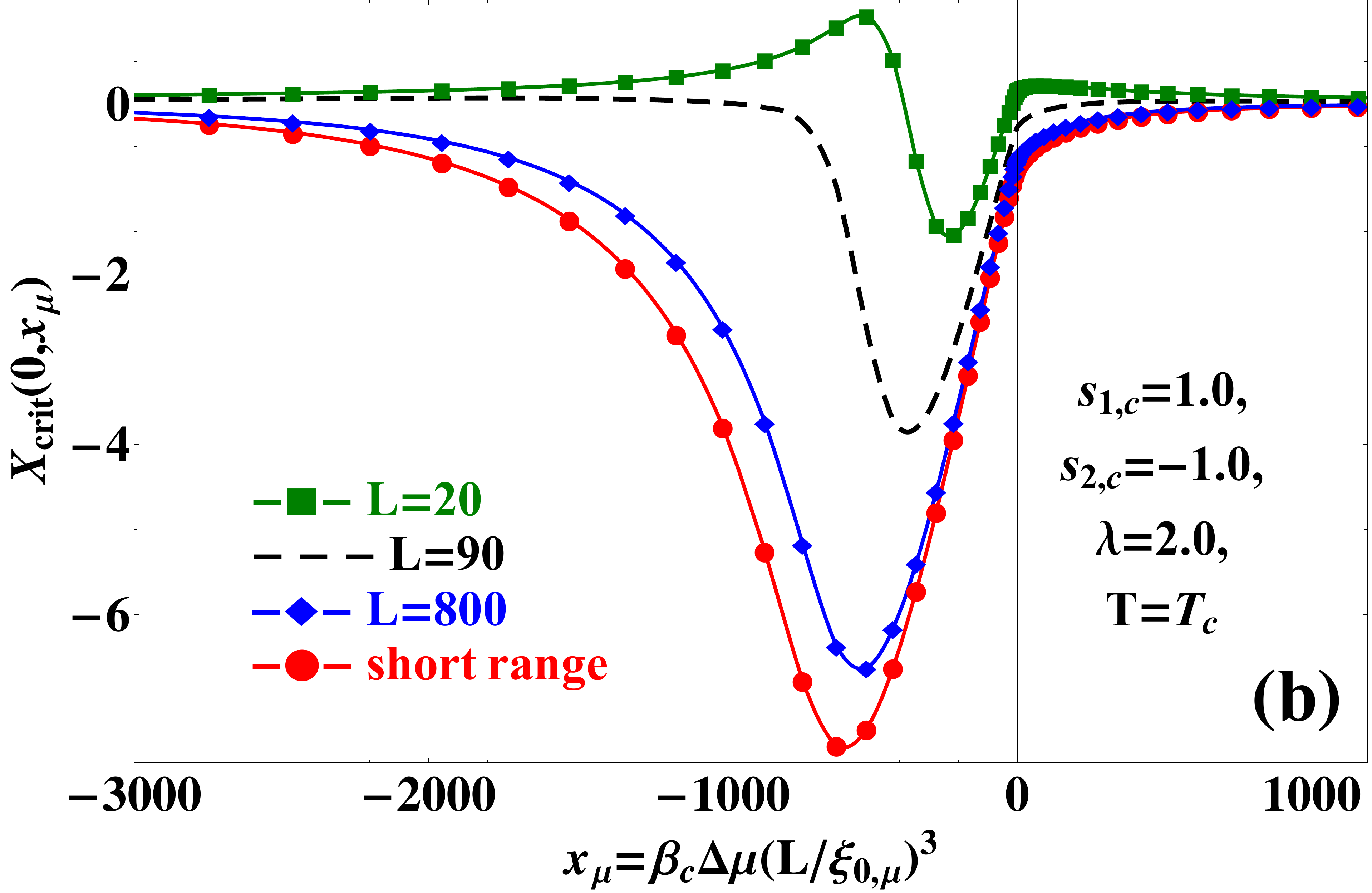}}}\\
\mbox{\subfigure{\includegraphics[width=\columnwidth]{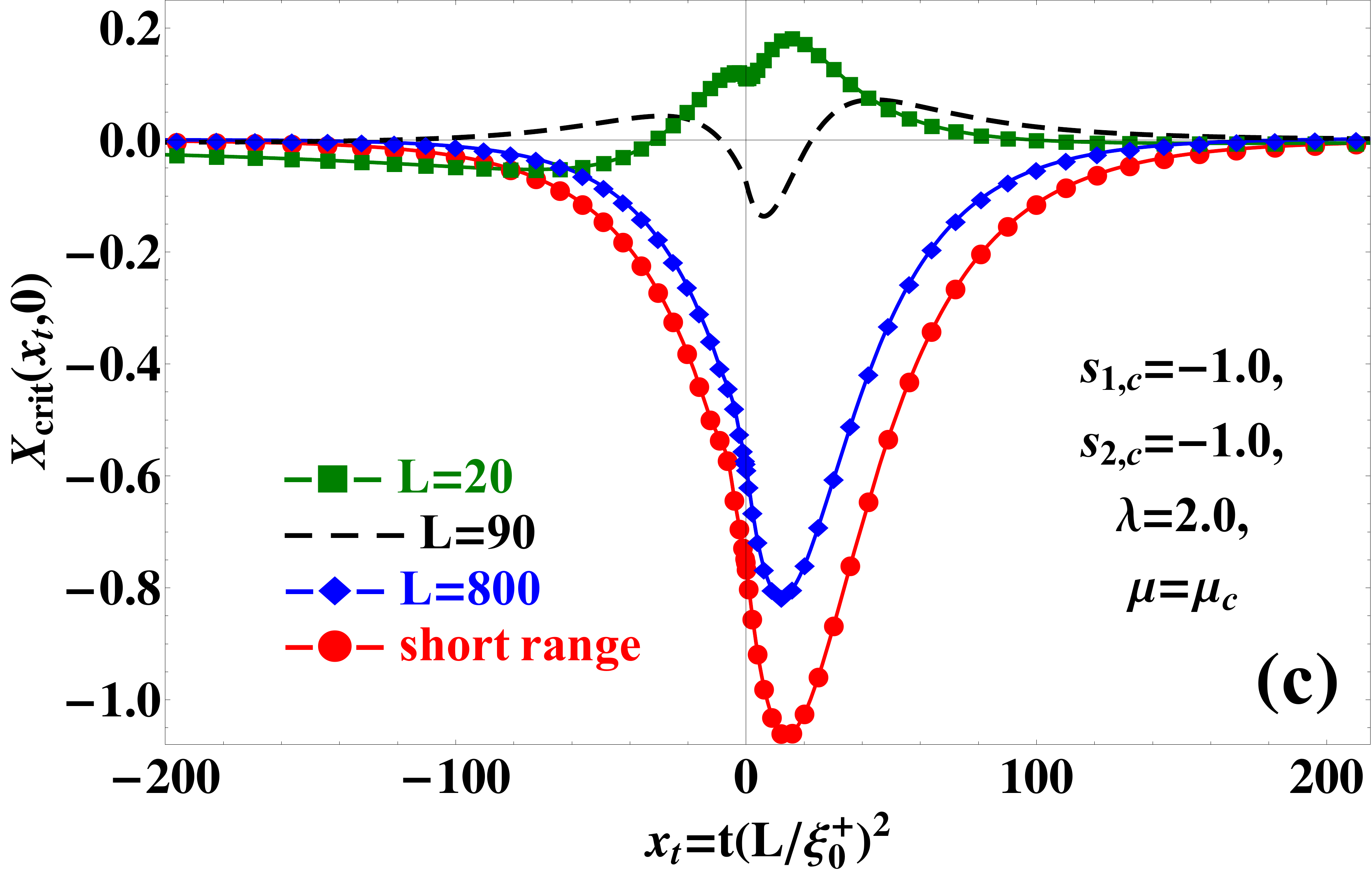}}\quad
      \subfigure{\includegraphics[width=\columnwidth]{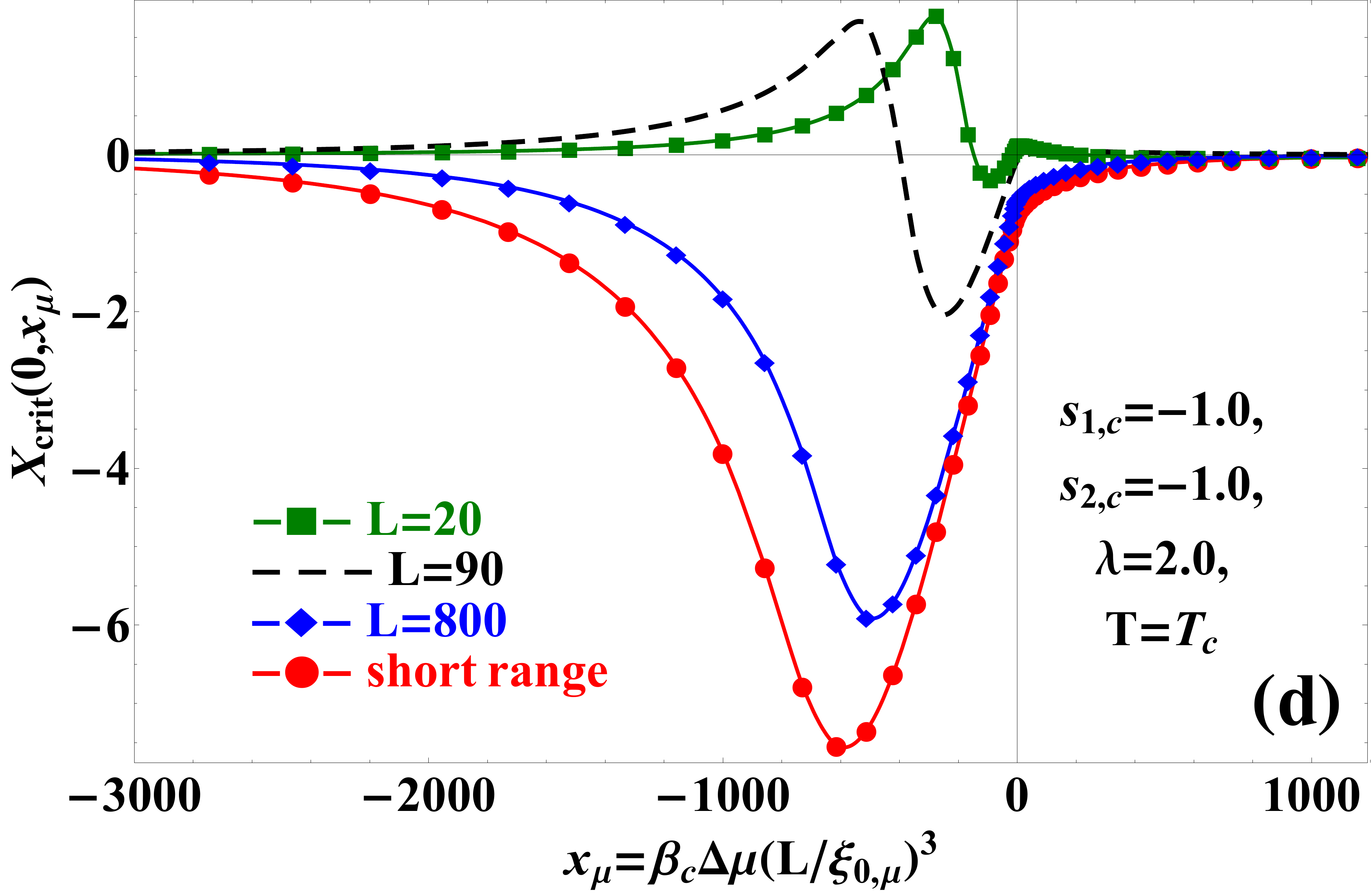}}}
  \caption{Behaviour of the scaling function $X_{\rm crit}(x_{t},x_{\mu})$ in a $d=3$ dimensional confined fluid system. On $\mathbf{(a)}$ and $\mathbf{(b)}$ the parameters reflecting the substrate-fluid interactions have values $s_{1,c}=1.0$ and $s_{2,c}=-1.0$, while on $\mathbf{(c)}$ and $\mathbf{(d)}$ we took $s_{1,c}=s_{2,c}=-1.0$. In all four cases the parameter reflecting the role of the long-ranged fluid-fluid interaction is $\lambda=2.0$. The considered separations between the confining walls are $L=20$ ({\color{dark_green}{{{\Large-}\hspace{-0.1cm}{\Large-}\hspace{-0.1cm}\small$\blacksquare$\hspace{-0.09cm}{\Large-}\hspace{-0.1cm}{\Large-}}}}), $L=90$ $(---)$ and $L=800$ ({\color{blue}{{\Large-}\hspace{-0.1cm}{\Large-}\hspace{-0.1cm}$\blacklozenge$\hspace{-0.09cm}{\Large-}\hspace{-0.1cm}{\Large-}}}). In  $\mathbf{(a)}$ and $\mathbf{(c)}$ the functional dependence on the temperature scaling variable $x_{t}$ at $\Delta\mu=0$ is visualized while in $\mathbf{(b)}$ and $\mathbf{(d)}$ the dependence on the field one $x_{\mu}$ at $T=T_{c}$ is presented. As it can be seen for $L\gg L_{{\rm crit}}\simeq 90$, the van der Waals interactions between the walls and the fluid medium become irrelevant and the scaling function for a fluid system with $L=800$ ({\color{blue}{{\Large-}\hspace{-0.1cm}{\Large-}\hspace{-0.1cm}$\blacklozenge$\hspace{-0.09cm}{\Large-}\hspace{-0.1cm}{\Large-}}}) is very similar in shape to a short-range one ({\color{red}{{\Large-}\hspace{-0.1cm}{\Large-}\hspace{-0.1cm}{\large$\bullet$}\hspace{-0.09cm}{\Large-}\hspace{-0.1cm}{\Large-}}}). Due to the positive substrate-fluid parameter in cases $\mathbf{(a)}$ and $\mathbf{(b)}$ this effect is more pronounced than in $\mathbf{(c)}$ and $\mathbf{(d)}$. For smaller separations $L=20$ ({\color{dark_green}{{\Large-}\hspace{-0.1cm}{\Large-}\hspace{-0.1cm}{\small$\blacksquare$}\hspace{-0.09cm}{\Large-}\hspace{-0.1cm}{\Large-}}}) the effect of the long-range van der Waals interactions becomes relevant everywhere, including the critical region of the system where the scaling function changes it's sign twice [see $\mathbf{(b)}$ and $\mathbf{(c)}$] or even three times [see $\mathbf{(d)}$]. In case $\mathbf{(a)}$ we observe that while for $L=20$ ({\color{dark_green}{{\Large-}\hspace{-0.1cm}{\Large-}\hspace{-0.1cm}{\small$\blacksquare$}\hspace{-0.09cm}{\Large-}\hspace{-0.1cm}{\Large-}}}) as a function of $x_t$ at $x_\mu=0$ the corresponding force is everywhere repulsive, it becomes under enlarging $L$ everywhere attractive, say, for $L=800$ ({\color{blue}{{\Large-}\hspace{-0.1cm}{\Large-}\hspace{-0.1cm}$\blacklozenge$\hspace{-0.09cm}{\Large-}\hspace{-0.1cm}{\Large-}}}), as in the case of a system with completely short-ranged interactions ({\color{red}{{\Large-}\hspace{-0.1cm}{\Large-}\hspace{-0.1cm}{\large$\bullet$}\hspace{-0.09cm}{\Large-}\hspace{-0.1cm}{\Large-}}}). When $\mu=\mu_c$ the curve for the short-ranged interactions depicts the numerical evaluation of the exact analytical result given by \eref{XCassranalytic}, while for $T=T_{c}$ this curve is evaluated within the presented mean-field theory.}
  \label{fig:short_range_tending}
\end{figure*}

The peculiarities of the scaling theory for systems with dispersion forces are described in Refs. \cite{DR2001,D2001,DRB2007,DRB2009,DDG2006,DSD2007}. One obtains that despite these systems do belong to the Ising universality class with short-ranged forces, the finite-size quantities decay algebraically with $L$ towards their bulk values, and not in an exponential in $L$ way, when $\xi \ln (\xi/a_{0})\gg L$ even for $\xi,L \gg a_{0}$, where $a_{0}$ is the characteristic distance between the molecules of the fluid system. In this regime the dominant finite-size contributions to the free energy and to the force between the walls bounding the system stem from the long-ranged algebraically decaying parts of the interaction potentials. One can formulate a criterion clarifying when the long-ranged tails of the interactions can not be disregarded even in the critical region of the finite system. Neglecting the thickness of the coating layers, for the system under consideration the corresponding criterion states that the long range tails of the interactions can be disregarded only when \cite{DSD2007,DRB2009,DDG2006}
\begin{eqnarray}\label{conditionL_l_ab}
2^{\sigma}\left(|s_{1}|+|s_{2}|\right)\left[L/\xi_{0,\mu}\right]^{\Delta/\nu-\sigma}\ll 1.
\end{eqnarray}
Using the scaling relations, the above can be formulated as a constraint on the thickness of the system under study. One obtains that when $L\gg L_{\rm crit}$, where
\begin{equation}\label{lcrit}
L_{{\rm crit}}=\xi_{0,\mu}\left[2^{\sigma}\left(|s_{1}|+|s_{2}|\right)\right]^{\nu/\beta},
\end{equation}
the effects of the long-ranged tails of the interaction can be neglected within the critical region of the finite system. For $d=\sigma=3$ and with $\beta$ and $\nu$  from Eq.  \eqref{3dIsing_exponents} for the three-dimensional Ising model one obtains
\begin{equation}\label{cL1}
L_{{\rm crit}}\simeq 55 \xi_{0,\mu}\left(|s_{1}|+|s_{2}|\right)^{1.923}.
\end{equation}
We conclude that for moderate values of $L$, i.e., when $L\lesssim L_{{\rm crit}}$ the behavior of the Casimir force can strongly depend on the details of the fluid-fluid [the amplitude $\xi_{0,\mu}$ depends on the details of the fluid-fluid interaction (see Eqs. (4.15) and (4.17) in Ref. [\onlinecite{DSD2007}]  and the text therein)] and substrate-fluid interactions which, as it turns out, can influence even the {\it sign} of the force. For large $L$, i.e., for $L\gg L_{\rm crit}$ the behavior of the force shall approach the one of the  short-ranged system. The last can be easily seen from \eref{CasimirF_l} by simply expanding then the scaling function $X_{\rm crit}$
\begin{eqnarray}\label{CasimirF_l_expand}
&&f_{\mathrm{tot}}(L|T,\mu)\simeq L^{-d}\left\{X_{\rm crit}^{\rm sr}(x_{t},x_{\mu})+
x_{l}X_{\rm crit}^{l}(x_{t},x_{\mu})\right.\nonumber\\
&&\left.+\dfrac{1}{2}\left[x_{s_{1}}X_{\rm crit}^{s_{1}}(x_{t},x_{\mu})+x_{s_{2}}X_{\rm crit}^{s_{2}}(x_{t},x_{\mu})\right]+x_{g}X_{\rm crit}^{g}(x_{t},x_{\mu})\right\}\nonumber\\
&&+(\sigma-1)\beta H_A(T,\mu) L^{-\sigma}\xi_{\mathrm{ret}}^{\sigma-d}.
\end{eqnarray}
Here $X_{\rm crit}^{\rm sr}$ originates from the short-range interactions [see Eqs. (\ref{XCassranalytic})]. It is well known that $X_{\rm crit}^{\rm sr}<0$, i.e., the force is attractive, under $(+,+)$ boundary conditions.   $X_{\rm crit}^{\rm sr}$  provides the {\it leading} behaviour of the force near the bulk critical point ($x_{t}=0,\ x_{\mu}=0$). There $X_{\rm crit}^{l},\ X_{\rm crit}^{s_{i}}$ and $X_{\rm crit}^{g}$ represent only {\it corrections} to the leading $L$ dependance. The reader can refer for a detailed comment on that matter to Ref. \cite{DSD2007} (see also Ref. \cite{DDG2006} for the properties of $X_{\rm crit}^{l}$). The validity of the proposed criterion as well as the statements made beneath it are well illustrated on Fig. \ref{fig:short_range_tending}.

The contribution of the dispersion forces to the total effective force $f_{\mathrm{tot}}$ can be distinguished from that of the critical Casimir force by their temperature dependence, because the leading temperature dependence of the former does not exhibit a singularity. Thus, one has
\begin{equation}
\label{decomp}
f_{\mathrm{tot}}(L|T,\mu)=f_{\mathrm{tot}}^{\mathrm{(reg)}}(L|T,\mu)+f_{\mathrm{tot}}^{\mathrm{(sing)}}(L|T,\mu),
\end{equation}
where
\begin{equation}
\label{def_vdW}
f_{\mathrm{vdW}} \equiv f_{\mathrm{tot}}^{\mathrm{(reg)}}(L|T,\mu),
\end{equation}
and
\begin{equation}
\label{def_Casimir}
  \beta F_{A,{\rm Cas}}(L|T,\mu)\equiv f_{\mathrm{tot}}^{\mathrm{(sing)}}(L|T,\mu).
\end{equation}
One expects that near the bulk critical point
\begin{equation}
\label{scaling_function}
\beta F_{A,{\rm Cas}}(L|T,\mu)=L^{-d}X_{\mathrm{Cas}}(x_t,x_\mu,\cdots),
\end{equation}
where $X_{\mathrm{Cas}}$ is a scaling function that for large enough $L$ (see below) approaches the scaling function of the short-ranged system $X_{\mathrm{Cas}}^{\mathrm{sr}}(x_t,x_\mu)$. From Eqs. (\ref{CasimirF_l}), (\ref{decomp}) -- (\ref{scaling_function}) it follows that the scaling function of the critical Casimir force $X_{\rm Cas}$ is proportional to the sum of $X_{\rm crit}$ and the singular part of the Hamaker term.

We will often compare the behavior of the system with subleading long-ranged interactions present with this one of a system with purely short-ranged interactions which will serve as a reference system. In such a purely short-range system one has $H_{A}=0$. Then, at the bulk critical point $(T=T_{c},\ \mu=\mu_{c})$ the leading term of the thermodynamic Casimir force between the slabs bounding the fluid has the form
\begin{equation}\label{ForceCasbulkcritpoint}
F_{A,{\rm Cas}}^{(\tau)}(L|T_{c},\mu_{c})=(d-1)\Delta^{(\tau)}(d)\dfrac{k_{B}T_{c}}{L^{d}},
\end{equation}
where $X_{\rm crit}^{\rm sr}(0)=(d-1)\Delta^{(\tau)}(d)$. Here $\Delta^{(\tau)}(d)={\cal O}(1)$ is an universal dimensionless quantity, called Casimir amplitude, which depends on the bulk and surface universality classes (specified by the boundary conditions $\tau$). Since the Casimir force is proportional to $k_{B}T_{c}$ the interaction between the walls can become rather strong in a system with high critical temperature such as, e.g., in classical binary liquid mixtures. Note that the {\it sign} of the force  depends on the sign of the Casimir amplitude $\Delta^{(\tau)}(d)$ which, on its turn, depends on the boundary conditions $\tau$. According to the usual convention negative sign corresponds to attraction, while positive sign means repulsion of the surfaces bounding the system.

The experimental and theoretical evidences accumulated till nowadays  support the statement that the Casimir force is attractive when the boundary conditions on both plates are the same, or similar, and is repulsive when they essentially differ from each other, e.g., when in the case of a one-component fluid one of the surfaces adsorbs the liquid phase of the fluid while the other prefers the vapor phase. It is instructive to go back to the explicit physical units in \eref{ForceCasbulkcritpoint} for the physically most relevant case of $d=3$. One has
\begin{equation}\label{DimensionsFCas}
F_{A,{\rm Cas}}^{(\tau)}(L)\simeq8.1\times10^{-3}
\dfrac{\Delta^{(\tau)}(d=3)}{(L/\mu{\rm m})^{3}}\dfrac{T_{c}}{T_{\rm roon}}\dfrac{{\rm N}}{{\rm m^{2}}},
\end{equation}
where $T_{\rm room}=20$ $^\circ$C (293.15 K). Sine as discussed above, for most systems and boundary conditions $\Delta^{(\tau)}(d)={\cal O}(1)$, when $T_{c}\simeq T_{\rm room}$ the thermodynamic Casimir force, for some space separation $L$, will be of the same order of magnitude as the quantum one
\begin{equation}\label{DimensionsFqCas}
F_{A,{\rm Cas}}^{\rm QED}(L)=-\dfrac{\pi^{2}}{240}\dfrac{\hbar c}{L^{4}}
\simeq-1.3\times10^{-3}
\dfrac{1}{(L/\mu{\rm m})^{4}}\dfrac{{\rm N}}{{\rm m^{2}}},
\end{equation}
and they both shall be significant and consequently measurable at or below the micrometer length scale.

We turn now to description of the model and the procedure under which our results have been obtained.
%
\section{The model}\label{sec:Model}
%
We are going to utilize the same type of model already used in Refs. \cite{DSD2007,DRB2007} but amended to take into account the specific features of the system considered in the current article. Among them is the role of the two competing substrate potentials. Here, in order to introduce the notations needed further, we briefly recall the basic expressions of that model paying a bit more attention only to difference of the current model with that one studied in Refs. \cite{DRB2007,DSD2007}.

We consider a lattice-gas model of a fluid confined between two planar walls, separated at a distance $L$ from each other, with grand canonical potential $\Omega\left[\rho(\mathbf{r})\right]$ given by
\begin{eqnarray}
&&\Omega\left[\rho({\bf r})\right]= k_{B} T \sum_{{\bf r}\in \mathbb{M}}\left\{
\rho({\bf r}) \ln\left[\rho({\bf r})\right]\right.\nonumber\\ &&\left.
+\left[1-\rho({\bf r})\right] \ln\left[1-\rho({\bf r})\right]\right\}
+\dfrac{1}{2}\sum_{{\bf r}, {\bf r}'\in \mathbb{M}}\rho({\bf
r})w^{l}({\bf r}-{\bf r}')\rho({\bf r}') \nonumber \\ &&
+\sum_{{\bf r} \in \mathbb{M}}\left[V^{(s_{1}|l|s_{2})}(z)-\mu\right] \rho({\bf r}),\label{Garand_pot_l_lattice_model}
\end{eqnarray}
where $\mathbb{M}$ is a simple cubic lattice in the region occupied by the fluid medium -- $\infty^{d-1}\times[0,L]$ and $V^{(s_{1}|l|s_{2})}(z)$ is an external potential that reflects the interactions between the confining walls and the constituents of the fluid, given by
\begin{eqnarray}\label{extrenalpot}
V^{(s_{1}|l|s_{2})}(z)&=&-\rho_{s_{1}}J_{\rm sr}^{s_{1},l}\delta(z)-\rho_{s_{2}}J_{\rm sr}^{s_{2},l}\delta(L-z)\nonumber\\&&
+v_{s_{1}}(z+1)^{-\sigma}+v_{s_{2}}(L+1-z)^{-\sigma},
\end{eqnarray}
where $v_{s_{i}}=-G(d,\sigma)\rho_{s_{i}}J^{s_{i},l},\ i=1,2$, with
\begin{equation}\label{v_s_l_def}
G(d,\sigma)=4\pi^{(d-1)/2}\dfrac{\Gamma\left(\frac{1+\sigma}{2}\right)}{\sigma\Gamma\left(\frac{d+\sigma}{2}\right)}.
\end{equation}
This type of functional can be viewed as a modification of the model utilized by Fisher and Nakanishi \cite{NF82,NF83} in their mean-field investigation of systems governed by short-range forces.

In \eref{Garand_pot_l_lattice_model} the terms in curly brackets multiplied by $k_{B}T$ correspond to the entropy contributions to the total energy, $w^{l}({\bf r}-{\bf r}')=-4J^{l}({\bf r}-{\bf r}')$ is the non-local coupling (interaction potential) between the constituents of the confined medium and $\mu$ is the chemical potential. Here and in the remainder of this paper, all length scales are taken in units of the lattice constant $a_{0}$ (for concrete values see Table. \ref{table_fluid_prop}), so that the particle number density $\rho(\bf r)$ becomes simply a number density which varies in the range $[0,1]$.

The variation of \eref{Garand_pot_l_lattice_model} with respect to $\rho(\bf r)$ leads to the equation of state for the equilibrium density $\rho^{*}(\bf r)$
\begin{eqnarray}
    2\rho^{*}({\bf r})-1&=&\tanh \left\{-\dfrac{\beta}{2}\sum_{{\bf r}'\in\mathbb{M}}
    w^{l}({\bf r}-{\bf r}')\rho^{*}({\bf r}')\right.\nonumber\\ && \left.+
    \dfrac{\beta}{2}\left[\mu-V^{(s_{1}|l|s_{2})}(z)\right]
    \right\}.\label{eqs_of_state_l}
    \end{eqnarray}
The advantage of this type of equation is that it lends itself to numerical solution by iterative procedures. For a given geometry and external walls-fluid potential $V^{(s_{1}|l|s_{2})}(z)$ its solution determines the equilibrium order parameter profile $\rho^{*}({\bf r})$ in the system. Inserting this profile into \eref{Garand_pot_l_lattice_model}, renders the grand canonical potential of the considered system.

Denoting, as in Refs. \cite{DRB2007,DSD2007} $\phi^{*}({\bf r})=2\rho^{*}({\bf r})-1$ and $\Delta\mu=\mu-\mu_{c}$, where $\mu_{c}=\frac{1}{2}\sum_{\mathbf{r}'}w^{l}(\mathbf{r}-\mathbf{r}')$, the equation of state \eref{eqs_of_state_l} can be rewritten in the standard form
\begin{eqnarray}\label{eq_state_standard_l_ab}
\phi^{*}({\bf r})=\tanh &&\left\{\beta\sum_{{\bf r}'\in\mathbb{M}} J^{l}({\bf r}-{\bf r}')\phi^{*}({\bf r}')\right.\nonumber\\&&\left.+\dfrac{\beta}{2}\left[\Delta\mu-\Delta V(z)\right]\right\}.
\end{eqnarray}
The bulk properties of the model are well known (see, e.g. \cite{Do96,B82} and the references therein). We recall that the order parameter $\phi^{*}$ of the system has a critical value $\phi^{*}=0$ which corresponds to $\rho_{c}=1/2$ so that $\phi^{*}=2(\rho^{*}-\rho_{c})$. The bulk critical point of the model is given by $\{\beta=\beta_{c}=[\sum_{\bf r}J^{l}({\bf r})]^{-1}, \mu=\mu_{c}=-2\sum_{\bf r}J^{l}({\bf r})\}$ with the sum running over the whole lattice. Within the mean-field approximation the critical exponents for the order parameter and the compressibility are $\beta = 1/2$ and $\gamma = 1$, respectively. The {\it effective} surface potential $\Delta V(z)$  in \eref{eq_state_standard_l_ab} is given by
\begin{eqnarray}\label{DeltaV_l_ab_thichlayers}
\Delta V(z)=
\dfrac{\delta v_{s_{1}}}{(z+1)^{\sigma}}+\dfrac{\delta v_{s_{2}}}{(L+1-z)^{\sigma}},
\end{eqnarray}
where $1\le z\le L-1$, and contributions of the order of $z^{-\sigma-1}$, $z^{-\sigma-2}$, etc., have been neglected,
\begin{eqnarray}
\delta v_{s_{i}}=-G(d,\sigma)
\left(\rho_{s_{i}}J^{s_{i},l}-\rho_c J^l\right),\ i=1,2\ \ \ \ \ \ \label{deltav_s_l_def}
\end{eqnarray}
are  ($T$- and $\mu$-independent) constants,
\begin{eqnarray}
J^{l}(\mathbf{r})=J_{\mathrm{sr}}^{l}\left\{\delta(|\mathbf{r}|)+\delta(|\mathbf{r}|-1)\right\}+
\dfrac{J^{l}\theta(|\mathbf{r}|-1)}{1+|\mathbf{r}|^{d+\sigma}}\label{Jldeftext},
\end{eqnarray}
is a proper lattice version of $-w^{l}(\mathbf{r})/4$ as the interaction energy between the fluid particles, and
\begin{eqnarray}
J^{s_{i},l}(\mathbf{r})=J_{\mathrm{sr}}^{s_{i},l}\delta(|\mathbf{r}|-1)+
\dfrac{J^{s_{i},l}\theta(|\mathbf{r}|-1)}{|\mathbf{r}|^{d+\sigma}},\ i=1,2\label{Jlsdeftext}
\end{eqnarray}
is the one between a fluid particle and a substrate particle, $\delta(x)$ is the discrete delta function and $\theta(x)$ is the Heaviside step function with the convention $\theta(0)=0$; in \eref{Jlsdeftext}  $\rho_{s_{i}},\ i=1,2$ are the number densities of the coated slabs in units of $a_{0}^{-d}$ (for concrete values see Table. \ref{table_fluid_sub}). Note that the effective potentials $\delta v_{s_{i}},\ i=1,2$ result from the difference between the relative strength of the substrate-fluid interactions for substrates with density $\rho_{s_{i}},\ i=1,2$ and that of the fluid-fluid interactions for a fluid with density $\rho_{c}$. In \eref{DeltaV_l_ab_thichlayers} the restriction $z\geq1$ holds because we consider the layers closest to the substrate to be completely occupied by the liquid phase of the fluid (which implies that we consider the strong adsorption limit), i.e., $\rho(0) = \rho(L) = 1$, which is achieved by taking the limits $J_{\mathrm{sr}}^{s_{i},l}\rightarrow\infty,\ i=1,2$; thus, the actual values of $\Delta V (0) = \Delta V (L)$ will play no role. In order to preserve the monotonic behavior of $w^{l}(\mathbf{r})$ as a function of the distance $r$ between the particles, in \eref{Jldeftext} we have to require that $J_{\mathrm{sr}}^{l}\geq J^{l}/(1 + 2^{d+\sigma})$, i.e., $\lambda< 1 + 2^{d+\sigma}$, where \begin{equation}\label{l_param_def_new}
\lambda=J^{l}/J_{\mathrm{sr}}^{l}.
\end{equation}
From Eqs. (\ref{eq_state_standard_l_ab}) -- (\ref{Jlsdeftext}) one can identify the dimensionless coupling constants $s_i$, $i=1,2$ appearing in \eref{CasimirF_l}
\begin{equation}\label{s_def_l}
s_{i}=-\dfrac{1}{2}\beta\delta v_{s_{i}},\ i=1,2.
\end{equation}
Here $s_{i}>0$, i.e., $\rho_{s_{i}}J^{s_{i},l}>\rho_{c}J^{l}$ corresponds to walls "preferring" the liquid phase of the fluid, while $s_{i}<0$, or $\rho_{s_{i}}J^{s_{i},l}<\rho_{c}J^{l}$ mirrors the one with affinity to its gas phase. The marginal case $s_{i}=0$ which corresponds to $\rho_{s_{i}}J^{s_{i},l}=\rho_{c}J^{l}$ will be commented further in the text.

In terms of $\phi$ the functional defined by  \eref{Garand_pot_l_lattice_model} takes the form
\begin{eqnarray}\label{Grand_pot_phi_l_ab}
&&\Omega[\phi({\bf r})]=k_{B}T\sum_{{\bf r}\in \mathbb{M}} \left\{\dfrac{1+\phi({\bf r})}{2} \ln\left[\dfrac{1+\phi({\bf r})}{2}\right]\right. \nonumber \\ && \left.+ \dfrac{1-\phi({\bf r})}{2}\ln\left[\dfrac{1-\phi({\bf r})}{2}\right]\right\}
-\dfrac{1}{2}\sum_{{\bf r} \in \mathbb{M}}[\Delta\mu-\Delta V(z)] \phi({\bf r}) \nonumber\\&&
-\dfrac{1}{2}\sum_{{\bf r}, {\bf r}'\in \mathbb{M}}J^{l}\left({\bf r}-{\bf r}'\right)\phi({\bf r})\phi\left({\bf r}'\right)+\Omega_{\rm reg},\ \ \ \ \ \ \
\end{eqnarray}
where
\begin{eqnarray}
\Omega_{\rm reg}=-&&\dfrac{1}{2}\sum_{{\bf r} \in \mathbb{M}}\left[\Delta\mu-\Delta V(z)-\sum_{{\bf r}'\in \mathbb{M}}J^{l}\left({\bf r}-{\bf
r}'\right)\right]\label{Omega_reg_l}
\end{eqnarray}
do not depend on $\phi({\bf r})$ and, therefore, is a regular background term carrying an $L-$dependence and thus showing up in the corresponding force.

The only quantity in \eref{CasimirF_l}, which still has to be identified for our model, is the value of the coupling constant $\Lambda$. According to Ref. \cite{DSD2007}
\begin{equation}\label{l_param_def}
\Lambda=v_{\sigma}/v_{2},
\end{equation}
where $v_{\sigma}$ and $v_{2}$ are coefficients in the Fourier transform $\hat{J}^{l}(\mathbf{k})=\sum_{\mathbf{r}}\exp(i\mathbf{k}\cdot \mathbf{r})J^{l}(\mathbf{r})$ of the interaction $J^{l}(\mathbf{r})$ [see \eref{Jldeftext}] (for more details see Ref. \cite{DSD2007}). It turns out that $\Lambda$ depends on $\lambda$.

In accordance with \eref{CasimirF_l}, for the finite-size behavior of the excess grand canonical potential per unit area $\mathcal{A}$ of a liquid film in the case when both confining surfaces strongly adsorb the liquid phase, one expects
\begin{eqnarray}
\omega_{\mathrm{ex}}(L|T,\mu)\simeq&&\gamma_{1}^{{\rm ns}}+\gamma_{2}^{\rm ns}+
\dfrac{k_{B}T}{L^{d-1}}X_\Omega(\ast) \nonumber \\
&&+H_{A}(T,\mu) L^{-(\sigma-1)}\xi_{\mathrm{ret}}^{\sigma-d},
\label{omega_ex_perA_l}
\end{eqnarray}
where $\gamma_{1}^{{\rm ns}}$ and $\gamma_{2}^{{\rm ns}}$ are the non-singular parts of the surface tensions at the surfaces of the confining walls, while the singular parts are incorporated in the scaling function $X_\Omega(\ast)$, which arguments are as those of the function $X_{\rm crit}$ in \eref{CasimirF_l}.

In \eref{omega_ex_perA_l} $H_{A}$ is the Hamaker term which explicit form is derived in Appendix \ref{sec:HamakerTerm}. The result is
\begin{equation}\label{Hamsumfinall}
\dfrac{H_A}{C(d,\sigma)}=-J^{s_1,s_2}\left(\rho_{s_{1}}-\dfrac{J^{l}}{J^{s_{1},l}}\rho_{b}\right)
\left(\rho_{s_{2}}-\dfrac{J^{l}}{J^{s_{2},l}}\rho_{b}\right).
\end{equation}
Note that this result is in full agreement with the Dzyaloshinskii-Lifshitz-Pitaevskii (DLP) theory \cite{L56,DLP61}. It provides, however, an easy possibility to study the $T$ and $\mu$ dependence of ${H_A}$ by studying the corresponding dependencies of $\rho_b$, say, near the critical point of the fluid system. In accord with the DLP theory it teaches us that the {\it sign} of the Hamaker term depends on the {\it contrast} of material properties of the two bounding substrates with respect to the substance which is in-between them. Coating the substrate surfaces of the system with some additional material does not change the leading-order $L$ dependence of the interactions between the substrates and, therefore, does not change the above property. Let us stress that when $s_1\equiv s_2=s$, i.e., when the bounding substances are made from same material, the Hamaker term will be {\it negative}, independent on the properties of the fluid in-between them, i.e.,
\begin{equation}\label{Ham_ocl_equal_s}
H_A=-C(d,\sigma)J^{s_1,s_1}\left(\rho_{s}-\dfrac{J^{l}}{J^{s,l}}\rho_{b}\right)^{2}<0,
\end{equation}
which corresponds to {\it attraction}  between the confining walls. Similar is the situation when $\rho_b\to 0$, i.e., when the fluid separating the substrates is replaced by vacuum. Then, again, $H_A<0$ {\it irrespective} of the material properties of the bounding substances, since  $\lim_{\rho_b\to 0}H_{A}=A_{s_1,s_2}<0$. Further details on the behavior of $H_{A}$ are given in Appendix \ref{sec:HamakerTerm}.

In the next section \ref{sec:FinSizeBehav} we are going to present results for $\omega_{\mathrm{ex}}(L|T,\mu)$ based on the model described in the current section.
%
\section{Finite-size behaviour of the model in a film geometry}\label{sec:FinSizeBehav}
%
Let us start by rewriting the equations presented in Sec. \ref{sec:Model} in the form suitable for studying  of a system in a film geometry. Because of the translational symmetry of the system along the bounding surfaces, the quantities of interest depend only on the spacial coordinate $z$ along which the system is finite. Thus, one can write: $\phi(\mathbf{r})\equiv\phi\left(\mathbf{r}_{\parallel},z\right)=\phi(z)$, where $\mathbf{r}=\left\{\mathbf{r}_{\parallel},z\right\}$, i.e., the local order parameter profile is given by $\{\phi(z),\ 0\leq z\leq L\}$, with $\phi(0)=\phi(L)=1$. Hence the equation \eref{eq_state_standard_l_ab} for the equilibrium profile  becomes
\begin{eqnarray}\label{order_parameter_equation_d4}
&&\mathrm{arctanh}\left[\phi^{*}(z)\right]=\dfrac{\beta}{2}\left[\Delta\mu-\Delta V(z)\right]\nonumber\\
&&+K\left\{\text{\ {a}}_{d,\sigma}\left(\lambda\right)\phi^{*}(z)+\text{\ {a}}_{d,\sigma}^{nn}\left(\lambda\right)\left[\phi^{*}(z+1)+\phi^{*}(z-1)\right]\right.\nonumber\\&&\left.+\lambda
\sum_{z'=0}^{L}g_{d,\sigma}(|z-z'|)\theta(|z-z'|-1)\phi^{*}(z')\right\},
\end{eqnarray}
where $\Delta\mu=\mu-\mu_{c}$, $\Delta V(z)$ is defined in \eref{DeltaV_l_ab_thichlayers}, $K=\beta J_{\mathrm{sr}}^{l}$, $\text{\ {a}}_{d,\sigma}\left(\lambda\right)=(2d-1)+\lambda(c_{d,\sigma}-d)$, $\text{\ {a}}_{d,\sigma}^{nn}\left(\lambda\right)=1.0+\lambda(c_{d,\sigma}^{nn}-0.5)$ with $c_{d,\sigma}^{nn}=g_{d,\sigma}(1)+g_{d,\sigma}^{nn}(\pm 1)$. The functions $c_{d,\sigma}$, $g_{d,\sigma}(|z-z'|)$ and $g_{d,\sigma}^{nn}(|z-z'|)$ are determined in Eqs. (C10), (C11) and (C12) of Ref. \cite{DSD2007}, respectively.

When $L\rightarrow\infty$ the equilibrium finite-size order parameter $\phi^{*}(z)$ tends to its bulk value $\phi_{b}=2(\rho_{b}-\rho_{c})\in[0,1]$. The corresponding equation for $\phi_b$, performing the limit $L\to\infty$ in \eref{order_parameter_equation_d4}, reads
\begin{eqnarray}\label{bulk_order_param_eq}
&&\phi_{b}=\tanh\left\{\dfrac{\beta}{2}\Delta\mu\right.\nonumber\\
&&\left.+K\left[\text{\ {a}}_{d,\sigma}\left(\lambda\right)+2\text{\ {a}}_{d,\sigma}^{nn}\left(\lambda\right)+2\lambda
\sum_{z\geq2}g_{d,\sigma}(z)\right]\phi_{b}\right\},
\end{eqnarray}
wherefrom one immediately identifies the coordinates of bulk critical point
\begin{eqnarray}\label{bulk_sys_crit_coupl_in_text}
K_{c}\left(\lambda\right)=\left[\text{\ {a}}_{d,\sigma}\left(\lambda\right)+2\text{\ {a}}_{d,\sigma}^{nn}\left(\lambda\right)+2\lambda
\sum_{z\geq 2}g_{d,\sigma}(z)\right]^{-1},\ \ \ \ \
\end{eqnarray}
and $\Delta\mu=0$. Note that the position of the bulk critical point depends on $\lambda$, i.e., on the presence and the strength of the long-ranged tails in the fluid-fluid interactions.

For a fluid confined to a film geometry the natural quantity to consider is the excess grand canonical potential per unit area $\mathcal{A}$: $\omega_{\rm ex}\equiv\lim_{\cal{A}\to\infty}[\Omega/{\cal{A}}]-L\omega_{\rm bulk}$. Using the result of Ref. \cite{DRB2009} [see Eq. (3.14) there], as well as the identity ${\rm arctanh} \, \phi=(1/2)[\ln(1+\phi)-\ln(1-\phi)]$, one can write $\beta \omega_{\rm ex}$ in the form
\begin{eqnarray}\label{deltaomegaexz_equilibrium}
\beta\omega_{\rm ex}&&=\sum_{z=0}^{L} \left\{\dfrac{1}{2}\ln\left[1-\phi^{*}(z)^{2}\right]-\dfrac{1}{2}\ln\left[1-\phi_{b}^{2}\right]\right.\nonumber\\&&\left.
+\dfrac{1}{4}\phi^{*}(z)\ln\left[\dfrac{1+\phi^{*}(z)}{1-\phi^{*}(z)}\right]-\dfrac{1}{4}\phi_{b}\ln\left[\dfrac{1+\phi_{b}}{1-\phi_{b}}\right]\right.
\nonumber\\&&\left.+\dfrac{1}{2}\beta\Delta V(z)\phi^{*}(z)-\dfrac{\beta\Delta\mu}{2}[\phi^{*}(z)-\phi_{b}]\right\}+\beta\omega_{\rm reg},\ \ \ \ \
\end{eqnarray}
where
\begin{eqnarray}\label{bethaomega_reg}
\beta\omega_{\rm reg}&&=\left[\dfrac{K}{(\sigma-1)K_{c}}(s_{1,c}+s_{2,c})-\dfrac{1}{4}C(d,\sigma)K\lambda\right]\nonumber\\&&
\times L^{-\sigma+1}\xi_{\rm ret}^{\sigma-d}.
\end{eqnarray}
As it is clear from \eref{Garand_pot_l_lattice_model}, the model presented above does not account for the direct wall-wall interaction, but only for the walls-fluid and fluid-fluid ones. Thus, in order to obtain the complete net force acting between the surfaces bounding the fluid one has to add to the force calculated from \eref{deltaomegaexz_equilibrium} via \eref{def} the one due to the direct wall-wall interaction. Then the resulting net force will be the {\it total} force $f_{\rm tot}(L|T,\mu)$ between the plates bounding the fluid. With the use of Eqs. (\ref{def}), (\ref{bethasigmaminusoneHAl}) and (\ref{deltaomegaexz_equilibrium}) one can write $f_{\rm tot}(L|T,\mu)$ in the form
\begin{eqnarray}\label{totalforcegeneralexpression_new}
f_{\rm tot}(L|T,\mu)&=&
-\dfrac{\beta}{2}\left[\omega_{\rm ex}(L+1|T,\mu)-\omega_{\rm ex}(L-1|T,\mu)\right]\nonumber\\
&&-\dfrac{4Ks_{1,c}s_{2,c}}{G(d,\sigma)K_{c}^{2}\lambda}L^{-\sigma}\xi_{\rm ret}^{\sigma-d},
\end{eqnarray}
where the last term represents the direct wall-wall interaction. On the other hand, if one subtracts from the potential $\omega_{\rm ex}$ its regular part $\omega_{\rm reg}$, i.e., if we consider the quantity
\begin{equation}
\label{Delta_omega_def}
\Delta\omega\equiv\lim_{\cal{A}\to\infty}[(\Omega-\Omega_{\rm reg})/{\cal{A}}]-L\omega_{\rm bulk},
\end{equation}
then, in accord with Eqs. (\ref{decomp}) -- (\ref{def_Casimir}), the $L$ dependence of $\Delta\omega$ via \eref{def} provides the singular part of the total force, i.e., the {\it critical} Casimir force -- $f_{\rm Cas}(L|T,\mu)$. Explicitly, one has
\begin{eqnarray}\label{critCasforcegeneralexpression_new}
\beta F_{A,{\rm Cas}}(L|T,\mu)&=&
-\dfrac{\beta}{2}\left[\Delta\omega(L+1|T,\mu)\right.\nonumber\\&&\left.-\Delta\omega(L-1|T,\mu)\right].
\end{eqnarray}
Obviously, near $T_c$ the total and the Casimir force are related via the expression
\begin{eqnarray}\label{totCasrelation_new}
&&f_{\rm tot}(L|T,\mu)=\beta F_{A,{\rm Cas}}-\dfrac{4Ks_{1,c}s_{2,c}}{G(d,\sigma)K_{c}^{2}\lambda}L^{-\sigma}\xi_{\rm ret}^{\sigma-d}\nonumber\\
&&+\left[\dfrac{K}{K_{c}}(s_{1,c}+s_{2,c})-\dfrac{1}{4}G(d,\sigma)K\lambda\right]L^{-\sigma}\xi_{\rm ret}^{\sigma-d}.
\end{eqnarray}

Eqs. (\ref{totalforcemeanfieldgeneral}), (\ref{order_parameter_equation_d4}) -- (\ref{totCasrelation_new}) provide the basis for our numerical treatment of the finite-size behavior of the total as well as of the Casimir force in fluid systems governed by van der Waals interactions. Let us briefly outline the main steps of this numerical procedure. We start by determining the equilibrium order-parameter profile [$\phi^{*}(z),\ 0\leq z\leq L$], solving iteratively \eref{order_parameter_equation_d4}. The solution of this equation depends, however, for a given range of parameters $T$ and $\Delta\mu$, on the choice of the initial state of the order-parameter profile. The two basic initial states are
\begin{description}
  \item[i] a "liquid-like" state in which all the sites of the lattice are occupied by a fluid particle, i.e. [$\phi^{*}(z)=1,\ 0\leq z\leq L$];
  \item[ii] a "gas-like" state where $\phi^{*}(0)=1,\ \phi^{*}(L)=1,\ \phi^{*}(z)=0,\ 1\leq z\leq L-1$.
\end{description}
Thus one needs to calculate the profile starting from both initial states. If the two final states coincide they provide the unique minimum of the functional -- see \eref{deltaomegaexz_equilibrium}. If they differ, one has to check which one of them provides the absolute minimum of the grand canonical potential. The simplest way to clarify that question is to calculate $\beta\omega_{\mathrm{ex}}$ via \eref{deltaomegaexz_equilibrium}. On a given line in the $(T,\Delta \mu)$ plane the values of the grand canonical potential of the two profiles coincide. The set of points on this line defines the phase diagram of the system. An alternative and practically more effective approach for determining of the phase diagram will be presented in Sec. \ref{sec:PhaseBehaviour}.

Heretofore for systems governed by dispersion interactions there are no closed-form analytical expressions about the scaling function of the Casimir force even within mean-filed theory. Such expressions exist, however, for systems governed by short-range interactions \cite{K97}. For $(+,+)$ boundary conditions at $\Delta\mu=0$ the corresponding results are
\begin{subequations}\label{XCassranalytic}
  \begin{equation}\label{ytgezero}
  \text{(i)}\ \ \ X_{\rm Cas}^{\rm sr}\left(x_{t}\geq0, x_\mu=0\right)=-\left[2K(k)\right]^{4}k^{2}\left(1-k^{2}\right),
  \end{equation}
  with $x_{t}=\left[2K(k)\right]^{2}\left(2k^{2}-1\right)$;
  \begin{equation}\label{zerogrytgeminpisq}
  \text{(ii)}\ \ \ X_{\rm Cas}^{\rm sr}\left(0\geq x_{t}\geq-\pi^{2}, x_\mu=0\right)=-4K^{4}(k),
  \end{equation}
  with $x_{t}=\left[2K(k)\right]^{2}\left(2k^{2}-1\right)$;
  \begin{equation}\label{ytleminpisq}
  \text{(iii)}\ \ \ X_{\rm Cas}^{\rm sr}\left(x_{t}\leq-\pi^{2}, x_\mu=0\right)=-4K^{4}(k)\left(1-k^{2}\right)^{2},
  \end{equation}
\end{subequations}
with $x_{t}=-\left[2K(k)\right]^{2}\left(k^{2}+1\right)$, where $K(k)$ is the complete elliptic integral of the first kind, $0\leq k<1$. In \eref{XCassranalytic} the scaling variable of $X_{\rm Cas}^{\rm sr}(x_{t}, x_\mu=0)$ is $x_{t}=t\left(L/\xi_{0}^+\right)^{1/\nu}=t\left(L/\xi_{0}^{+}\right)^{2}$. We note that $X_{\rm Cas}^{\rm sr}(x_{t}, x_\mu=0)$ is \textit{analytic} for \textit{all} values of $x_{t}$, because the film critical point $\left[T=T_{c,L},\mu=\mu_{c,L}\right]$ is located off coexistence at $\mu_{c,L}-\mu_{c}\sim L^{-3}$ \cite{FN81,NF82,NF83,PE92}. Obviously one has $x_{t}\geq0$ if $k\geq1/\sqrt{2}$ (with $k=1/\sqrt{2}$ corresponding to the bulk critical point), $0\geq x_{t}\geq-\pi^{2}$ if $1/\sqrt{2}\geq k\geq0$ (with $k=0$ corresponding to the actual film critical point), and $x_{t}\leq-\pi^{2}$ if $-1<k<0$ (negative $k$ describes the region below the bulk critical point). For the field dependent scaling function of the Casimir force $X_{\rm Cas}(x_t,x_{\mu}\ne 0)$ there are \textit{no} analytical expressions even for short-range systems.

\begin{figure*}[t!]
\centering
\mbox{\subfigure{\includegraphics[width=8.8 cm]{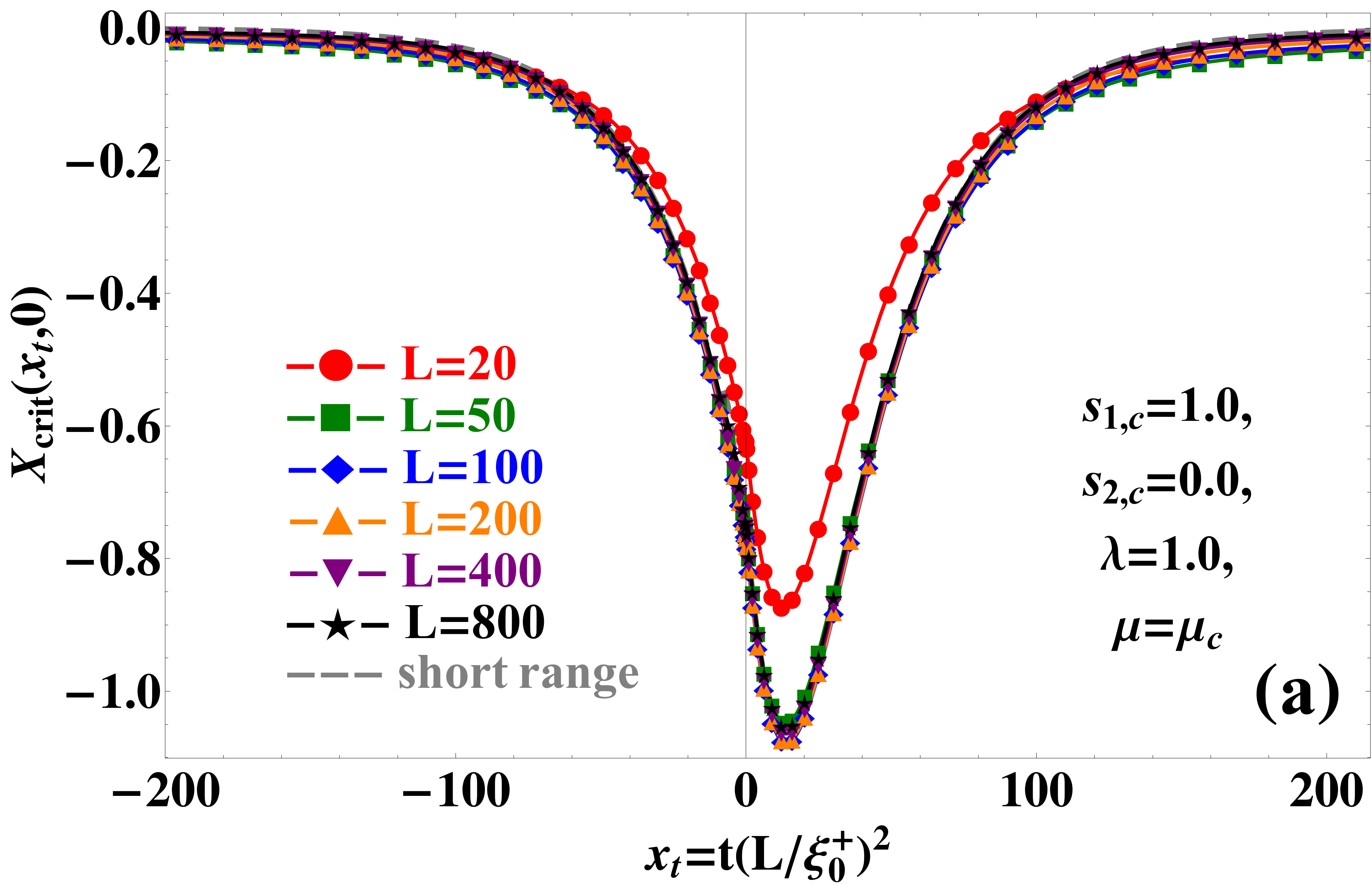}}\quad
      \subfigure{\includegraphics[width=8.6 cm]{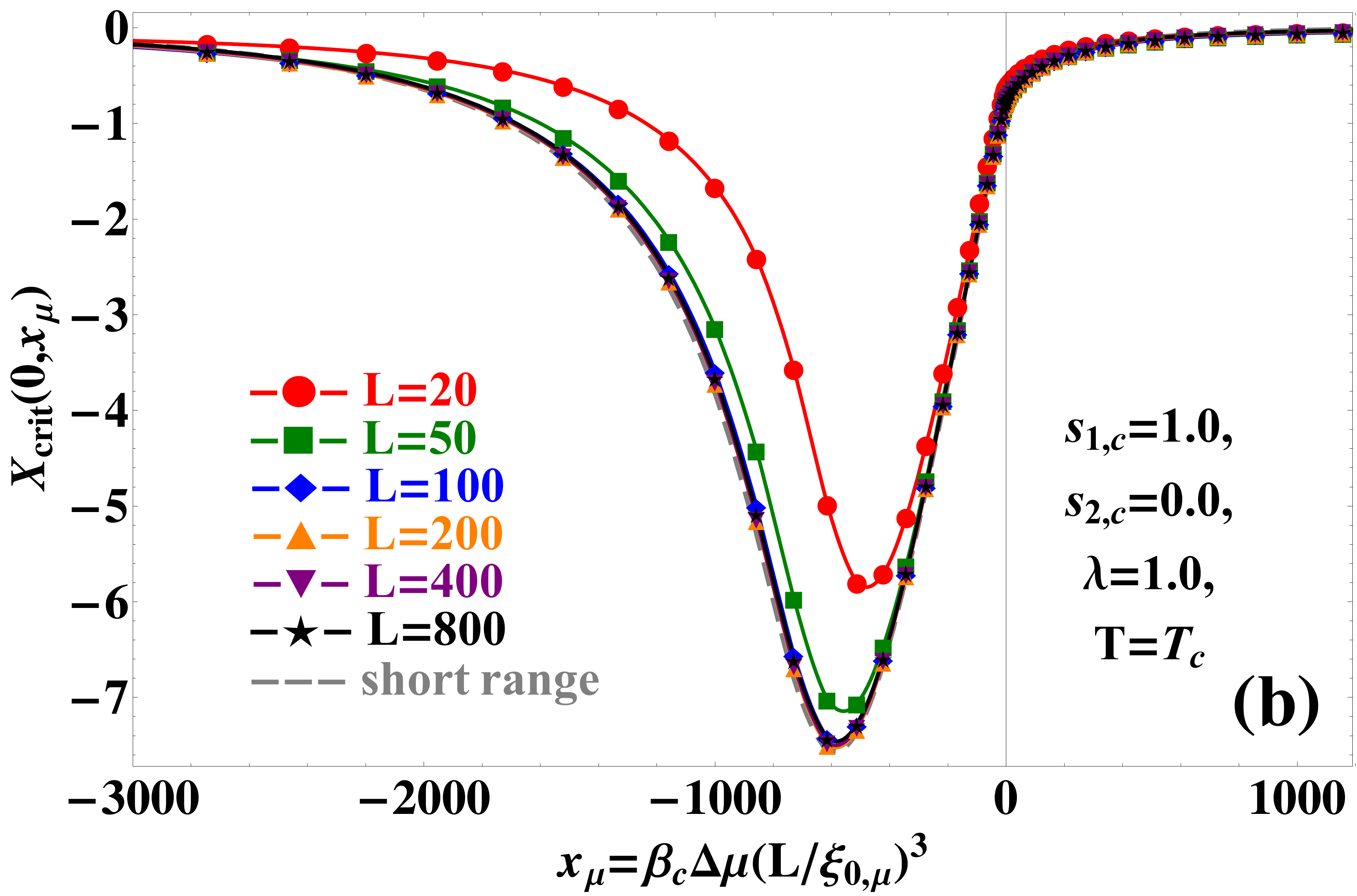}}}\\
\mbox{\subfigure{\includegraphics[width=8.8 cm]{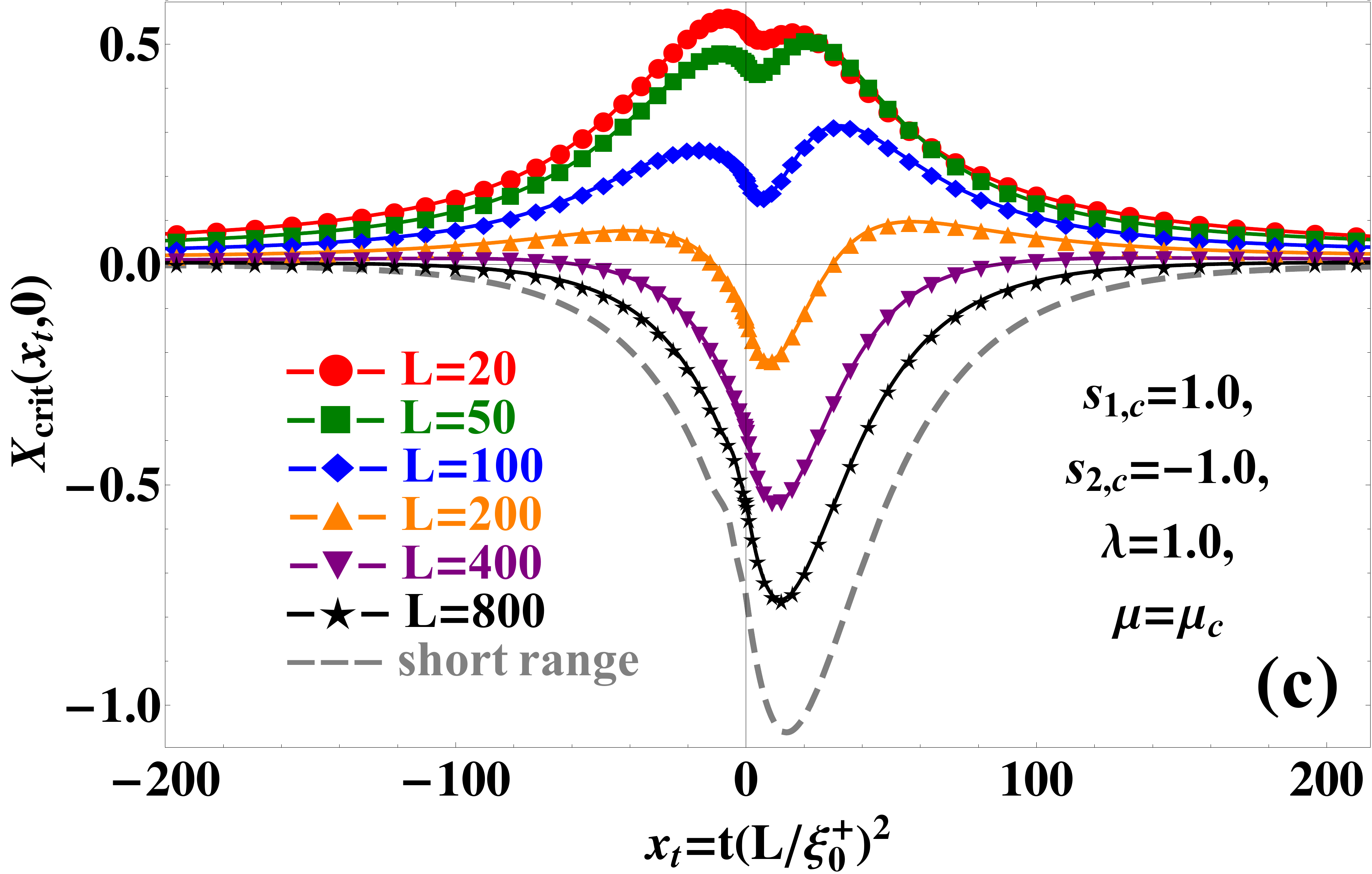}}\quad
      \subfigure{\includegraphics[width=8.6 cm]{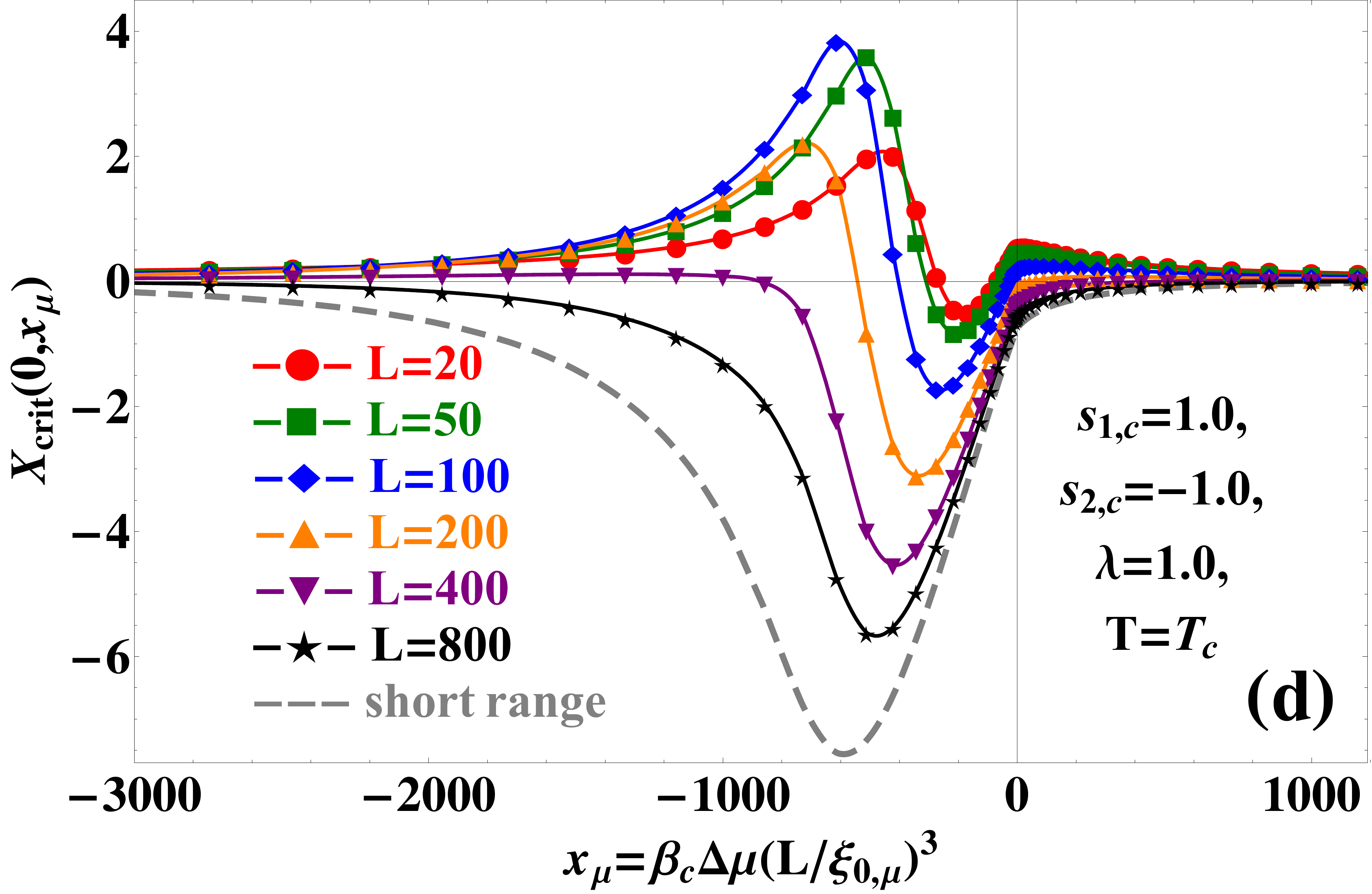}}}
  \caption{Behavior of the scaling function $X_{\rm crit}(x_{t},x_{\mu})$ in a $d=3$ dimensional confined fluid system. On $\mathbf{(a)}$ and $\mathbf{(c)}$  the temperature dependence of $X_{\rm crit}(x_{t},x_{\mu})$ is shown at $\Delta\mu=0.0$ (i.e., at $x_{\mu}=0.0$), while on $\mathbf{(b)}$ and $\mathbf{(d)}$  -- the field one at $T=T_{c}$,(i.e., at $x_{t}=0.0$). In $\mathbf{(a)}$ and $\mathbf{(b)}$ the parameters characterizing the interactions in the systems have values $s_{1,c}=1.0$, $s_{2,c}=0.0$ and $\lambda=1.0$, while in $\mathbf{(c)}$ and $\mathbf{(d)}$ one has $s_{1,c}=1.0$, $s_{2,c}=-1.0$ and $\lambda=1.0$. The considered separations between the confining walls are $L=20$ ({\color{red_n}{{\Large-}\hspace{-0.1cm}{\Large-}\hspace{-0.1cm}{\large$\bullet$}\hspace{-0.09cm}{\Large-}\hspace{-0.1cm}{\Large-}}}), $50$ ({\color{dark_green}{{\Large-}\hspace{-0.1cm}{\Large-}\hspace{-0.1cm}{\small$\blacksquare$}\hspace{-0.09cm}{\Large-}\hspace{-0.1cm}{\Large-}}}), $100$ ({\color{blue_n}{{\Large-}\hspace{-0.1cm}{\Large-}\hspace{-0.1cm}$\blacklozenge$\hspace{-0.09cm}{\Large-}\hspace{-0.1cm}{\Large-}}}), $200$ ({\color{orange_n}{{\Large-}\hspace{-0.1cm}{\Large-}\hspace{-0.1cm}$\blacktriangle$\hspace{-0.09cm}{\Large-}\hspace{-0.1cm}{\Large-}}}), $400$ ({\color{purple_n}{{\Large-}\hspace{-0.1cm}{\Large-}\hspace{-0.1cm}$\blacktriangledown$\hspace{-0.09cm}{\Large-}\hspace{-0.1cm}{\Large-}}}) and $800$ ({\color{black}{{\Large-}\hspace{-0.1cm}{\Large-}\hspace{-0.1cm}$\bigstar$\hspace{-0.09cm}{\Large-}\hspace{-0.1cm}{\Large-}}}) layers. On every sub-figure the scaling functions are compared to one for a system with $L=800$ layers ({\color{gray_n}{$---$}}) governed by pure short-range interactions $(\lambda=s_{i,c}=0.0,\ i=1,2)$. One observes that when $s_{2,c}=0.0$ all scaling functions of systems with $L>20$ are indistinguishable from the one for a short-range system [see $\mathbf{(a)}$ and $\mathbf{(b)}$] and correspond to attractive force. When $s_{2,c}=-1.0$ the scaling functions do trace separate curves for different film thicknesses $L$.  For $\Delta\mu=0.0$ [see $\mathbf{(c)}$] one notices that for $L\leq 100$ the scaling functions correspond to repulsive forces. For $L=200$ the force changes sign twice. When $L$ is further increased the force becomes entirely attractive for $L\geq800$. At $T=T_{c}$ [see $\mathbf{(d)}$] some of the scaling functions change sign twice in the "gas" phase of the fluid medium (those for $L\leq100$). Upon increasing the separation the change occurs only once, and is not observed for $L\geq800$. When $\mu=\mu_c$ the curve for the short-ranged interactions depicts the numerical evaluation of the exact analytical result given by \eref{XCassranalytic}, while for $T=T_{c}$ this curve is evaluated within the presented mean-field theory.}
  \label{fig:CFspsm}
\end{figure*}
\begin{figure*}[t!]
\centering
\mbox{\subfigure{\includegraphics[width=8.8 cm]{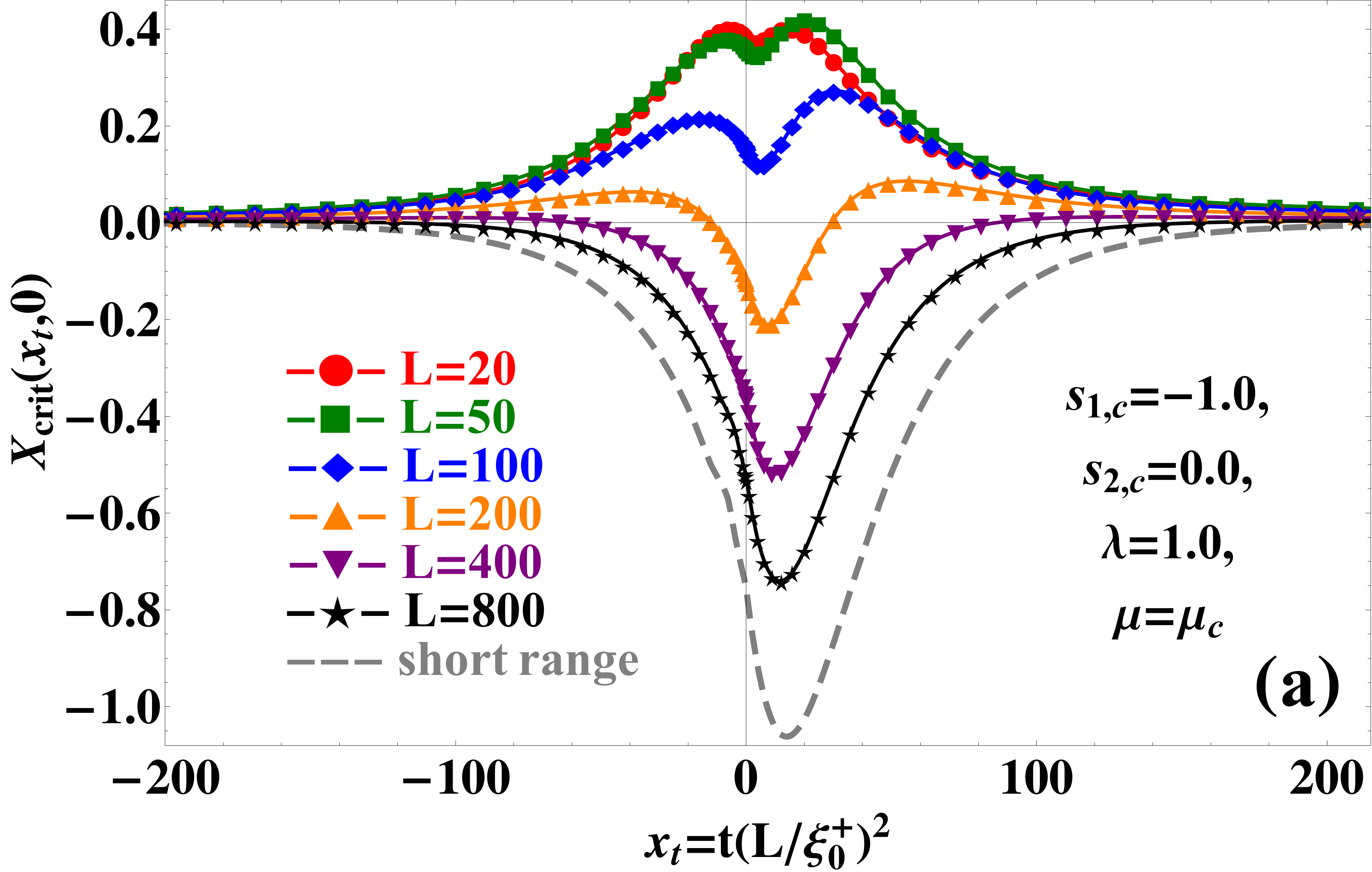}}\quad
      \subfigure{\includegraphics[width=8.6 cm]{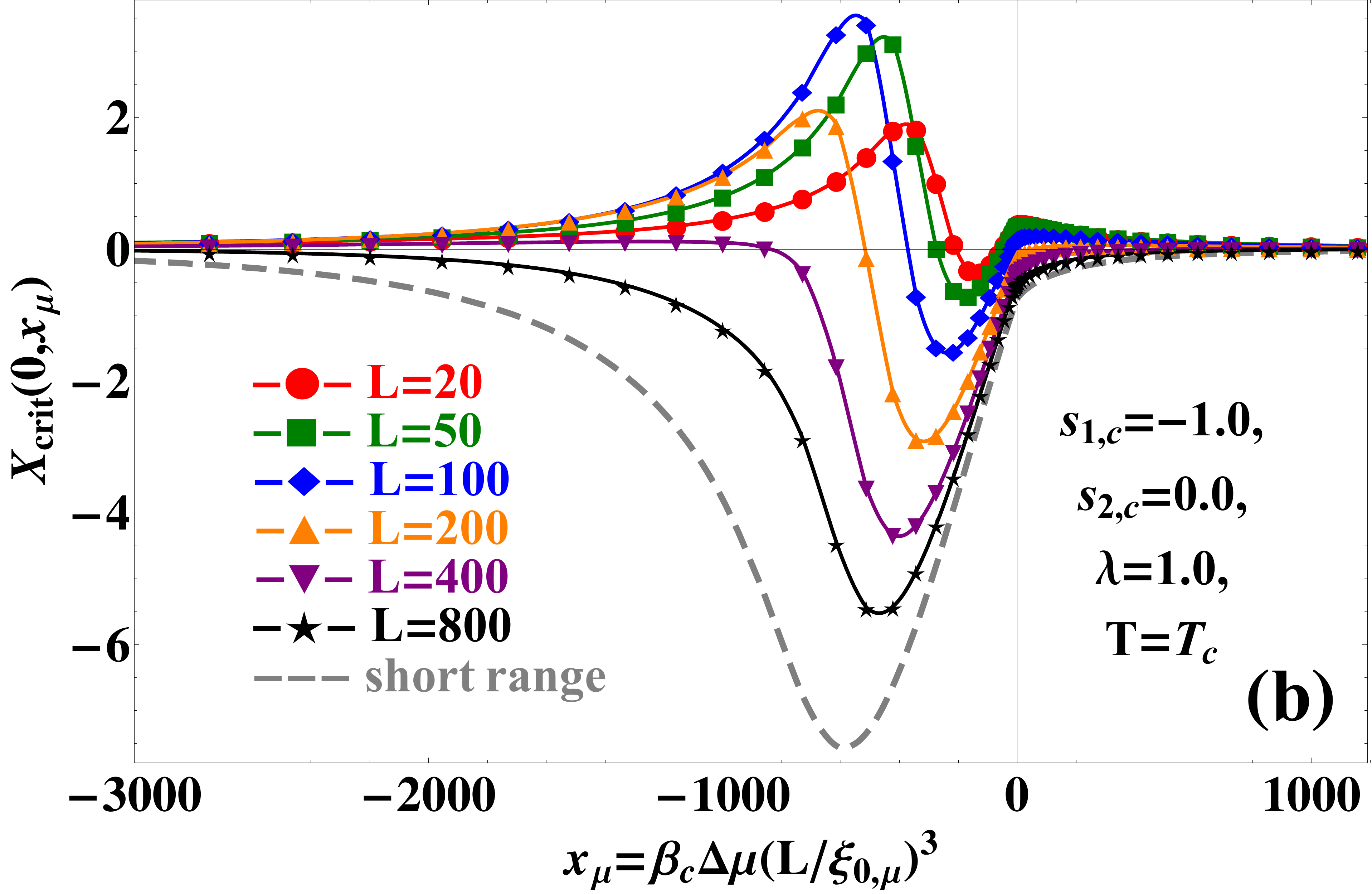}}}\\
\mbox{\subfigure{\includegraphics[width=8.8 cm]{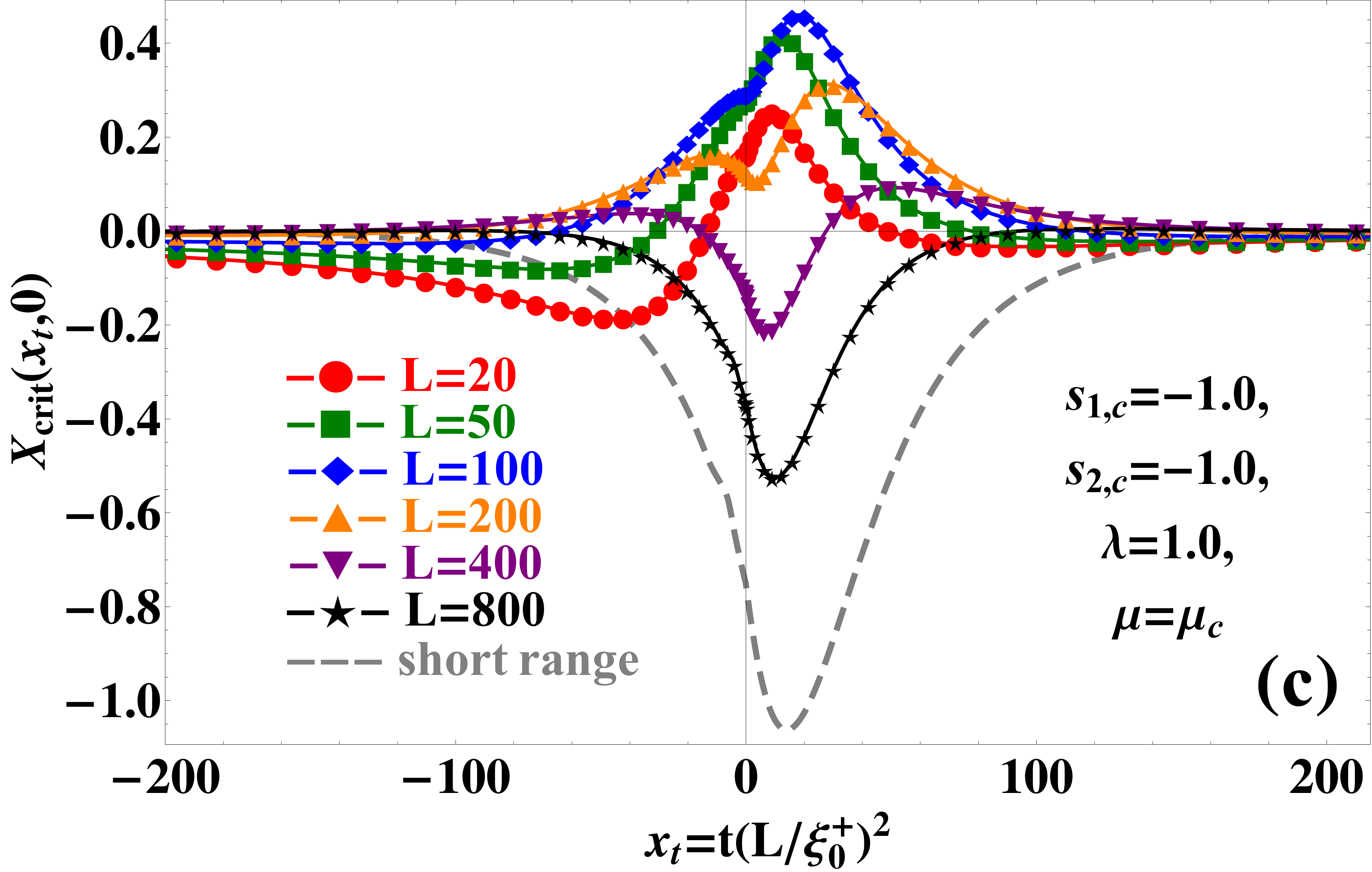}}\quad
      \subfigure{\includegraphics[width=8.6 cm]{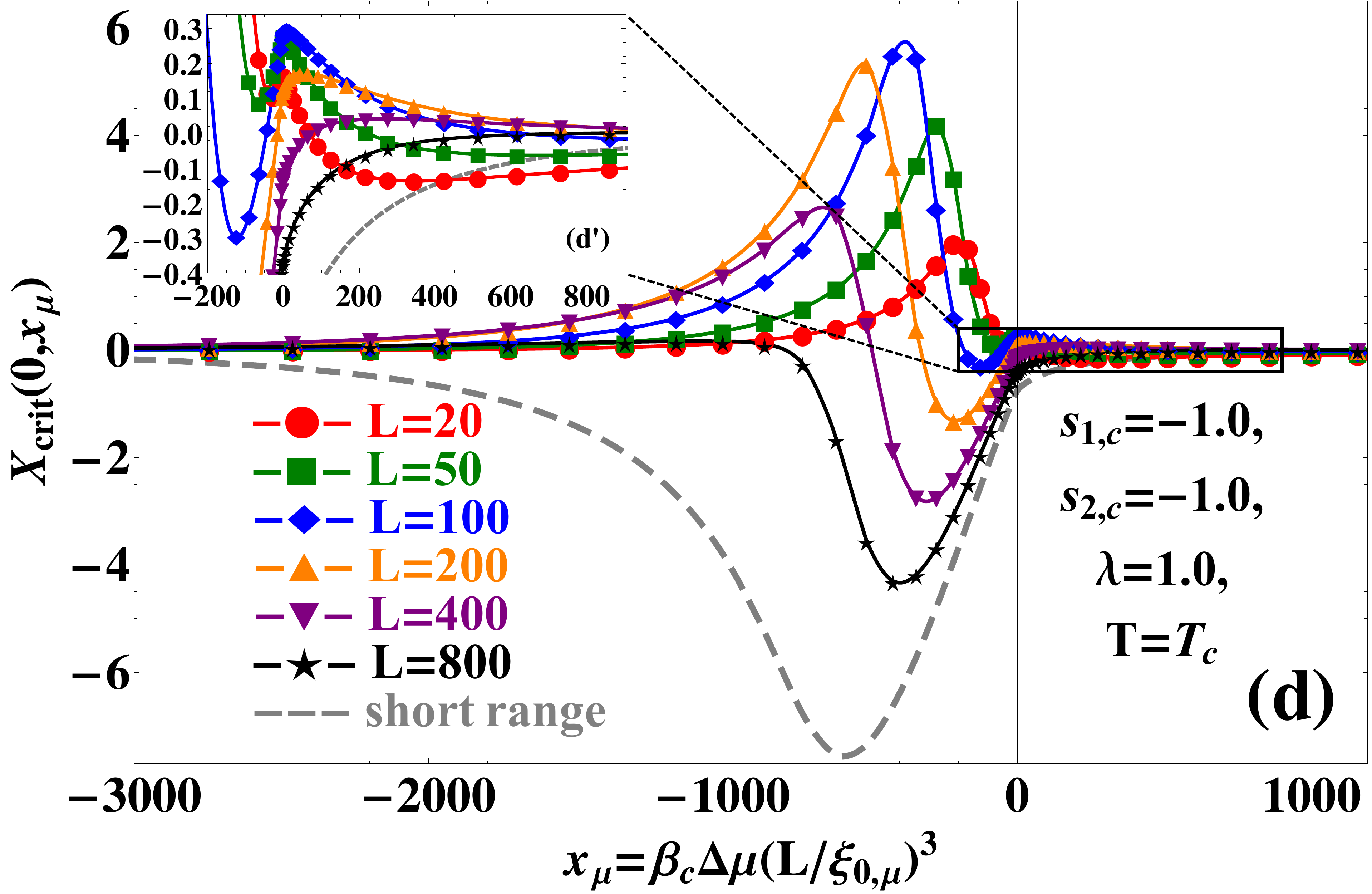}}}
  \caption{Behavior of the scaling function $X_{\rm crit}(x_{t},x_{\mu})$ in a $d=3$ dimensional confined fluid system. On $\mathbf{(a)}$ and $\mathbf{(c)}$ the temperature dependence of $X_{\rm crit}(x_{t},x_{\mu})$ is shown at $\Delta\mu=0.0$, (i.e., $x_{\mu}=0.0$), while on $\mathbf{(b)}$ and $\mathbf{(d)}$ -- the field one at $T=T_{c}$, (i.e., at $x_{t}=0.0$). In $\mathbf{(a)}$ and $\mathbf{(b)}$ the parameters characterizing the interactions in the systems have values $s_{1,c}=-1.0$, $s_{2,c}=0.0$ and $\lambda=1.0$, while in $\mathbf{(c)}$ and $\mathbf{(d)}$ one has $s_{1,c}=-1.0$, $s_{2,c}=-1.0$ and $\lambda=1.0$. The behavior of the scaling functions in a system characterized by $\lambda=1.0$, $s_{1,c}=-1.0$ and $s_{2,c}=0.0$ is similar to that of such with $\lambda=1.0$, $s_{1,c}=1.0$ and $s_{2,c}=-1.0$ [compare wit Fig. \ref{fig:CFspsm} $\mathbf{(c)}$ and $\mathbf{(d)}$]. For $\lambda=1.0$ and $s_{1,c}=s_{2,c}=-1.0$ at $\Delta\mu=0.0$ [see $\mathbf{(c)}$] one observes that for $L=20$ ({\color{red_n}{{\Large-}\hspace{-0.1cm}{\Large-}\hspace{-0.1cm}{\large$\bullet$}\hspace{-0.09cm}{\Large-}\hspace{-0.1cm}{\Large-}}}) the scaling function changes sign twice, having two minima and a maximum. When the separation $L$ is increased the values of the minima decrease rapidly towards zero, while that of the maximum increases, being highest for $L=100$ ({\color{blue_n}{{\Large-}\hspace{-0.1cm}{\Large-}\hspace{-0.1cm}$\blacklozenge$\hspace{-0.09cm}{\Large-}\hspace{-0.1cm}{\Large-}}}). When $L=200$ ({\color{orange_n}{{\Large-}\hspace{-0.1cm}{\Large-}\hspace{-0.1cm}$\blacktriangle$\hspace{-0.09cm}{\Large-}\hspace{-0.1cm}{\Large-}}}) the scaling function corresponds to repulsive force, now having two maxima and a single minimum. Reaching $L=400$ ({\color{purple_n}{{\Large-}\hspace{-0.1cm}{\Large-}\hspace{-0.1cm}$\blacktriangledown$\hspace{-0.09cm}{\Large-}\hspace{-0.1cm}{\Large-}}}) the scaling function changes sign twice, and corresponds to attractive force for $L\geq800$ ({\color{black}{{\Large-}\hspace{-0.1cm}{\Large-}\hspace{-0.1cm}$\bigstar$\hspace{-0.09cm}{\Large-}\hspace{-0.1cm}{\Large-}}}). At $T=T_{c}$ [see $\mathbf{(d)}$] for $L=20$ and $50$ the scaling functions change sign once for $x_{\mu}>0$ [see $\mathbf{(d')}$]. Then for $L=100$ one observes triple sign change, with a shallow minimum in the region $x_{\mu}>0$ and pronounced one in $x_{\mu}<0$. For $L=100$ the maximum of the scaling function has its highest value. Upon increasing the separation sign change of the scaling function occurs twice, and is not observed for $L>800$. Note that for $L=800$ the scaling function still changes sign in the region $x_{\mu}<0$. When $\mu=\mu_c$ the curve for the short-ranged interactions depicts the numerical evaluation of the exact analytical result given by \eref{XCassranalytic}, while for $T=T_{c}$ this curve is evaluated within the presented mean-field theory.}
  \label{fig:CFsmsm}
\end{figure*}
\begin{figure*}[t!]
\centering
\mbox{\subfigure{\includegraphics[width=8.85 cm]     {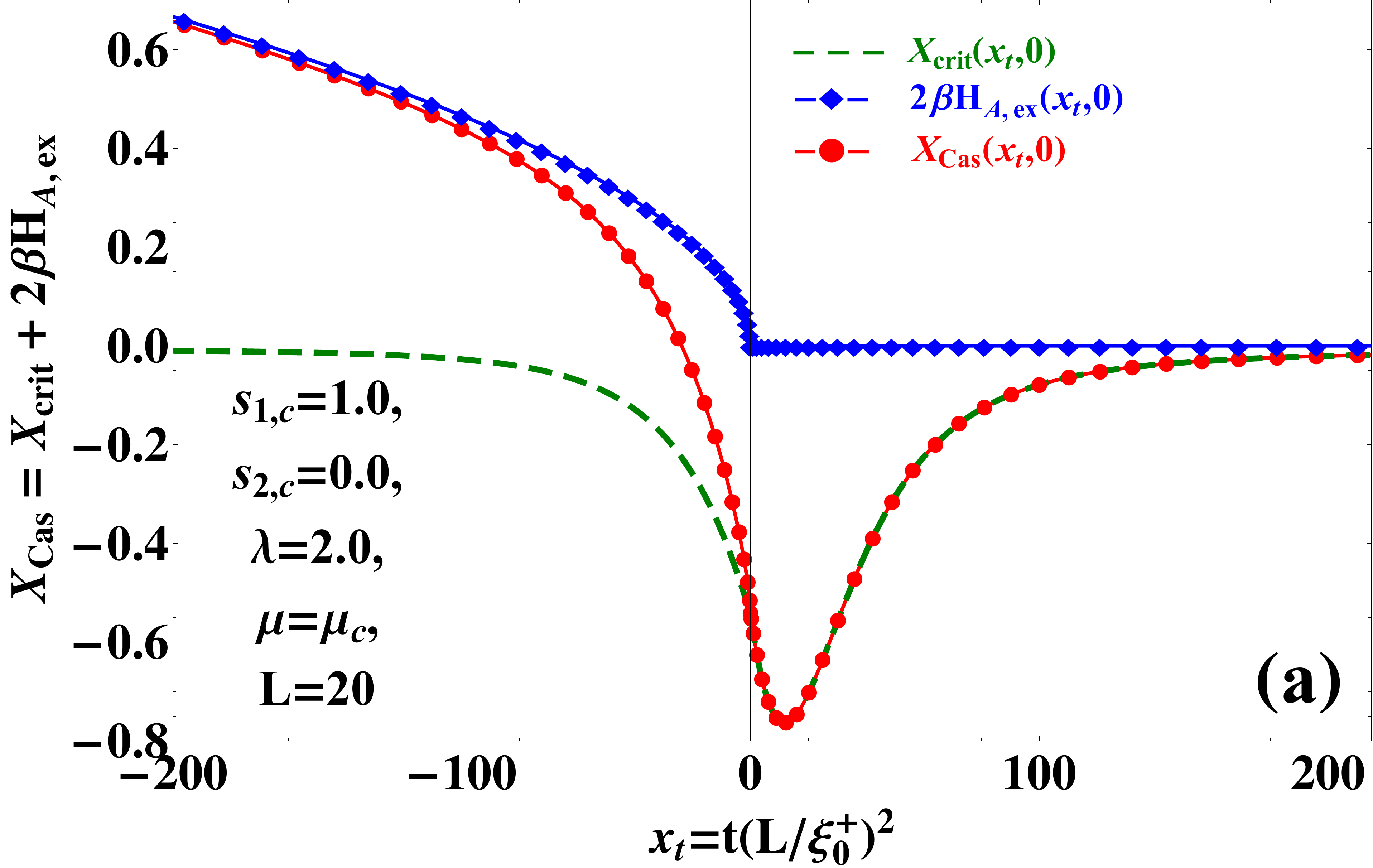}}\quad
      \subfigure{\includegraphics[width=\columnwidth]{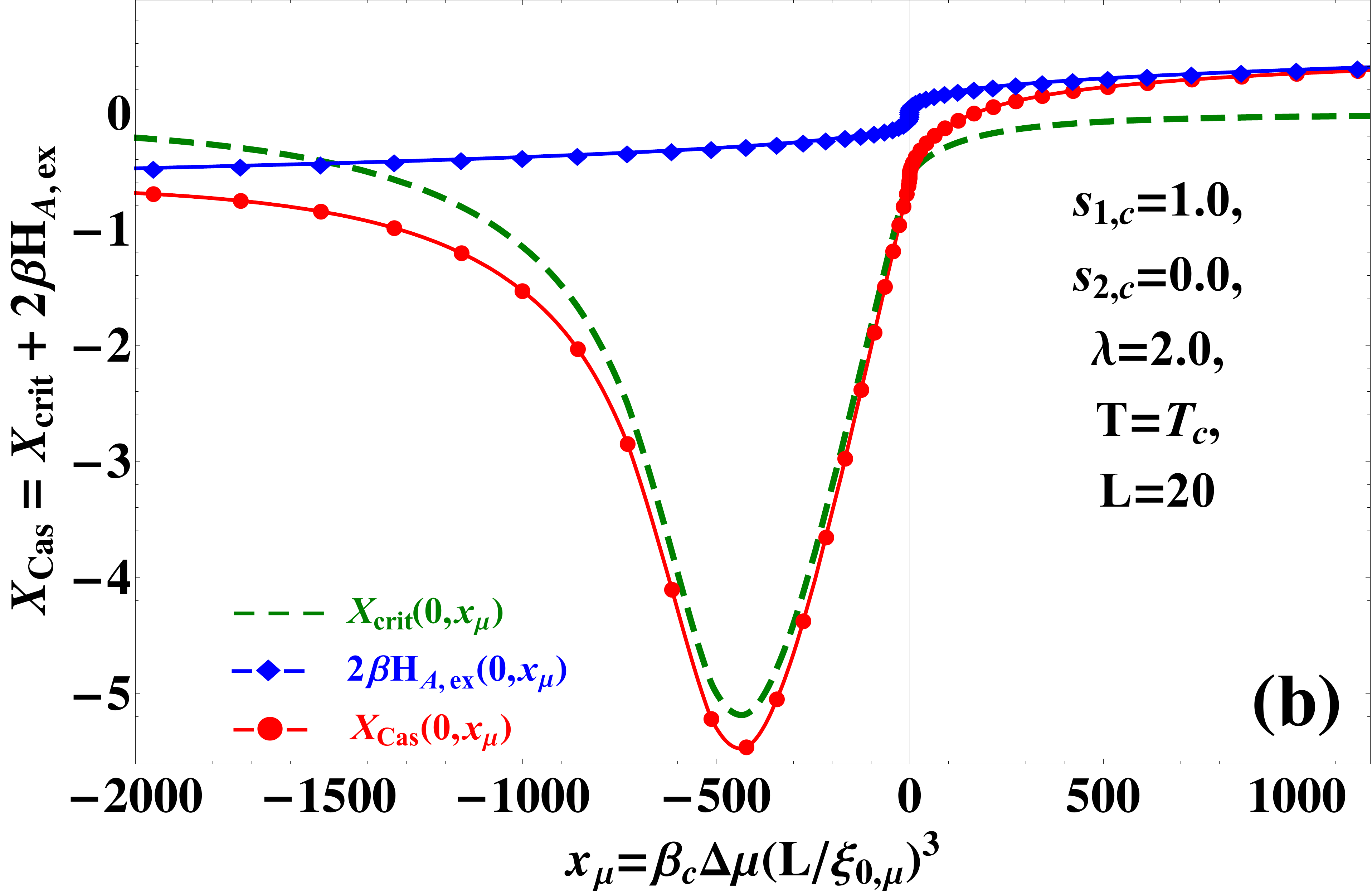}}}\\
\mbox{\subfigure{\includegraphics[width=\columnwidth]{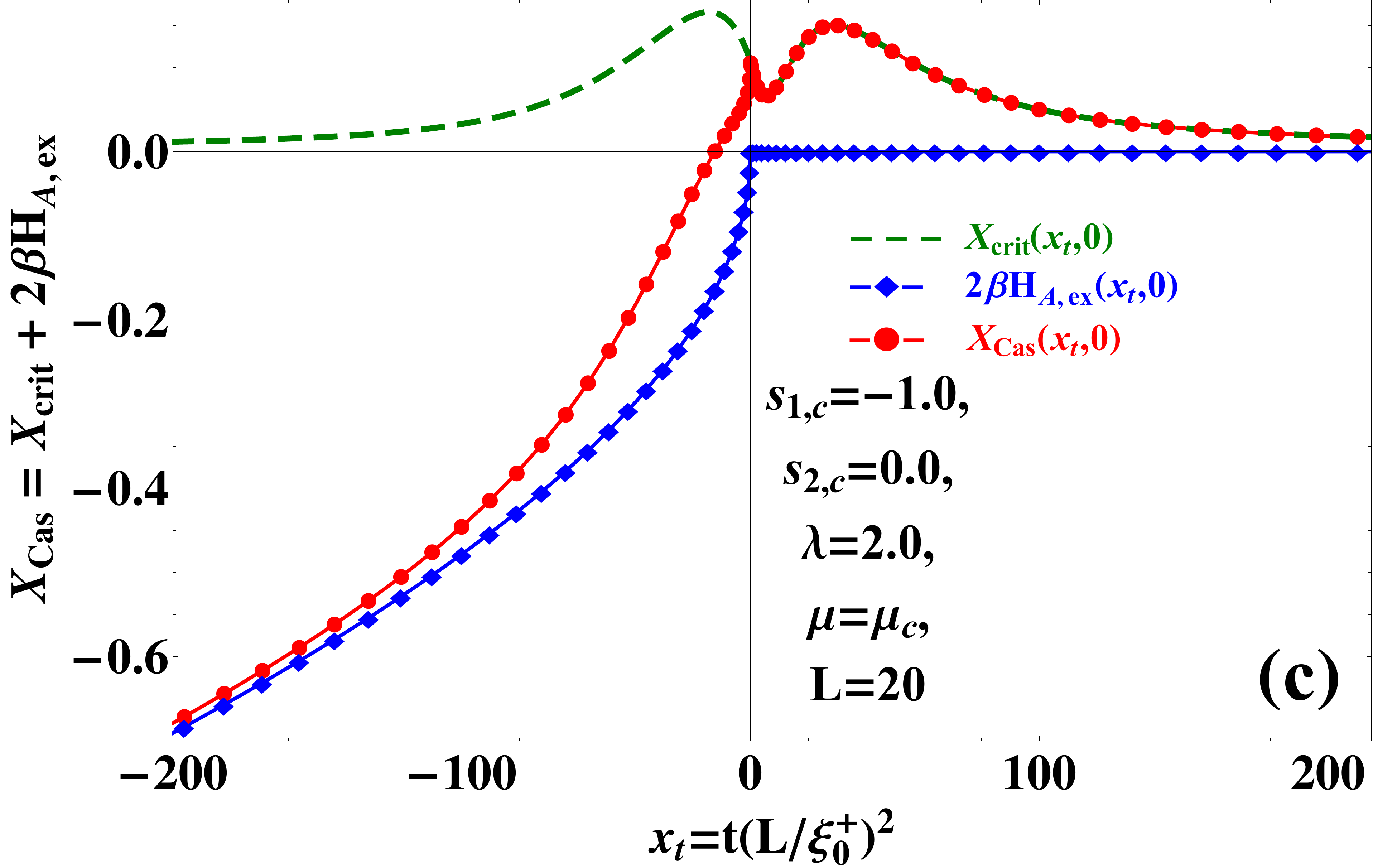}}\quad
      \subfigure{\includegraphics[width=\columnwidth]{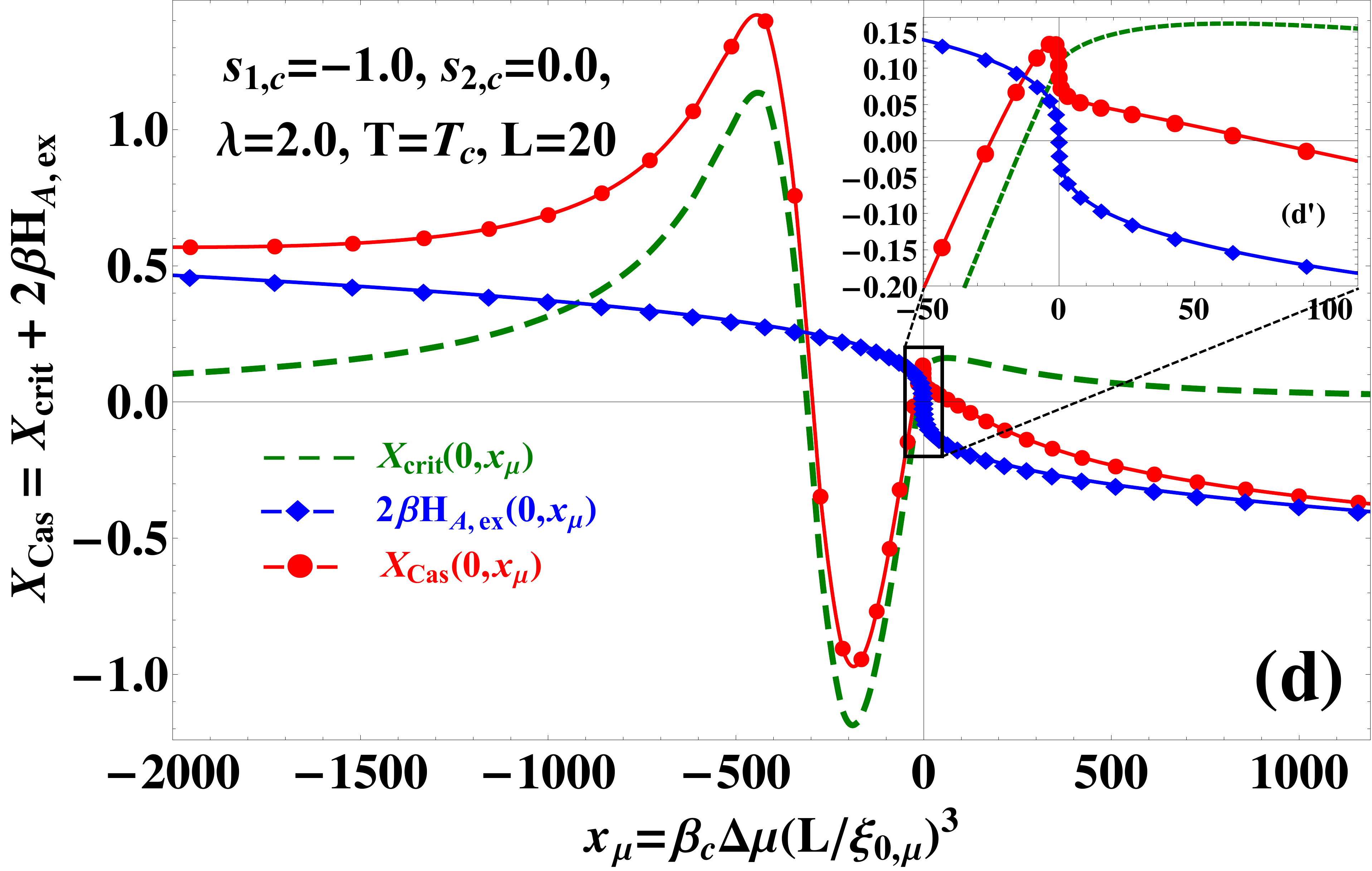}}}
  \caption{Interplay between the scaling function $X_{\rm crit}(x_{t},x_{\mu})$ ({\color{dark_green}{$---$}}) and the van der Waals term in excess of its regular contributions $2\beta H_{A,\ {\rm ex}}(x_{t},x_{\mu})$ ({\color{blue_n}{{\Large-}\hspace{-0.1cm}{\Large-}\hspace{-0.1cm}$\blacklozenge$\hspace{-0.09cm}{\Large-}\hspace{-0.1cm}{\Large-}}}), resulting in the scaling function of the critical Casimir force $\beta F_{A,{\rm Cas}}(x_{t},x_{\mu})L^{3}\equiv X_{\rm Cas}\simeq X_{\rm crit}(x_{t},x_{\mu})+2\beta H_{A,\ {\rm ex}}(x_{t},x_{\mu})$ ({\color{red_n}{{\Large-}\hspace{-0.1cm}{\Large-}\hspace{-0.1cm}{\large$\bullet$}\hspace{-0.09cm}{\Large-}\hspace{-0.1cm}{\Large-}}}) in a $d=3$ dimensional confined fluid system at $\Delta\mu=0.0,\ (x_{\mu}=0.0)$ [$\mathbf{(a)}$ and $\mathbf{(c)}$] and $T=T_{c},\ (x_{t}=0.0)$ [$\mathbf{(b)}$ and $\mathbf{(d)}$]. The separation between the walls confining the fluid medium is taken $L=20$ and the parameters characterizing the interactions in the systems have the following values: $\lambda=2.0$, $s_{2,c}=0.0$; [$\mathbf{(a)}$ and $\mathbf{(b)}$] $s_{1,c}=1.0$; [$\mathbf{(c)}$ and $\mathbf{(d)}$] $s_{1,c}=-1.0$.}
  \label{fig:ForceInterplay}
\end{figure*}

In the next section, based on the results reported in sections \ref{sec:Model} -- \ref{sec:FinSizeBehav}, we present numerical results for the behaviour of the scaling functions of the critical Casimir force and of the van der Waals term for the cases $d=\sigma=4$ and $d=\sigma=3$.
\section{Results for the behavior of the forces in $d=3$ confined fluid system with $\sigma=3$ dispersion potentials}\label{sec:Results}
In the current section, using the results for $d=\sigma=4$ from the mean-field type numerical study of the system introduced in sections \ref{sec:Model} -- \ref{sec:FinSizeBehav}, we will present some approximate results for the behavior $X_{\rm crit}$, the Hamaker term and the resulting critical Casimir force, between the two surfaces confining a van der Waals type fluid in $d=3$.

Within the mean-field theory the dependance of the corresponding forces from the temperature $T$ and the chemical potential difference $\Delta\mu$ near the bulk critical point $(T=T_c,\Delta\mu=0)$ is given through the dimensionless temperature and field scaling variables $x_{t}=t\left(L/\xi_{0}^{+}\right)^{2}$ and $x_{\mu}=\beta_{c}\Delta\mu\left(L/\xi_{0,\mu}\right)^{3}$. In our numerical treatment we take these variables to  range in the intervals: $x_{t}\in\left[-24^{2};24^{2}\right]\equiv[-576;576]$ and $x_{\mu}\in\left[-24^{3};24^{3}\right]\equiv[-13824;13824]$. We consider systems with fluid layers $L$ being $20,\ 50,\ 100,\ 200,\ 400\ \text{and}\ 800$. We fix one of the walls-fluid coupling parameters (say $s_{1,c}$) to have a value either $1.0$ or $-1.0$, while the other one, $s_{2,c}$, is varied from $0.0$ to $-1.0$ with a step of $-0.2$. The fluid-fluid coupling parameter $\lambda$ is supposed to be either $1.0$ or $2.0$.

In order to determine the behaviour of the above mentioned quantities in a $d=3$ confined fluid system, using the presented approximation, we proceed in the following way. First, within the mean-field treatment of the critical behaviour $(d=\sigma=4)$ we solve the equation for the equilibrium order parameter profile -- \eref{order_parameter_equation_d4}. Next, with the use of Eqs. (\ref{deltaomegaexz_equilibrium}) and (\ref{totalforcegeneralexpression_new}) we obtain the total force of interaction between the plates. Subtracting from \eref{totalforcegeneralexpression_new} the Hamaker term \eref{3bethaHA_ocf} we end up with the scaling function $X_{\rm crit}$ [see also \eref{totalforcemeanfieldgeneral}]. In order that this function contributes properly to the critical Casimir and hence to the total force of interaction in $d<4$, one must normalize it accordingly. The need of such normalization is explained in details in Ref. \cite{DSD2007} (see there Sections IV.A.1 and IV.A.3). For boundary conditions $\tau$ one has
\begin{eqnarray}\label{Xbar}
X^{(\tau)}_{\rm crit}(\cdot)=\frac{2 \Delta^{(\tau)}(d=3)}{X_{\rm crit,sr}^{(\tau),\mathrm{MF}}(t=\Delta\mu=0)} \left[\frac{\xi_0^{+}(0)}{\xi_0^{+}\left(\lambda\right)}\right]^4 X^{(\tau),\mathrm{MF}}_{\rm crit}(\cdot), \ \ \ \ \ \
\end{eqnarray}
where $X_{\rm crit,sr}^{(\tau,\mathrm{MF})}$ is the value of the scaling function for a system within mean-field treatment governed by short-range interactions at its corresponding bulk critical point, and $X^{(\tau),\mathrm{MF}}_{\rm crit}(\cdot)$ is the scaling function of the critical Casimir force, calculated for $d=\sigma=4$, with $\lambda \neq 0,\ s_{1,c}\neq 0,\ s_{2,c}\neq 0$.  Here $\Delta^{(\tau)}(d=3)$ is the Casimir amplitude for the $d=3$ Ising universality class with boundary conditions $\tau$, while $\xi_0^{+}(0)$ and $\xi_0^{+}(\lambda)$ are the amplitudes of the bulk correlation length in mean-field systems with, correspondingly, short-ranged ($\lambda=0$) and long-ranged ($\lambda\ne 0$) fluid-fluid interactions, as $\xi_0^{+}\left(\lambda\right)=\sqrt{v_{2}}$ (see Eqs. (4.15) and (4.17) in Ref. [\onlinecite{DSD2007}]). Therefore for $\lambda=0,\ 1\ \text{and}\ 2$ one has $v_{2}=1/9,\ 0.1640,\ \text{and}\ 0.1998$, and hence $\xi_{0}^{+}=1/3,\ 0.4050,\ \text{and}\ 0.4470$ respectively. Taking into account that the value of the Casimir amplitude for the $d=3$ Ising universality class with $(+,+)$ boundary conditions \cite{Has2010} is
\begin{equation}\label{Casamppp}
\Delta^{(+,+)}=-0.410(29)
\end{equation}
and that $X_{\mathrm{Cas},\mathrm{sr}}^{(\tau),\mathrm{MF}}(t=\Delta\mu=0)=-1.7315$, for the normalizing coefficients one obtains:
$0.217$ for $\lambda=1$ and  $ 0.147$ when $\lambda=2$.

The dependence of the scaling function $X_{\rm crit}$ on the temperature $x_{t}$ and field $x_{\mu}$ scaling variables is summarized in Figs. \ref{fig:CFspsm} and \ref{fig:CFsmsm}  for different separations $L$.

Once having a good approximation of $X_{\rm crit}$ in $d=3$, obtained in the way described above,  one can determine the scaling function of the  critical Casimir force of the fluid system by adding to $X_{\rm crit}$ the singular part of the Hamaker term given by \eref{2bethaHA_ocf}, i.e., that part of it that depends on $\phi_{b}$.  When performing this procedure one must pay attention to the use of the coupling parameters $s_{i,c},\ i=1,2$, because they also depend on the dimensionality $d$ of the system. Using Eqs. (\ref{v_s_l_def}), (\ref{deltav_s_l_def}) and (\ref{s_def_l}) one has
\begin{equation}\label{sd3sd4relation}
s_{i,c}(d=\sigma=3)=\dfrac{G(3,3)}{G(4,4)}s_{i,c}(d=\sigma=4),\ i=1,2.
\end{equation}

As it is clear from the above, in system governed by short-range interactions $X_{\rm crit}$ coincides with $X_{\rm Cas}^{\rm sr}$; $X_{\rm Cas}^{\rm sr}$ is \textit{negative}, which corresponds to {\it attractive} critical Casimir force, for any value of the scaling variables $x_t$ and $x_\mu$ under $(+,+)$ boundary conditions -- see \eref{XCassranalytic} for $X_{\rm Cas}^{\rm sr}$ in the case $x_\mu=0$, as well as Fig. \ref{fig:short_range_tending}$\mathbf{(a)}$ and $\mathbf{(b)}$. Note that under such boundary conditions one observes an enhancement of the order parameter due to the confinement in comparison to the order parameter at the same distance from an individual wall. On that ground in systems with long-ranged interactions such that $s_{1,c}>0$ and $-s_{1,c}\ll s_{2,c}\leq0$ we expect  $X_{\rm crit}$ to remain again negative for any separation $L$ and at any value of the scaling variables, irrespective of the value of $\lambda$ (which is always non-negative). Indeed, this turns out to be true and is depicted in Fig. \ref{fig:CFspsm}(\textbf{a}) and (\textbf{b}). However when $s_{2,c}\approx-s_{1,c}$ and the separation between the walls is relatively small a significant part of the system is disordered which results in non-negative or sign-changing scaling function [see Fig. \ref{fig:CFspsm}(\textbf{c}) and (\textbf{d})].  As the distance $L$ is increased the influence of the effective surface potential $\Delta V(z)$ [see \eref{DeltaV_l_ab_thichlayers}] quickly decreases and only the additional ordering effect of the fluid-fluid interactions influences the behavior of the order parameter and hence of $X_{\rm crit}$. Along with the same line of arguing, when both wall-fluid coupling parameters are negative, $X_{\rm crit}$ is negative for any $x_{t}$ and $x_{\mu}$ only for very large separations $L$ where the effect of the long-ranged interactions on the behavior of the system is negligible. Naturally, since the short-ranged surface potentials do support $(+,+)$ boundary conditions, the role of the negative substrate potentials, which oppose the order near the boundary, will be stronger than that of the positive substrate potentials which try to reinforce the phase preferred near the boundary. For example, we observe that the behavior of $X_{\rm crit}$ in a system with $s_{1,c}=1.0,\ s_{2,c}=-1.0,\ \lambda=1.0$ and in such with $s_{1,c}=0.0,\ s_{2,c}=-1.0,\ \lambda=1.0$ is almost identical for any $L$, both as a function of $x_{t}$ and $x_{\mu}$ [compare Fig. \ref{fig:CFspsm}(\textbf{c}) and (\textbf{d}) and Fig. \ref{fig:CFsmsm}(\textbf{a}) and (\textbf{b})] \cite{note2}. Thus, for a fixed $\lambda$ the behavior of the scaling function is mainly determined by the interplay between the short-ranged surface fields and the strong negative wall-fluid coupling $s_{2,c}$. If $s_{1,c}=s_{2,c}=-1.0$ and $\lambda<2.0$ the scaling function  $X_{\rm crit}$ exhibits an unexpected behavior as a function of $L$: for moderate values of $L$ the maximum of the {\it repulsive} part of the force increases with increasing $L$ both as a function of $x_{t}$ and $x_{\mu}$ [see Fig. \ref{fig:CFsmsm}(\textbf{c}) and (\textbf{d})]; for larger values of $L$ the maximum decreases, as expected, and the overall behavior of the scaling function approaches that one of the system with completely short-ranged interactions.

The behavior of the critical Casimir force $X_{\rm Cas}$ is depicted on Fig. \ref{fig:ForceInterplay}. In order to illustrate only the main idea, here we restrict ourself to the  choice of parameters $\lambda=2.0$, $s_{2,c}=0.0$ while $s_{1,c}=\pm 1.0$. The data for the Casimir force and the Hamaker term are presented for a system with $L=20$ layers.  First, let us recall that for $T>T_{c}$ at $\Delta\mu=0$ one has $\phi_{b}=0$ and, hence, the behavior of $X_{\rm Cas}$ and that of $X_{\rm crit}$ coincide in this region [see Fig. \ref{fig:ForceInterplay}(\textbf{a}) and (\textbf{c})]. On the other hand, if $T<T_{c}$ one has $\phi_{b}\neq0$, with the  singular (i.e., the $\phi_b$ dependent) part of the Hamaker term corresponding to repulsion for $s_{1,c}>0$ and attraction otherwise, see \eref{2bethaHA_ocf}. For $s_{1,c}=1.0$ the scaling function $X_{\rm crit}$ is negative, but the singular part of the Hamaker term is positive, i.e., repulsive in the low-temperature region. Thus, the Casimir force is attractive near and above $T_c$ but becomes repulsive below $T_c$. For $s_{1,c}=-1.0$ one has that $X_{\rm crit}$ is non-negative, but with the singular part of the Hamaker term being negative, i.e., attractive for $T<T_c$. Thus, the resulting critical Casimir force $X_{\rm Cas}$ changes sign from being repulsive above $T_c$ to becoming attractive  slightly below $T_c$. The  behaviour of the Casimir force as a function of $x_{\mu}$ at $T=T_{c}$ for the two sub-cases  of $s_{1,c}=\pm 1.0$  is depicted on Fig. \ref{fig:ForceInterplay}(\textbf{b}) and (\textbf{d}). For $s_{1,c}=1.0$ one observes that $X_{\rm crit}<0$, while  the singular part of the Hamaker term changes sign from positive to negative with $\Delta \mu$ decreasing. Thus, the resulting Casimir force changes sign once from being slightly positive (repulsive) for $\Delta \mu \gtrsim 0$  to negative (attractive) for $\Delta \mu \lesssim 0$. The case $s_{1,c}=-1.0$ is much more interesting. Since then $X_{\rm crit}$ changes sign twice for $\Delta \mu<0$, while the singular part of the Hamaker term changes sign ones, the resulting Casimir force happens to change its sign tree times as a function of $x_\mu$.

One might pose the question about the phase behavior of the thin fluid systems with such competing surface and substrate potentials. We consider that question in the next section.

\section{Phase behaviour of systems with dispersion forces}\label{sec:PhaseBehaviour}
\begin{figure}[h]
\centering
\includegraphics[width=\columnwidth]{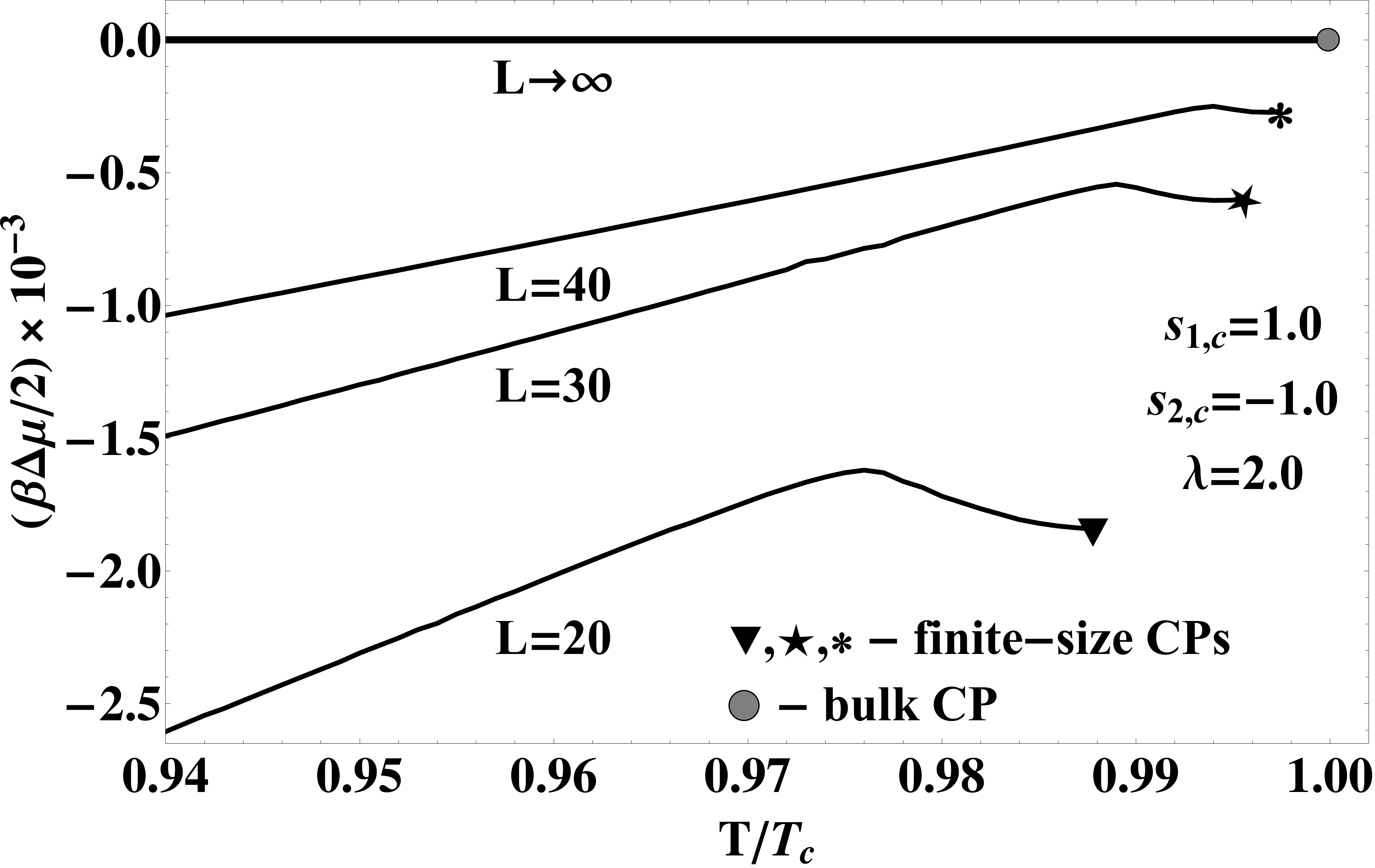}
  \caption{Phase diagrams of a $d=4$ confined finite-size fluid system with strong adsorption of the "liquid" phase of the fluid medium on the confining walls, for different separations $L$ between them. The values of the coupling parameters characterizing the interactions in the system are $s_{1,c}=1.0,\ s_{2,c}=-1.0\ \text{and}\ \lambda=2.0$, and the coordinates of the finite-size critical points for the different separations are: bottom-up $L=20$ ({{\large$\blacktriangledown$}}) $(0.98779783, -1.8421\times10^{-3})$; $L=30$ ({{\large$\star$}}) $(0.99553852, -0.5957\times10^{-3})$; $L=40$ ({{\large$\ast$}}) $(0.99743470, -0.2684\times10^{-3})$.}
  \label{fig:PDdiffL}
\end{figure}
The phase behaviour of a confined fluid medium between parallel walls exerting identical surface adsorption potentials $\delta\mu_{1}\equiv J_{\mathrm{sr}}^{s_{1},l}=\delta\mu_{2}\equiv J_{\mathrm{sr}}^{s_{2},l}\neq0$ on both surface layers has been studied extensively both theoretically and experimentally \cite{OG78,FN81,NF82,NAFI83,NF83,REUM86,BME87,E90,DSD2007,DRB2007,DRB2009}. So far the considerations have been made either for fluid systems governed by pure short-range interactions \cite{FN81,NF82,NAFI83,NF83,REUM86,BME87,E90}, or for systems with surface potentials strongly preferring the same phase of the fluid \cite{OG78,DSD2007,DRB2007,DRB2009}. Here we are going to consider the case when at least one of the wall-fluid potentials favor a phase of the fluid different from the one preferred by the short-ranged surface potentials.

\begin{figure}[t!]
\centering
\includegraphics[width=\columnwidth]{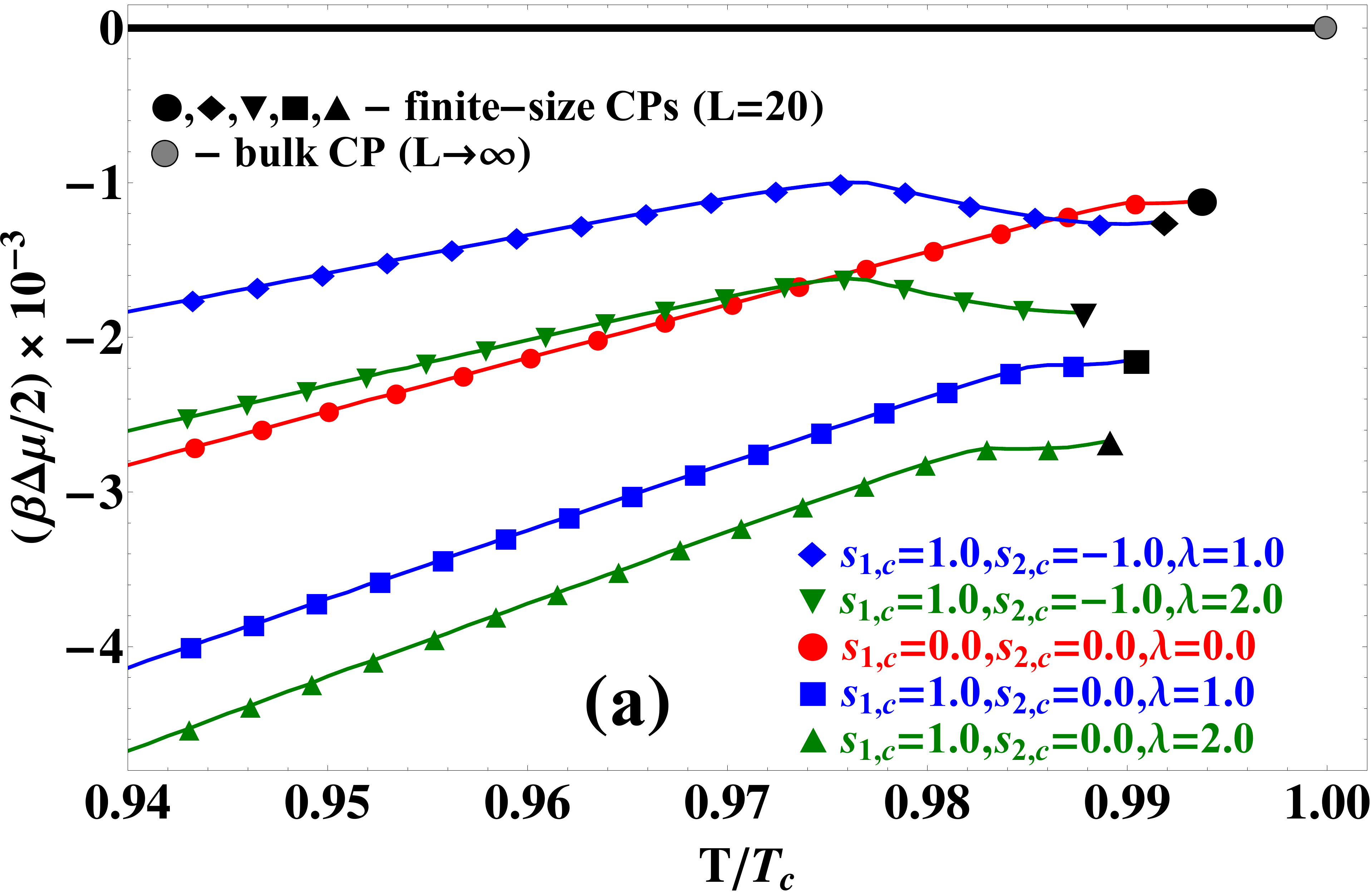}
\includegraphics[width=\columnwidth]{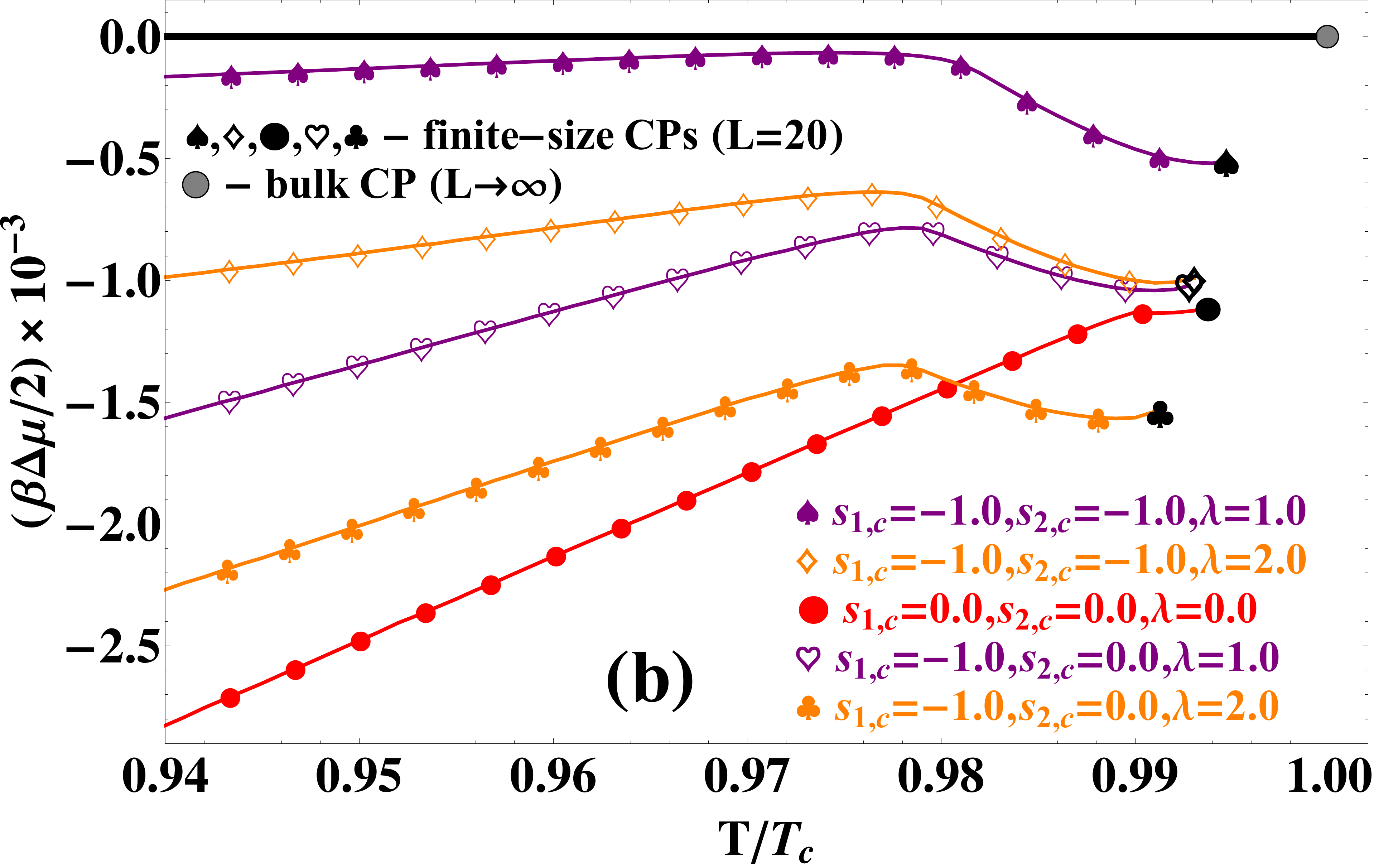}
  \caption{Comparison between the phase diagrams of $d=4$ confined finite-size fluid systems with strong adsorption of the "liquid" phase of the fluid medium on the confining walls, obtained for $L=20$ and various values of the coupling parameters: \textbf{(a)} bottom-up ({\color{green_n}{\large$\blacktriangle$}}) $s_{1,c}=1.0,\ s_{2,c}=0.0,\ \lambda=2.0$; ({\color{blue_n}{$\blacksquare$}}) $s_{1,c}=1.0,\ s_{2,c}=0.0,\ \lambda=1.0$; ({\color{red_n}{\large$\bullet$}}) $s_{1,c}=s_{2,c}=\lambda=0.0$;  ({\color{green_n}{\large$\blacktriangledown$}}) $s_{1,c}=1.0,\ s_{2,c}=-1.0,\ \lambda=2.0$; ({\color{blue_n}{\large$\blacklozenge$}}) $s_{1,c}=1.0,\ s_{2,c}=-1.0,\ \lambda=1.0$; \textbf{(b)} bottom-up ({\color{red_n}{\large$\bullet$}}) $s_{1,c}=s_{2,c}=\lambda=0.0$; ({\color{orange_n}{\large$\clubsuit$}}) $s_{1,c}=-1.0,\ s_{2,c}=0.0,\ \lambda=2.0$; ({\color{purple_n}{\large$\heartsuit$}}) $s_{1,c}=-1.0,\ s_{2,c}=0.0,\ \lambda=1.0$;  ({\color{orange_n}{\large$\diamondsuit$}}) $s_{1,c}=-1.0,\ s_{2,c}=-1.0,\ \lambda=2.0$; ({\color{purple_n}{\large$\spadesuit$}}) $s_{1,c}=-1.0,\ s_{2,c}=-1.0,\ \lambda=1.0$. The coordinates of the finite-size critical points $[T_{c,L}/T_{c},(\mu_{c,L}-\mu_{c})/(k_{B}T_{c,L})]$ of the different systems are: \textbf{(a)} bottom-up ({{\large$\blacktriangle$}}) $(0.98912146, -2.6691\times10^{-3})$; ({{$\blacksquare$}}) $(0.99046179, -2.1410\times10^{-3})$; ({{\large$\blacktriangledown$}}) $(0.98779783, -1.8421\times10^{-3})$; ({\large{$\bullet$}}) $(0.99376528, -1.1101\times10^{-3})$; ({{\large$\blacklozenge$}}) $(0.99185806, -1.2482\times10^{-3})$; \textbf{(b)} bottom-up ({{\large$\clubsuit$}}) $(0.99130666, -1.5386\times10^{-3})$; ({\large{$\bullet$}}) $(0.99376528, -1.1101\times10^{-3})$; ({\large{$\heartsuit$}}) $(0.99279091, -1.0222\times10^{-3})$; ({\large{$\diamondsuit$}}) $(0.99303773, -0.9915\times10^{-3})$; ({\large{$\spadesuit$}}) $(0.99468863, -0.5085\times10^{-3})$. The phase diagram of a bulk system $(L\rightarrow\infty)$ is shown as thick black line ending at the bulk critical point ({\color{gray_n}{\large$\bullet$}}).}
  \label{fig:PDdiffsandlambda}
\end{figure}

\begin{figure*}[t!]
\centering
\mbox{\subfigure{\includegraphics[width=\columnwidth]{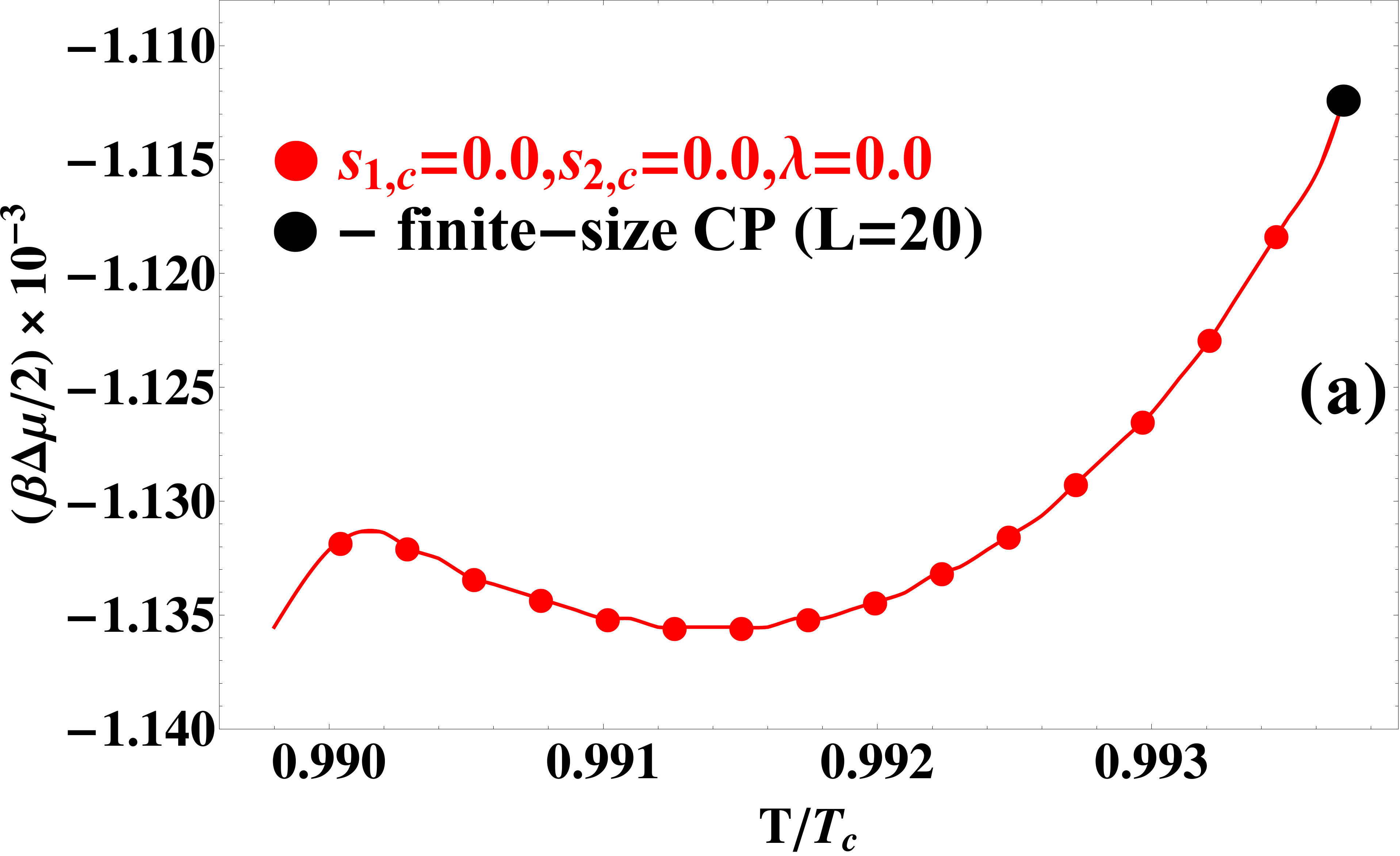}}\quad
      \subfigure{\includegraphics[width=\columnwidth]{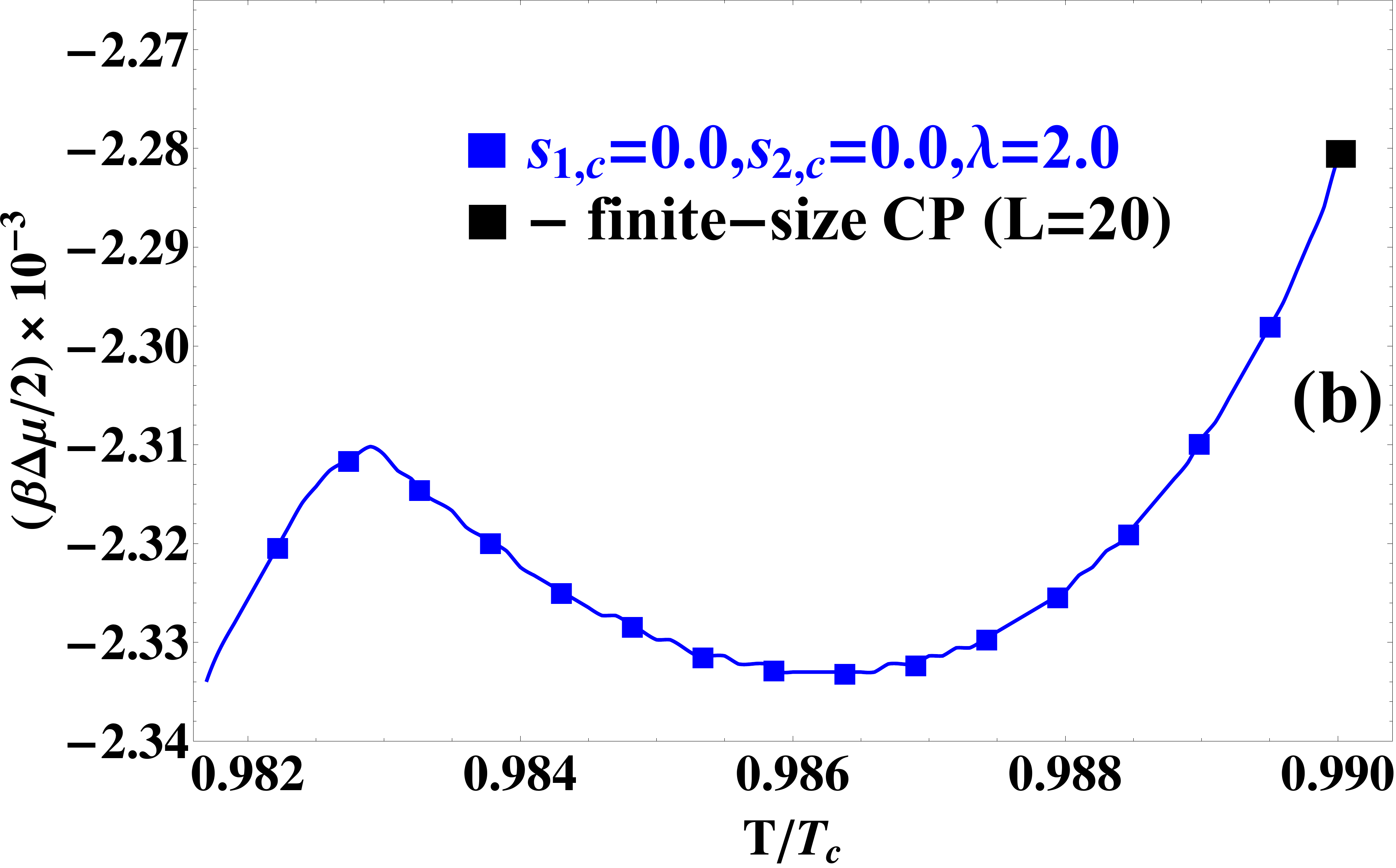}}}\\
\mbox{\subfigure{\includegraphics[width=\columnwidth]{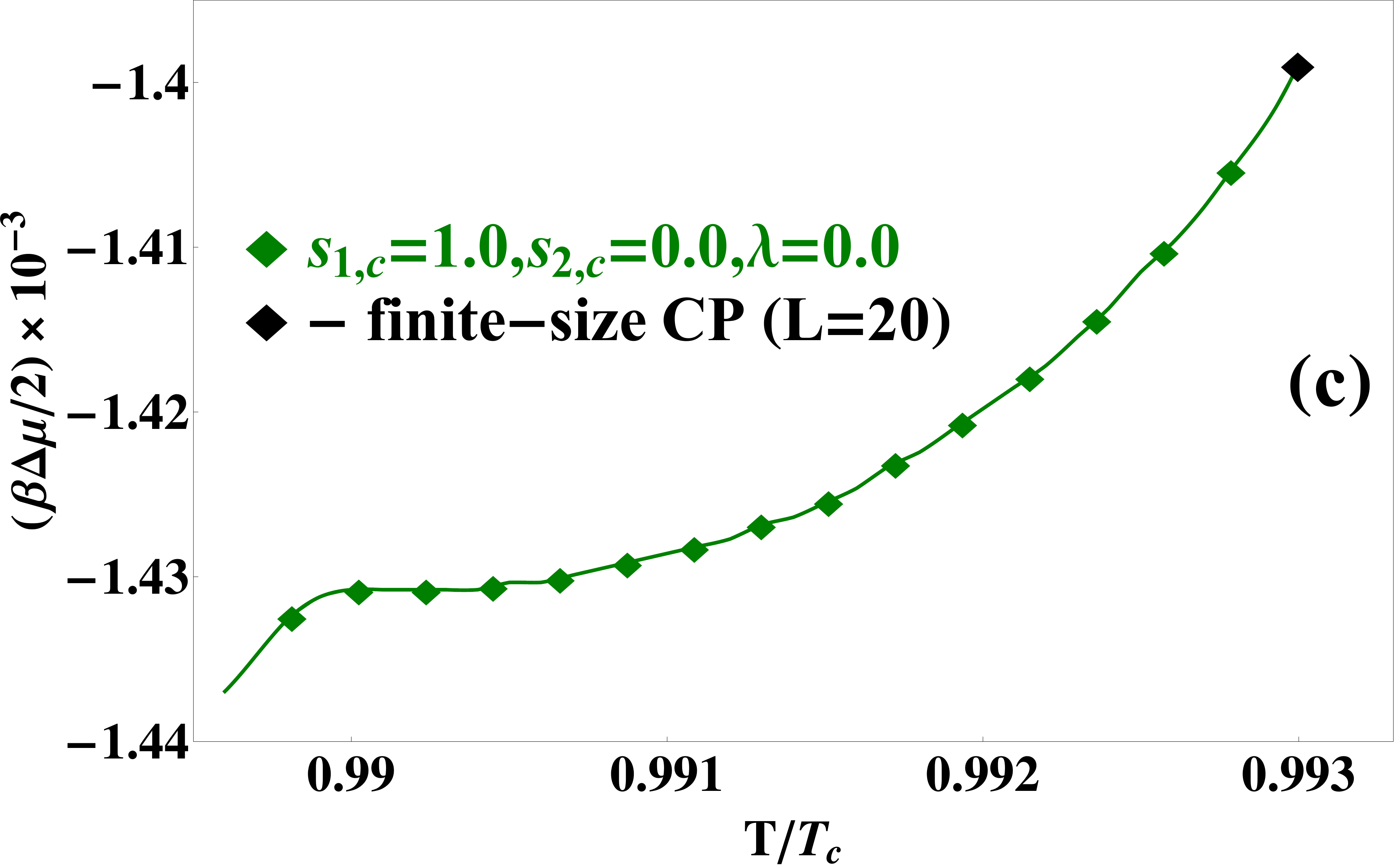}}\quad
      \subfigure{\includegraphics[width=\columnwidth]{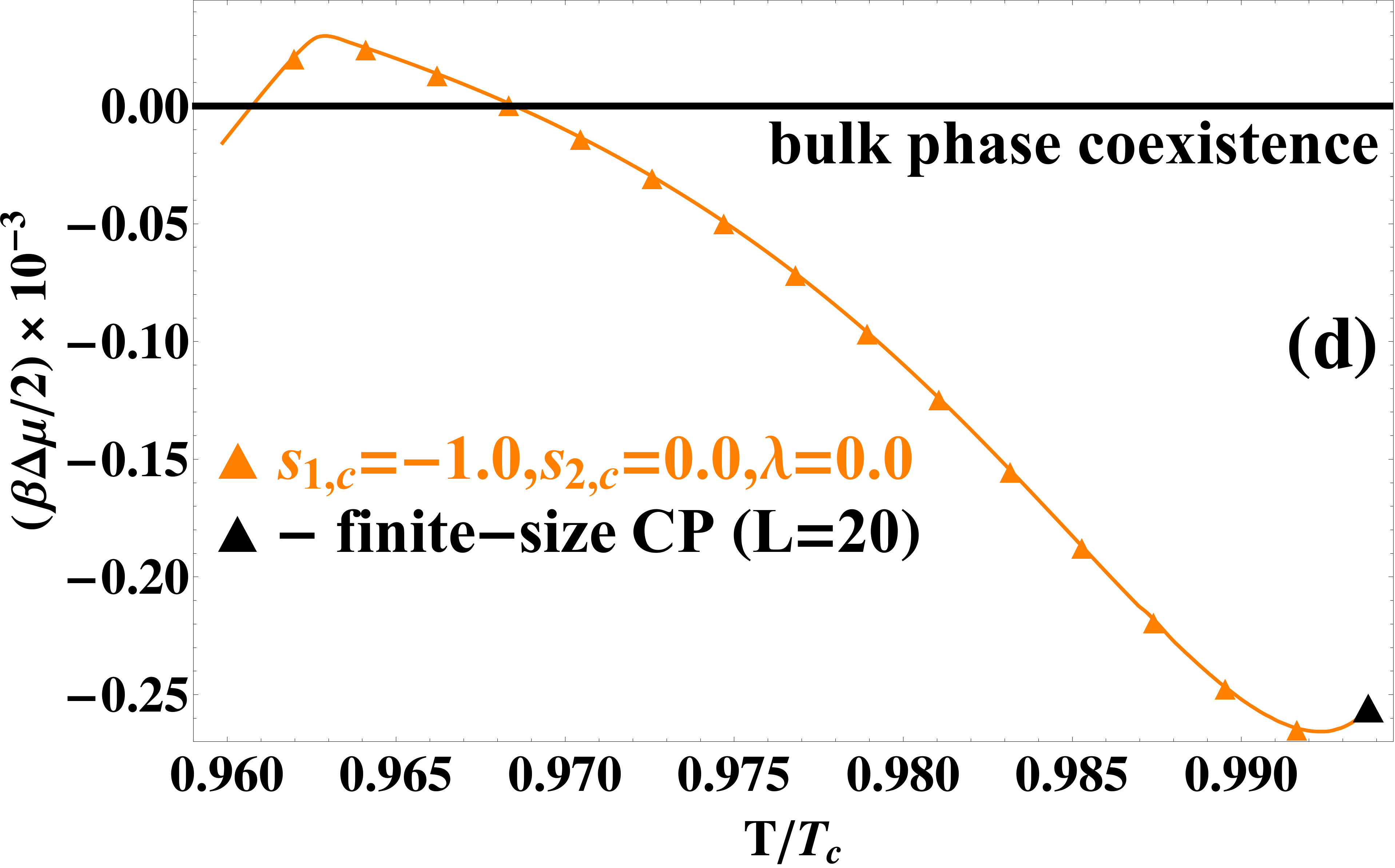}}}
  \caption{Blow-up views of some of the phase diagrams at $d=4$ very close to the capillary condensation critical point $T_{c,L}$ for $L=20$. The influences of the different dispersion interactions (walls-fluid and fluid-fluid) is visualized.  The phase diagram of a short-range system is given in \textbf{(a)} in order to serve as a reference on which the changes due to the influence of fluid-fluid interaction, i.e., on $\lambda$ -- see \textbf{(b)}, or on the wall-fluid interaction, i.e., on $s_{i}$ -- see \textbf{(c)} and \textbf{(d)} is studied.  In \textbf{(b)} only the influence of the long-range fluid-fluid interactions is shown $(\lambda\neq0,\ s_{1,c}=s_{2,c}=0.0)$, while on \textbf{(c)} and \textbf{(d)}, that of a single wall-fluid is investigated $(s_{1,c}\neq0,\ \lambda=s_{2,c}=0.0)$.}
  \label{fig:NCPD}
\end{figure*}

The phase behavior of systems with $L=20, 30, 40$ and of the bulk system with $s_{1,c}=1.0$, $s_{2,c}=-1.0$ and $\lambda=2.0$, obtained within our mean-field model, is presented in Fig. \ref{fig:PDdiffL}. A relatively detailed study of the influence of the different values of $s_{1,c}$, $s_{2,c}$ and $\lambda$ on the phase behavior of thin fluid films (in the case considered $L=20$) is illustrated in Figs. \ref{fig:PDdiffsandlambda} and \ref{fig:NCPD}, where Fig. \ref{fig:NCPD} represents blow-up views of some of the phase diagrams very close to the capillary condensation critical points $T_{c,L}$.
\begin{figure}[h!]
\centering
\includegraphics[width=\columnwidth]{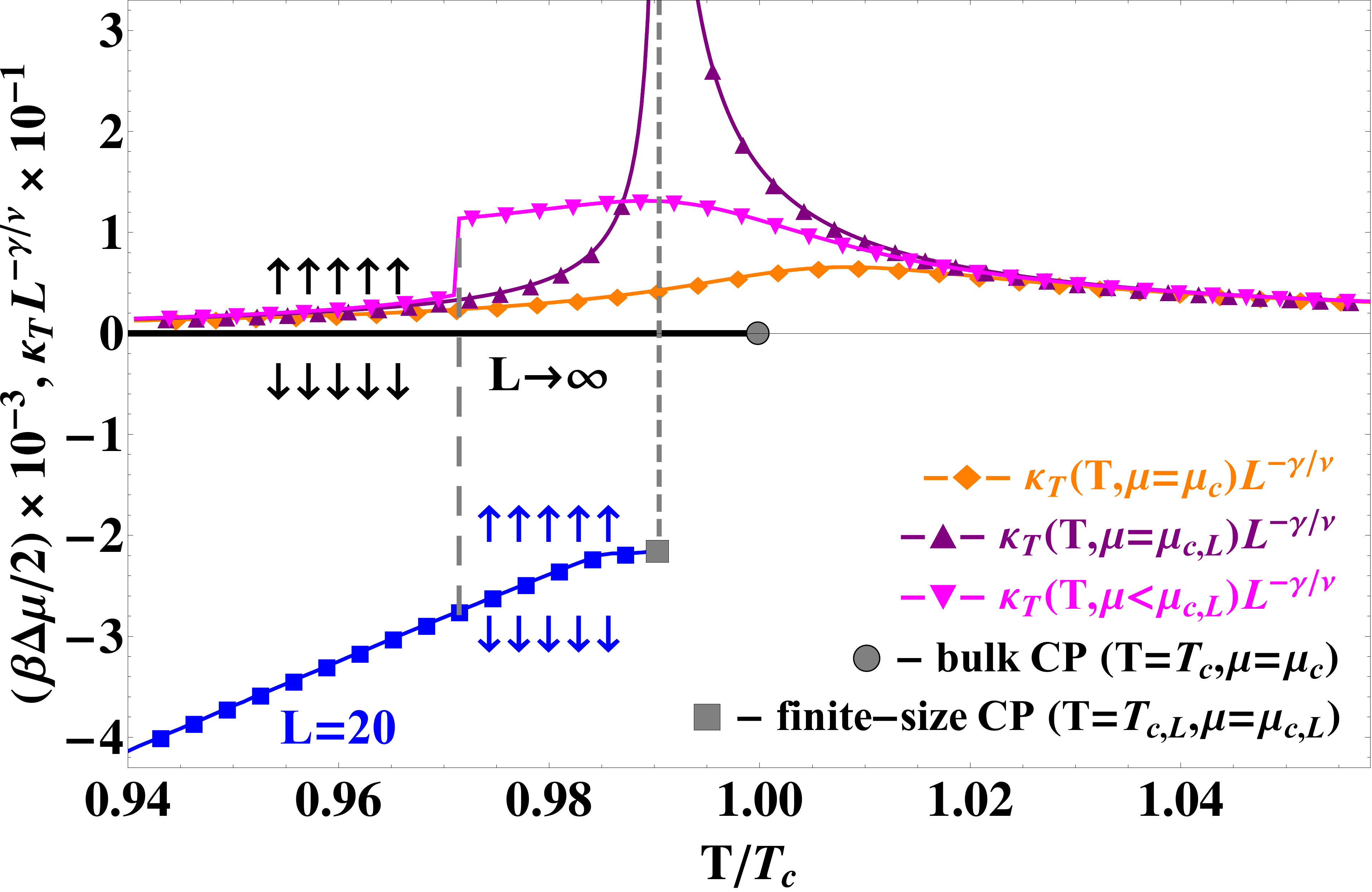}
  \caption{The scaling functions of the total isothermal compressibility $\kappa_{T} L^{-\gamma/\nu}$ [({\color{orange_n}{\large$\blacklozenge$}}), ({\color{purple_n}{\large$\blacktriangle$}}) and ({\color{magenta_n}{\large$\blacktriangledown$}})] and the phase diagram ({\color{blue_n}{$\blacksquare$}}) of a $d=4$ finite-size fluid system, confined between parallel walls both strongly adsorbing the "liquid" phase of the fluid medium, for $L=20$ and values of the coupling parameters $s_{1,c}=1.0,\ s_{2,c}=0.0\ \text{and}\ \lambda=1.0$. The bulk critical point is denoted by the symbol {\color{gray_n}{\large$\bullet$}}, while the finite-size critical point $(T=T_{c,L},\mu=\mu_{c,L})$ is marked by the symbol {{$\blacksquare$}}.  The curve designated by ({\color{orange_n}{\large$\blacklozenge$}}) shows the behavior of $\kappa_{T} L^{-\gamma/\nu}$  at $\mu=\mu_{c}$. This behavior at the finite-size critical point  $(T=T_{c,L},\mu=\mu_{c,L})$ is depicted with the symbol   ({\color{purple_n}{\large$\blacktriangle$}}). When $T<T_{c,L}$ and $\mu<\mu_{c,L}$, the compressibility $\kappa_{T}$ changes with a finite jump ({\color{magenta_n}{\large$\blacktriangledown$}}), being an indicator of a first-order phase transition. The points of first-order phase transition form the coexistence curve, i.e., determine the phase diagram in the phase plane.}
  \label{fig:PDSusceptibility}
\end{figure}
We observe that the overall phase behavior of the system is similar to that one obtained for a short-ranged system by Fisher and Nakanishi \cite{FN81} -- one has a line of a capillary condensation shifted towards the gas bulk phase which terminates at its own finite-size critical point $T_{c,L}$. One can identify the following properties of the phase diagrams presented in Figs. \ref{fig:PDdiffL}, \ref{fig:PDdiffsandlambda} and \ref{fig:NCPD}:

{\it i)} The precise position of the phase diagram depends on $L$.

This property is well known also for systems with short-ranged interaction.

{\it ii)} For a fixed $L$ the precise position of the phase diagram depends on $s_{1,c}$, $s_{2,c}$ and $\lambda$ -- smaller  $L$ stronger the corresponding influence of these parameters.

Negative wall-fluid couplings $s_{1,c}$ and/or $s_{2,c}$ have disordering effect on the system which manifests in increase of the gas phase and decrease of the liquid one near the surfaces. Generally the phase diagram, as then expected, appears above the one of a short-range systems [see Fig.  \ref{fig:PDdiffsandlambda}(\textbf{b})] since in order to "liquify" the system  one now needs higher "pressure".

{\it iii)} The position of $T_{c,L}$, thus the shift of the critical point of the finite system with respect to that one of the bulk one, depends not only on $L$, as in the short-ranged systems, but also on $s_{1,c}$, $s_{2,c}$ and $\lambda$.

For $\lambda\neq 0$ the ordering in the system increases and,  hence, one observes increase in both the critical temperature shifts as well as in the chemical potential difference compared to systems governed by pure short-range interactions. When $s_{1,c}$ and $s_{2,c}$ are both positive this effect is even further enhanced. If, on the other hand, for fixed $\lambda\neq0$ the parameters $s_{1,c}$ and $s_{2,c}$ are both negative or, at least, have different signs, the shift of the finite-size critical point with respect to that one in short-range systems can be completely canceled or, for strong enough potentials, this shift can even further decrease via $T_{c,L}$ moving toward the corresponding bulk critical point.

{\it iv)} The competing effect of $T$, $\Delta\mu$, $s_{1,c}$, $s_{2,c}$ and $\lambda$ on the ordering in the system leads near the capillary condensation critical point for thin films to a non-monotonic behavior of the phase diagram  -- see Fig. \ref{fig:NCPD}. The behavior of the system with at least one of the wall-fluid potentials being negative is especially interesting -- see Fig. \ref{fig:NCPD}(\textbf{d}), where in a small temperature interval at phase coexistence when $T$ decreases $\Delta \mu$ increases.

Below we briefly explain how the above phase diagrams have been calculated.

On Fig. \ref{fig:PDSusceptibility} we illustrate the method used to obtain the phase diagram and the finite-size critical point at which the coexistence curve ends of finite systems characterized by long-ranged fluid-wall potentials. We do that by studying within our mean-field model the scaling functions of the total isothermal compressibility $\kappa_{T} L^{-\gamma/\nu}$. When  $\mu=\mu_{c}$ one has that $\kappa_{T} L^{-\gamma/\nu}$ is a smooth function of the temperature, even in the vicinity of the bulk critical temperature, reaching its maximum for $T>T_{c}$. For $\mu_{c,L}<\mu<\mu_{c}$ the value of the maximum increases, being still finite, and  appearing at $T<T_{c}$. Upon reaching the finite-size critical point $(T=T_{c,L},\mu=\mu_{c,L})$, $\kappa_{T} L^{-\gamma/\nu}$ diverges, giving rise to power-law singularity. At temperatures $T<T_{c,L}$ and chemical potentials $\mu<\mu_{c,L}$, the value of $\kappa_{T} L^{-\gamma/\nu}$ changes with a finite jump, being an indicator of a first-order phase transition. The points of first-order phase transition form a line -- the coexistence curve in the phase plane, separating the observed phases: spin-"up" phase in magnetic systems, liquid phase in one-component fluid systems, A-rich phase in binary liquid mixtures and, correspondingly, spin-"down" phase, gas phase, B-rich phase. At the critical point and beyond the differences between these phases disappear. The resulting phase diagrams are shown in Fig. \ref{fig:PDdiffsandlambda}, where (\textbf{a}) summarizes the curves with $s_1=1.0$, and (\textbf{b}) -- those with $s_1=-1.0$, where different combinations of values of $s_2$ and $\lambda$ are considered. The line marked with filled (red) circles represents the phase diagram for a system with completely short-ranged interaction.

As it was shown in Refs. \cite{DRB2007,DRB2009}, for the total isothermal compressibility per particle (or susceptibility, if one considers magnetic systems), one has
\begin{equation}\label{totalcomp}
\kappa_{T}=\dfrac{1}{L+1}\sum_{z}\kappa_{T}(z)=\dfrac{1}{L+1}\sum_{z,z^{*}}\left({\bf\rm{R}}^{-1}\right)_{z,z^{*}},
\end{equation}
where ${\bf\rm{R}}^{-1}$ is the inverse of the matrix ${\bf\rm{R}}$ with elements
\begin{equation}\label{Rmatel}
R_{z,z'}=\dfrac{\delta_{z,z'}}{1-\phi^{2}(z')}-\beta J^{l}(z-z'),
\end{equation}
and
\begin{eqnarray}\label{localcomp}
&&\kappa_{T}(z)\equiv\sum_{z^{*}}G(z,z^{*})\equiv\sum_{z^{*}}\left({\bf\rm{R}}^{-1}\right)_{z,z^{*}}\nonumber\\
&&=\dfrac{1}{4}\sum_{\rm{\bf{r}}^{*}}\left[\langle \phi({\rm{\bf{0}}},z)\phi({\rm{\bf{r}}}_{\parallel}^{*},z^{*})\rangle-\langle \phi({\rm{\bf{0}}},z)\rangle\langle \phi({\rm{\bf{r}}}_{\parallel}^{*},z^{*})\rangle\right],
\end{eqnarray}
is the "local" isothermal compressibility, which reflects the response of the system from a given layer due to the change of the external field in that layer. More precisely, in \eref{localcomp} $G(z,z^{*})$ is the density-density correlation function, $G(z,z^{*})=\delta\phi(z)/[2\delta h(z^{*})]$, where the functional derivative is taken with respect to the field $h(z^{*})=\beta[\Delta\mu-\Delta V(z^{*})]/2$.

In order to determine $\kappa_{T}(x_{t},x_{\mu}|L,\{s_{i},\ i=1,2\},\lambda)$ or its "scaling function" $X_{\kappa}\equiv L^{-\gamma/\nu}\kappa_{T}$ in a fluid film with thickness $L$, we first solve numerically \eref{order_parameter_equation_d4} within our mean-field model, which allows us to determine the matrix ${\bf\rm{R}}$ with the use of \eref{Rmatel}. After that, using \eref{localcomp}, we obtain the "local" isothermal compressibility $\kappa_{T}(z)$, and summing over $z$, the total isothermal compressibility. We recall that within the mean-field treatment one has $\nu=1/2$ and $\gamma=1$.

The phase diagrams depicted on Fig. \ref{fig:PDdiffsandlambda} result from the interplay between the fluid-fluid and walls-fluid interactions. Note that the increase of $\lambda$ at fixed $s_{i,c},\ i=1,2$ lowers $\beta\Delta\mu$ at coexistence, while the decrease of the wall-fluid coupling at one of the walls (say $s_{2,c}$) keeping $\lambda$ and $s_{1,c}$ constants, increases it.
\section{Experimental feasibility of the predicted effect}\label{sec:ExperimentalReal}
In the fast emerging field of micro- and nano- devices, the fluctuation induced and dispersion forces like the Casimir and van der Waals ones, associated with the interaction between the individual components of such devices, play an essential role. Due to the quantum origin of these forces, currently there are no theoretical as well as experimentally suggested control parameters that can be used to conveniently and reversibly govern the sign and the strength of these forces if the interacting objects are in vacuum. The situation changes if these interactions take place in a fluid. Very recently in Ref. \cite{DLBBP2014} the authors suggested controlled quantum Casimir levitation caused by the introduction of thin film surface coating on a porous planar substrates both immersed in a suitable fluid. More specifically, they have shown that the Casimir force between Teflon and cassiterite (${\rm SnO_2}$) nanosheet immersed in cyclodecane is attractive at large separations but repulsive at small separations, resulting in a stable equilibrium distance where the total force is zero. In Ref. \cite{BEKK2011} the authors reported the use of aerogels, yielding repulsion down to submicron distances at realistic porosities. The described so far set-ups are quite close in spirit to the considerations presented in the current article with the basic difference that the force will be due to the critical fluctuations of the fluid, instead of the fluctuation of the electromagnetic field. We suppose that the usage of low density substrates like aerogels can be also applied in the thermodynamic Casimir effect since the parameters $s_{i}$ directly depend, as shown in Eqs. (\ref{deltav_s_l_def}) and (\ref{s_def_l}), on the densities of the slabs.
\begin{table}[h!]
\centering
\caption{\label{table_fluid_prop} Physical characteristics of the considered confined fluids (column 1) -- the critical density $\rho_{c}$ in units $\rm{g/cm^{3}}$ (column 2), temperature $T_{c}$ measured in $\rm{K}$ (column 3), pressure $P_{c}$ in $\rm{MPa}$ (column 4), the lattice constant $a_{0}$ in {\AA} (column 5), the distance $r_{0}^{l}$ also in {\AA} (column 6) at which the inter-particle potential is zero and the potential well depth $J_{\rm sr}^{l}$ in units $k_{B}T_{c}$ (column 7). The values of $\rho_{c},\ T_{c}$ and $P_{c}$ for argon are taken from Ref. [\onlinecite{HuFr1981}], for krypton from Ref. [\onlinecite{ND2003}] and for xenon from Ref. [\onlinecite{SK94}]. The value of $a_{0}$ is calculated based on the data for $\rho_{c}$ (see the text) and the presented values of $r_{0}^{l}$ and $J_{\rm sr}^{l}$ are taken from Ref. [\onlinecite{RCCGS92}].}
\begin{tabular}{cccccccccccccccccccccc}
\hline \hline
1  &&& 2 &&& 3   &&& 4   &&& 5 &&& 6 &&& 7\\
\hline \hline
$\text{fluid}$  &&& $\rho_{c}$ &&& $T_{c}$   &&& $P_{c}$   &&& $a_{0}$ &&& $r_{0}^{l}$ &&& $\beta_{c}J_{\rm sr}^{l}$\\
\hline \hline
$\rm{Ar}$          &&& 0.536 &&&  150.663   &&&  4.86 &&& 4.98 &&& 3.867 &&& 0.798\\
$\rm{Kr}$          &&& 0.908 &&&  209.480   &&&  5.53 &&& 5.35 &&& 4.165 &&& 0.790\\
$\rm{Xe}$          &&& 1.113 &&&  289.765   &&&  5.84 &&& 5.81 &&& 4.512 &&& 0.794\\
\hline \hline
\end{tabular}
\end{table}
\begin{table*}[th!]
\centering
\caption{\label{table_fluid_sub} Physical characteristics of the confining silver (Ag), gold (Au) and carbon based (C) substances and of their interactions with the fluid media -- the distances $r_{0}^{s}$ and $r_{0}^{l,s}$ in {\AA} (columns 2 and 4) at which the inter-particle potential within the substrate and between it and the fluid is zero, the corresponding potential well depths $J_{\rm sr}^{s}$ and $J_{\rm sr}^{l,s}$ in units $k_{B}T_{c}$ (columns 3 and 5), the density $\rho$ of the substrates in $\rm{g/cm^{3}}$ (column 6), the number density $\rho_{\rm nd}$ (column 7), the substrate-fluid coupling parameter evaluated at the critical temperature -- $s_{c}$ (column 8) and the Young modulus $E$ of the corresponding substrate in GPa (column 9). The values of $r_{0}^{s}$ and $J_{\rm sr}^{s}$ are taken from Ref. [\onlinecite{RCCGS92}], while those of $r_{0}^{l,s}$ and $J_{\rm sr}^{l,s}$ are calculated via Eqs. (\ref{Kong_rules}). The density of the CAS (the bottom value in column 6) as well as its Young modulus (the bottom value in column 9) are taken from Ref. [\onlinecite{KOI2012}].}
\centering
\begin{tabular}{ccccccccccccccccccccccccc}
\hline \hline
1  &&& 2 &&& 3 &&& 4 &&& 5   &&& 6   &&& 7 &&& 8 &&& 9\\
\hline \hline
$\text{fluid/substrate}$  &&& $r_{0}^{s}$ &&& $\beta_{c}J_{\rm sr}^{s}$ &&& $r_{0}^{l,s}$ &&& $\beta_{c}J_{\rm sr}^{l,s}$ &&& $\rho$ &&& $\rho_{\rm nd}$ &&& $s_{c}$ &&& $E$\\
\hline \hline
$\rm{Ar/Ag}$           &&&       &&& 0.120 &&& 3.597 &&&0.258  &&&         &&& 3.62 &&&1.12   &&&     \\
$\rm{Kr/Ag}$           &&& 3.148 &&& 0.087 &&& 3.804 &&&0.194  &&& 10.49   &&& 4.49 &&&1.00   &&&83  \\
$\rm{Xe/Ag}$           &&&       &&& 0.063 &&& 4.061 &&&0.142  &&&         &&& 5.74 &&&0.89   &&&     \\
\hline
$\rm{Ar/Au}$           &&&       &&& 0.130 &&& 3.628 &&&0.278  &&&         &&& 3.65 &&&1.29   &&&     \\
$\rm{Kr/Au}$           &&& 3.239 &&& 0.094 &&& 3.831 &&&0.211  &&& 19.30   &&& 4.52 &&&1.17   &&&79  \\
$\rm{Xe/Au}$           &&&       &&& 0.068 &&& 4.083 &&&0.156  &&&         &&& 5.78 &&&1.06   &&&     \\
\hline
$\rm{Ar/C}$            &&&       &&& 0.351 &&& 3.864 &&&0.525  &&&         &&& 0.043&&&-0.788 &&&      \\
$\rm{Kr/C}$            &&& 3.851 &&& 0.252 &&& 4.033 &&&0.428  &&& 0.014   &&& 0.054&&&-0.779 &&&3.86  \\
$\rm{Xe/C}$            &&&       &&& 0.182 &&& 4.245 &&&0.341  &&&         &&& 0.069&&&-0.782 &&&      \\
\hline \hline
\end{tabular}
\end{table*}

In order to show with concrete examples that the experimental observation of the effects theoretically predicted in the current article  is possible, we consider the occurrence of critical fluctuations in simple non-polar fluids such as argon (Ar), krypton (Kr) or xenon (Xe), confined between graphene coated carbon nanotube aerogel substrate (CAS) \cite{KOI2012} and a silver (Ag) or gold (Au) one. Both confining surfaces are assumed coated by monatomic (or very thin) lead (Pb) film, ensuring the (+,+) boundary conditions. The basic data needed for the calculations concerning the considered fluids are presented in Table \ref{table_fluid_prop}. Furthermore, from the data reported in Ref. [\onlinecite{RCCGS92}] we have that $\beta_{c}J_{\rm sr}^{\rm Ar,Pb}=1.79,\ \beta_{c}J_{\rm sr}^{\rm Kr,Pb}=1.32$ and $\beta_{c}J_{\rm sr}^{\rm Xe,Pb}=0.96$, where $\beta_{c}\equiv1/(k_{B}T_{c})$, see Table \ref{table_fluid_prop}, is specific for each fluid medium. As far as the geometry of the experimental set-up matters, one can choose to study the predicted behavior of the force either in plane parallel geometry \cite{BCOR2002,ABBCGMR2009,NLBdeBMEW2012}, or consider such where one of the plates is flat and the other has the shape of a spherical lens with large $(>1\ {\rm cm})$ curvature radius \cite{SKDL2011}. As pointed out in Ref. [\onlinecite{ABBCGMR2009}], the difficulties in the usage of the plane parallel geometry are bigger, which leads to lower accuracy of the results in comparison to the other configuration. The use of plates in the form of spherical lens to study effects predicted for systems composed out of flat parallel surfaces can be mathematically justified, as usual, by the Derjaguin \cite{D34,DV2012} or the gradient expansion \cite{BEK2011,BEK2014b,FLM2014} approaches, which are applicable for calculations of the interactions between curved objects.

In what follows we  evaluate the coupling parameters $s_{c}$ and $\lambda$ which are the basic parameters reflecting the material properties in the theory for the behavior of the force presented above -- see Figs. \ref{fig:CFspsm} -- \ref{fig:ForceInterplay}. From Eqs. (\ref{deltav_s_l_def}) and (\ref{s_def_l}) we have that $s_{c}\equiv0.5G(d,\sigma)[\rho_{\rm nd}\beta_{c}J^{l,s}-\rho_{c}\beta_{c}J^{l}]$, and thus, in order to determine $s_{c}$, one needs to know the long-range inter-particle interaction energies $J^{l,s}$ and $J^{l}$, as well as the number density $\rho_{\rm nd}$ (see column 7 in Table \ref{table_fluid_sub}) of the confining substrate relative to that of the fluid medium $\rho_{c}$. Since within the  mean-field theory the number density of the fluid is $\rho_{c}=1/2$ one has $\rho_{\rm nd}=0.5\rho[\rm mol/cm^{3}]/\rho_{c}[\rm mol/cm^{3}]$, with $\rho[{\rm mol/cm^{3}}]=\rho[{\rm g/cm^{3}}]/(A_{{\rm r}}u N_{A})$, where $A_{{\rm r}}$ is the standard atomic weight of the substance, $u$ -- the atomic mass unit and $N_{A}$ is the Avogadro constant. From here one can also estimate the lattice constant of the fluid medium at the critical point;  one has $a_{0}\ \text{[{\AA}]}=10^{10}\times\{\rho[{\rm g/cm^{3}}]/(A_{{\rm r}}u)\}^{-1/3}$ (see column 5 in Table \ref{table_fluid_sub}). We will make the assumption that all interactions between the constituents of the system are of Lennard-Jones type, i.e., the interaction potential can be written in the form
\begin{eqnarray}\label{LJpot}
w_{\rm LJ}(r)&&=4J_{\rm sr}\left[\left(\dfrac{r_{0}}{r}\right)^{12}-\left(\dfrac{r_{0}}{r}\right)^{6}\right]\nonumber\\
&&=J_{\rm sr}\left[\left(\dfrac{r_{m}}{r}\right)^{12}-2\left(\dfrac{r_{m}}{r}\right)^{6}\right],
\end{eqnarray}
where $r_{m}=2^{1/6}r_{0}$ is the distance at which the potential reaches its minimum, $r_{0}$ is such that $w_{\rm LJ}(r_{0})=0$ and $J_{\rm sr}$ is the depth of the potential well. Therefore one has that $J^{l,s}=2J_{\rm sr}^{l,s}$ (see columns 5 in Table \ref{table_fluid_sub}) and $\lambda\equiv J^{l}/J_{\rm sr}^{l}=2$ (see column 7 in Table \ref{table_fluid_prop}). We evaluate the cross interaction potential parameters $J_{\rm sr}^{l,s}$ and $r_{0}^{l,s}$ based on Kong's combining rules, because of their appreciated accuracy \cite{DelMil2001}
\begin{subequations}\label{Kong_rules}
\begin{equation}\label{Kong_one}
J_{\rm sr}^{l,s}\left(r_{0}^{l,s}\right)^{6}=\left[J_{\rm sr}^{l}\left(r_{0}^{l}\right)^{6}J_{\rm sr}^{s}\left(r_{0}^{s}\right)^{6}\right]^{1/2},
\end{equation}
\begin{equation}\label{Kong_one}
J_{\rm sr}^{l,s}\left(r_{0}^{l,s}\right)^{12}=\dfrac{J_{\rm sr}^{l}\left(r_{0}^{l}\right)^{12}}{2^{13}}\left\{1+\left[\dfrac{J_{\rm sr}^{s}\left(r_{0}^{s}\right)^{12}}{J_{\rm sr}^{l}\left(r_{0}^{l}\right)^{12}}\right]^{1/13}\right\}^{13}.
\end{equation}
\end{subequations}
After all that we are ready to compute the key parameter $s_c$ that influences the behavior of the force for moderate values of $L$. The obtained results are reported in column 8 of Table \ref{table_fluid_sub}). One observes that, depending on the choice of materials, one can indeed have both positive and {\it negative} parameters $s_c$. The existence of combination of materials for which $s_c$ is negative represents the main prerequisite for the experimental relevance of the effects predicted in the current article.

Although the mean-field theory gives poor quantitative estimation of the behavior of the scaling function $X_{\rm Cas}$, it is tempting to, nevertheless, evaluate the corresponding force $F_{\rm Cas}$ based on it. To do so we consider the interaction between a CAS planar substrate and a gold disk with radius $R=4\ \mu{\rm m}$ and thickness $h=2\ \mu{\rm m}$, immersed in xenon. We will assume that the separation between the confining surfaces is $L=20a_{0}\simeq12\ {\rm nm}$. Then for $\mu=\mu_{c}$ one shall observe double sign change of the force with the occurrence of two maxima, at which the force is repulsive and a minimum, at which it is attractive. The magnitude of the force at the maxima is predicted to be $F_{\rm Cas,\ \max_{1}}\simeq16\ {\rm nN}$ and $F_{\rm Cas,\ \max_{2}}\simeq7.6\ {\rm nN}$, observed at $T\simeq283.5\ {\rm K}$ and $T\simeq300\ {\rm K}$ respectively. The minimum is found at $T\simeq291\ {\rm K}$ and its value is $F_{\rm Cas,\ \min}\simeq-24\ {\rm nN}$. Finally we predict that the two sign changes occur at $T\simeq288.9\ {\rm K}$ and $T\simeq295\ {\rm K}$ respectively. Note that the predicted values of $F_{\rm Cas}$ are significantly larger than the weight of the disk, which is approximately $2\times10^{-2}\ {\rm nN}$. We stress  that the predicted magnitude of the force is well in the range of the one reported in Ref. [\onlinecite{GMHNHBD2009}], where the authors estimated that an object with an effective interaction area ${\cal A}=1\ \mu {\rm m}^{2}$ at a distance $L=10\ {\rm nm}$ to a wall  experiences a force $F_{\rm Cas}\simeq 4\ {\rm nN}$ when immersed in a fluid with $T_{c}=300\ {\rm K}$ (about the critical one of a water-2,6-lutidine mixture). Let us also mention that although the presented study is limited to the interaction between parallel planar surfaces through some fluid medium, it can be straightforwardly extended to objects of various geometries, e.g., via using say the Derjaguin or the gradient expansion approaches to calculate the interactions between curved objects. Then one can study also the behavior of colloidal particles immersed in a critical fluid or in a binary mixture close to its consulate point. Thus, another possible application of our findings is for governing the behavior of colloids in a critical solvent -- imagine coated colloid particles. Then, under proper choice of the core material of the colloid and of the coating the particles will have a stable equilibrium distance from the bounding plate that will depend on the temperature and the pressure of the fluid. Thus, we hope that our result can be used as a guide to the design of new experiments on colloid systems.
\section{Summary and Concluding Remarks}\label{sec:DisandConcRem}
In the current article, in view of future experiments exploring and potential devices utilizing  the critical Casimir forces in classical one- or two-component fluids confined by parallel substrates at a distance $L$ from each other,  we studied the behavior of the critical and the near critical Casimir force and its interplay with the van der Waals force, as well as the net force due to both of them. In the envisaged set-up the walls of the slabs are considered coated by thin layers exerting strong preference to a given phase of the fluid modeled by strong adsorbing local surface potentials, while the slabs, on the other hand, influence the fluid by long-range dispersion potentials at least one of which supports the opposite phase of the fluid. The fluid-fluid interactions are assumed to decay as $r^{-(d+\sigma)}$, with $r$ being the distance between the particles of the medium, while the walls-fluid interactions decay as $z^{-\sigma}$ where $z$ is the fluid particle-single wall separation. The strengths of the dispersion interactions in the system are depicted via the dimensionless coupling parameters -- $\lambda$ and $s_i,\ i=1,2$, accounting for the fluid-fluid and slabs-fluid interactions, respectively. While $\lambda$ is always non-negative, for the slabs-fluid coupling parameters $s_1$ and $s_2$ we assumed that $s_1\lessgtr 0$ while $s_{2}\leq 0$. The confined fluid medium is modeled within the framework of the lattice gas model defined in Sec. \ref{sec:Model}. After developing a theory for arbitrary dimension $d$ and long-range decay exponent $\sigma$ of the dispersion interactions -- see Appendix \ref{sec:HamakerTerm} and \ref{sec:FinSizeBehav}, we studied in Sec. \ref{sec:Results} the forces in $d=3$ systems, assuming that the dispersion interactions are of "genuine" van der Waals type, i.e., $d=\sigma=3$. Away from the critical region the total force $f_{\rm tot}$ between the plates of the bounded fluid is simply proportional to the Hamaker constant -- see \eref{way} and \eref{HamConstant}, while near $(T=T_c, \Delta \mu=0)$ one has additional contribution to the force $f_{\rm tot}$ which is proportional to $L^{-d}X_{\rm crit}$, where $X_{\rm crit}$ is a scaling function reflecting the role of the critical fluctuations of the order parameter -- see \eref{CasimirF_l}.

Using general scaling arguments, as well as explicit mean-field model calculations, we have obtained the following main results:

(1) In terms of some critical thickness $L_{\rm crit}$ of the thin films -- see Eqs. (\ref{lcrit}) and (\ref{cL1}), we have established a criterion for the importance of dispersion forces within the critical region of the system. For $L\lesssim L_{\rm crit}$ the contributions of the dispersion forces are  important and cannot be neglected neither within the critical region of the system, including the bulk critical point, nor outside of that region -- see Figs. \ref{fig:short_range_tending}, \ref{fig:CFspsm} and  \ref{fig:CFsmsm} where the validity of these statements is clearly visualized. In the opposite case, i.e., when the separation $L$ between the walls is much larger than  $L_{\rm crit}$ the dispersion forces provide only corrections to the leading behavior of the critical Casimir and the total force within the critical region of the system. Outside the critical region, however, i.e., for $T\neq T_{c},\ \Delta\mu\neq0$, the influence of the dispersion (van der Waals) interactions becomes essential for the behavior of the total force between the bounding surfaces. This is of experimental importance because it is difficult to thermodynamically position the system right at the critical point.

(2) In Appendix \ref{sec:HamakerTerm} we have derived expressions for the behavior of the Hamaker term in the total force between the plates -- see \eref{Hamsumfinall} and \eref{bethasigmaminusoneHAl}, that allow for a study of the temperature and the field dependencies of the Hamaker constant including that ones within the critical region of the system. The behavior of this part of the total force as a function of the scaling variables $x_t$ and $x_\mu$ for different values of $s_{1}$, $s_{2}$ and $\lambda$ is shown in Fig. \ref{fig:ForceInterplay}.

(3) We demonstrated that for a suitable set of slabs-fluid and fluid-fluid coupling parameters the competition between the effects due to the coatings and the slabs can result in {\it sign change} of $X_{\rm crit}$ -- see Figs. \ref{fig:CFspsm} and \ref{fig:CFsmsm}, as well as of the Casimir force -- see Fig. \ref{fig:ForceInterplay}, when one changes the temperature $T$, the chemical potential of the fluid $\mu$, or $L$. The last can be used in designing nano-devices  and for governing behavior of objects at small, below micrometer, distances.

(4) In addition to the critical behavior of the considered forces, in Sec. \ref{sec:PhaseBehaviour} we discussed and gave detailed explanation for the near critical behavior of the phase diagram, see Figs. \ref{fig:PDdiffL}, \ref{fig:PDdiffsandlambda} and \ref{fig:NCPD}, of finite-size systems governed by dispersion interactions. The influence of these interactions on the shift of the finite-size critical point has also been commented.

(5) In the final Sec. \ref{sec:ExperimentalReal} in scope of possible experimental realization of the theoretical set-up, we explored the usage of concrete substances. For each pair of substrate and fluid we calculated the coupling parameters $s_{c}$ and $\lambda$, and used this knowledge to make estimations for the critical Casimir force $F_{\rm Cas}$ at several temperatures where it exhibits extrema or changes its sign. We have demonstrated that by proper choice of the fluid and substrate materials one can indeed achieve $s_c>0$ and $s_c<0$, which is the main prerequisite for the experimental feasibility of the effects predicted in the current article.
	
Let us stress that the problem of quantitative description of the mutual influence of the fluctuation of the electromagnetic field and the order parameter fluctuations of a medium when it is close to its critical point is extremely complicated and currently there is no general theory available to scope with it. The Lifshitz theory -- which is the basic one for studying the QED Casimir effect has never been meant to nor can deal with the problem of a critical medium between two other media. The main quantity which knowledge is required for practical applications of this theory is the dielectric permittivity $\varepsilon(\omega)$. It is normally tabulated at some temperature, usually the room  temperature.  It shall be noted, however, that in a critical fluid $\varepsilon(\omega)$ is itself a singular function of the temperature \cite{SBMG80,MRDJ2003}. We are not aware even of a theory that can reliably predict quantitatively how $\varepsilon(\omega)$ will depend on the temperature and $\omega$ near the critical point of the medium for a specific material characterized by some characteristic spectrum. On the other side -- when studying critical phenomena one normally starts with some effective Hamiltonian where only few basic features of the critical medium are reflected. The current article provides a uniform treatment of the contributions due both to the van der Waals forces and the critical Casimir forces in the framework of an, unavoidably, relatively simple model in which, however, all the calculations have been done on equal footing. The expression that we derived for the Hamaker term is, nevertheless, in full agreement with the Dzyaloshinskii-Lifshitz-Pitaevskii theory \cite{L56,DLP61}. On the other side the mean-field theory is considered as a reliably theoretical horse for the qualitative description of the critical phenomena.

\acknowledgments
G. V. gratefully acknowledges the support of the Bulgarian NSF under contract No. DMU 03/37. D. D. thanks S. Dietrich for helpful comments and the critical reading of the manuscript of this article.  The discussions with A. Maci{\`o}{\l}ek are also gratefully acknowledged. 
\appendix
\section{The Hamaker term for a van der Waals system of two different substrates separated by fluid medium}\label{sec:HamakerTerm}

In this Appendix we derive an expression for the Hamaker term for a van der Waals system of two different substrates separated by fluid medium. We follow the same lines of action when performing the calculations as in Ref. \cite{DSD2007} with the important generalization that now, within our model (see Sec. \ref{sec:Model}), the substrates on both sides of the fluid are allowed to be, in general, different from each other.

For the constituent components of the Hamaker constant one explicitly derives
\begin{subequations}\label{Aconst_ocl}
  \begin{eqnarray}
  &&A_{l}(T,\mu)=-C(d,\sigma)J^{l}\rho_{b}^{2}(T,\mu)<0,\label{Al}\\
  &&A_{s_{1},s_{2}}(T)=-C(d,\sigma)J^{s_{1},s_{2}}\rho_{s_{1}}(T)\rho_{s_{2}}(T)<0,\ \ \ \ \ \ \label{As1s2}\\
  &&A_{s_{i},l}(T,\mu)=C(d,\sigma)
  J^{s_{i},l}\rho_{s_{i}}(T)\rho_{b}(T,\mu)>0,\ i=1,2,\ \ \ \ \ \ \ \ \ \ \ \ \ \  \label{Alsi}
\end{eqnarray}
\end{subequations}
where $C(d,\sigma)=G(d,\sigma)/(\sigma-1)$, and for the direct substrate-substrate potential we have assumed, in accordance with Eqs. (\ref{Jldeftext}) and (\ref{Jlsdeftext}), that it is given again by van der Waals type interaction
\begin{equation}\label{Jssdeftext}
J^{s_{1},s_{2}}(\mathbf{r}-\mathbf{r}')=\dfrac{J^{s_{1},s_{2}}}{|\mathbf{r}-\mathbf{r}'|^{d+\sigma}}.
\end{equation}
Let us note that $A_l<0$ and $A_{s_1,s_2}<0$, i.e., the direct substrate-substrate interaction as well as the fluid-fluid interaction lead to attraction between the plates bounding the system while the fluid-substrate interactions are contributing to a {\it repulsion} between them. Furthermore, let us stress that $A_l$ and $A_{s_{i},l},\ i=1,2$, are {\it singular} functions on the temperature near $T_c$ being dependent on the bulk order parameter $\rho_b$, while $A_{s_1,s_2}$ is an analytic function of $T$ near $T=T_c$.

Using Eqs. ({\ref{Al}}) -- (\ref{Alsi}),  for $H_{A}$ one immediately derives from \eref{Ham_l}  that
\begin{eqnarray}\label{Ham_ocl}
  H_{A}(T,\mu&)&=-C(d,\sigma)
  \left\{J^{l}\rho_{b}^{2}(T,\mu)+
  J^{s_{1},s_{2}}\rho_{s_{1}}(T)\rho_{s_{2}}(T)\right.\nonumber\\
  &&-\left.\rho_{b}(T,\mu)\left[J^{s_{1},l}\rho_{s_{1}}(T)+J^{s_{2},l}\rho_{s_{2}}(T)\right]\right\}.
  \end{eqnarray}
For the sake of simplicity in the next lines of this section we are going to omit giving explicitly the arguments of $H_{A}$, $\rho_{b}$ and $\rho_{s_{i}},\ i=1,2$, as they do not change and are of no relevance for the following discussion. Within our model  $A_l$ and $A_{s_i,l}$, $i=1,2$ being a part of the grand canonical potential $\Omega$ have to be proportional to $k_B T$. Obviously, similar statement has to hold for $A_{s_1,s_2}$. Thus, one has that this shall be also true  for the pair potentials $w^{X,Y}$ between any two substances $X$ and $Y$ belonging to the system. According to the general theory of the dispersion forces between molecules \cite{J2011}, neglecting the quantum effects, for the "zero frequency contribution" to $w^{X,Y}(r)$ one indeed has (see Eq. (6.24) on p. 121 in Ref. \cite{J2011})
\begin{equation}
\label{wI}
w^{X,Y}(r)=-\frac{3k_B T}{2(4\pi\varepsilon_0)^2r^6} \alpha_{0X}\alpha_{0Y},
\end{equation}
where $\alpha_{0X}$ and $\alpha_{0Y}$ are the static electronic polarizabilities of each of the two substances $X$ and $Y$.
The last implies that the constants $A_l$, $A_{s_i,l},\ i=1,2$, and $A_{s_1,s_2}$ are not independent. From \eref{wI} one derives that
\begin{equation}\label{reljsl}
J^{s_1,s_2}J^l=J^{s_1,l}J^{s_2,l}.
\end{equation}
Using this equation one can rewrite \eref{Ham_ocl} in the form given in Eq. \eqref{Hamsumfinall} in the main text. There, when calculating the critical component of the force, we often characterize the system by the values of the parameters $\lambda$ and  $s_{i,c},\ i=1,2$ of the parameters $s_{i},\ i=1,2$ at the bulk critical point of the fluid medium $(\beta=\beta_{c}, \Delta\mu=0)$ [see Eqs. (\ref{deltav_s_l_def}), (\ref{l_param_def_new}) and (\ref{s_def_l})]. Thus, it is convenient to express also $H_{A}$ in terms of these variables.   One obtains that
\begin{eqnarray}
&&\beta(\sigma-1)H_{A}=\nonumber\\&&
=-\dfrac{4K}{G(d,\sigma)K_{c}^{2}\lambda}
\prod_{i=1,2}\left[s_{i,c}-\dfrac{1}{4}G(d,\sigma)K_{c}\lambda\phi_{b}\right].\ \ \ \ \ \ \ \ \ \label{bethasigmaminusoneHAl}
\end{eqnarray}
Since $s_{i}=(K/K_{c}) s_{i,c},\ i=1,2$, with $K=\beta J_{\mathrm{sr}}^{l}\ \text{and}\ K_{c}=\beta_{c} J_{\mathrm{sr}}^{l}$ it is clear that $s_{i}$, $i=1,2$ are temperature dependent, while  $s_{i,c},\ i=1,2$ are not.
When $d=\sigma=4$ \eref{bethasigmaminusoneHAl} simplifies to
\begin{eqnarray}
3\beta H_{A}
&=&-\dfrac{32K}{\pi^{2}K_{c}^{2}\lambda}s_{1,c}s_{2,c}\nonumber\\&&+\dfrac{K}{K_{c}}
(s_{1,c}+s_{2,c})\phi_{b}-\dfrac{\pi^{2}}{32}\lambda K\phi_{b}^{2},\ \ \ \ \ \ \ \ \ \ \ \ \label{3bethaHA_ocf}
\end{eqnarray}
where we have taken into account that in mean-filed approximation $\rho_{c}=1/2$. In \eref{3bethaHA_ocf} the term proportional to $\lambda^{-1}$ corresponds to the direct interaction between the confining walls, the second (proportional to $\lambda^{0}$) corresponds to the interaction between the confining walls and the fluid medium, while the last one (proportional to $\lambda$) reflects the interaction between the constituents of the fluid medium. The corresponding result for a $d=\sigma=3$ system is
\begin{eqnarray}
2\beta H_{A}
&=&-\dfrac{6K}{\pi K_{c}^{2}\lambda}s_{1,c}s_{2,c}\nonumber\\&&+\dfrac{K}{K_{c}}
(s_{1,c}+s_{2,c})\phi_{b}-\dfrac{\pi}{6}\lambda K\phi_{b}^{2},\ \ \ \ \ \ \ \ \ \ \ \ \label{2bethaHA_ocf}
\end{eqnarray}
which, taking into account \eref{HamConstant}, means that the Hamaker constant is equal to
\begin{equation}
\beta A_{\mathrm{Ham}}=\pi^{2}\lambda K
\prod_{i=1,2}\left[\dfrac{6s_{i,c}}{\pi\lambda K_{c}}-\phi_{b}\right]. \label{AHam_d3sigma3_l}
\end{equation}
One must pay attention when using Eqs. (\ref{3bethaHA_ocf}) and (\ref{2bethaHA_ocf}), because the substrate-fluid coupling parameters $s_{i,c},\ i=1,2$ depend on the dimensionality $d$ of the system and the decay exponent $\sigma$ [see Eqs. (\ref{deltav_s_l_def}) and (\ref{s_def_l})]. The behaviour of \eref{2bethaHA_ocf} as a function of the temperature and field scaling variables $x_{t}$ and $x_{\mu}$ is depicted in Fig. \ref{fig:ForceInterplay}.


\begin{thebibliography}{168}
\bibitem[{not({\natexlab{a}})}]{note1}
\bibinfo{note}{As it is customary in texts on critical phenomena in the
  remainder of the current text we use the same symbol $\beta$ also to indicate
  the {\it critical exponent} associated with the critical behavior of the
  order parameter; the meaning of the symbol is determined by the context.}
\bibitem[{not({\natexlab{b}})}]{note2}
\bibinfo{note}{Let us note that due to the symmetry of the $(+,+)$ boundary
  conditions it is irrelevant $s_{c}$ near which boundary is negative; in Fig.
  \ref{fig:CFsmsm}(\textbf{a}) and (\textbf{b}) the negative wall-fluid
  coupling parameter is $s_{1,c}$.}
\end{thebibliography}
\end{document}